\documentclass[twocolumn]{aastex62}
\usepackage{amsmath}
\usepackage{graphicx}
\newcommand{\Rs}{$\rm{R}_{\odot}$}

\graphicspath{{./}{figures/}}
\accepted{\today}
\submitjournal{ApJ}
\shorttitle{AWSoM validation}
\shortauthors{Sachdeva et al.}

\begin{document}
\title{Validation of the Alfv\'{e}n Wave Solar atmosphere Model (AWSoM) with Observations from the Low Corona to 1 AU }
\correspondingauthor{Nishtha Sachdeva}
\email{nishthas@umich.edu}

\author{Nishtha Sachdeva}
\affil{Department of Climate and Space Sciences and Engineering, University of Michigan, Ann Arbor, MI 48109, USA }

\author{Bart van der Holst}
\affil{Department of Climate and Space Sciences and Engineering, University of Michigan, Ann Arbor, MI 48109, USA }

\author{Ward B. Manchester}
\affil{Department of Climate and Space Sciences and Engineering, University of Michigan, Ann Arbor, MI 48109, USA }

\author{Gabor T\'{o}th}
\affil{Department of Climate and Space Sciences and Engineering, University of Michigan, Ann Arbor, MI 48109, USA }

\author{Yuxi Chen}
\affil{Department of Climate and Space Sciences and Engineering, University of Michigan, Ann Arbor, MI 48109, USA }

\author{Diego G. Lloveras}
\affiliation{Instituto de Astronom\'{i}a y F\'{i}sica del Espacio, CONICET-University of Buenos Aires, Ciudad de Buenos Aires,  CC 67-Suc 28, Argentina}

\author{Alberto M. Vasquez}
\affil{Instituto de Astronom\'{i}a y F\'{i}sica del Espacio, CONICET-University of Buenos Aires, Ciudad de Buenos Aires,  CC 67-Suc 28, Argentina}
\affil{Departamento de Ciencia y Tecnolog\'{i}a, Universidad Nacional de Tres de Febrero, Buenos Aires, Argentina}

\author{Philippe Lamy}
\affil{Laboratoire Atmosph\`eres, Milieux et Observations Spatiales, CNRS
\& UVSQY, Guyancourt, France}

\author{Julien Wojak}
\affiliation{Laboratoire d'Astrophysique de Marseille, UMR 7326, CNRS \& Aix-Marseille Universit\'{e}, Marseille France}

\author{Bernard V. Jackson}
\affiliation{Center for Astrophysics and Space Sciences, University of California at San Diego, La Jolla, CA, USA}

\author{Hsiu-Shan Yu}
\affiliation{Center for Astrophysics and Space Sciences, University of California at San Diego, La Jolla, CA, USA}

\author{Carl J. Henney}
\affiliation{AFRL/Space Vehicles Directorate, Kirtland AFB, NM, USA}
%\affil{Institute of Scientific Research, Boston College, Chestnut Hill, MA, USA}
%\affil{National Solar Observatory, Tucson, AZ, USA}

\begin{abstract}
We perform a validation study of the latest version of the Alfv\'{e}n Wave Solar atmosphere Model (AWSoM) within the Space Weather Modeling Framework (SWMF). To do so, we compare the simulation results of the model with a comprehensive suite of observations for Carrington rotations representative of the solar minimum conditions extending from the solar corona to the heliosphere up to the Earth. In the low corona ($r < 1.25$ \Rs), we compare with EUV images from both STEREO-A/EUVI and SDO/AIA and to three-dimensional (3-D) tomographic reconstructions of the electron temperature and density based on these same data. We also compare the model to tomographic reconstructions of the electron density from SOHO/LASCO observations ($2.55 < r < 6.0$\Rs). In the heliosphere, we compare model predictions of solar wind speed with velocity reconstructions from InterPlanetary Scintillation (IPS) observations. For comparison with observations near the Earth, we use OMNI data. Our results show that the improved AWSoM model performs well in quantitative agreement with the observations between the inner corona and 1 AU. %In the lower corona, the model and the tomographic reconstructions match with a $20\%$ accuracy.
The model now reproduces the fast solar wind speed in the polar regions. Near the Earth, our model shows good agreement with observations of solar wind velocity, proton temperature and density. AWSoM offers an extensive application to study the solar corona and larger heliosphere in concert with current and future solar missions as well as being well suited for space weather predictions.
\end{abstract}


\section{Introduction}\label{sec:Intro}
Predicting space weather events and their geomagnetic effects requires accurate physics-based modeling of the solar atmosphere, extending from the upper chromosphere, into the corona and including the heliosphere. In the last few decades, extensive resources have been used to develop both analytic and numerical modeling techniques \citep{Mik1999,Gro2000, Rou2003,Coh2007,Fen2011,Eva2012}. In addition, a wealth of observational data are now available (Air Force Data Assimilation Photospheric flux Transport - Global Oscillation Network Group (ADAPT-GONG), \citealp{Arg2010}; Solar Dynamics Observatory (SDO)/Atmospheric Imaging Assembly (AIA), \citealp{Lem2012}; Solar-Terrestrial Relations Observatory (STEREO), \citealp{How2008}; Solar and Heliospheric Observatory (SOHO)/Large Angle and Spectrometric COronagraph (LASCO), \citealp{Bru1995}; IPS, \citealp{Jac1998}) to both drive and validate these models. The state-of-the-art three-dimensional (3D) extended MHD models that have been developed, improved and validated with observations over time provide a comprehensive understanding of the coronal structure, heating and solar wind acceleration in the context of a fluid description. 

%(highly debatable since the cone model applied to ENLIL begins at 30 Rs) 
Modern global models incorporate Alfv\'{e}n wave turbulence, a physical mechanism for which the measurements of the {\it Mariner 2,4,5} spacecraft established firm evidence of occurrence in the solar wind and the heliosphere \citep{Col1968,Bel1971}. Based on this discovery, one-dimensional (1D) models incorporating Alfv\'{e}n waves were developed \citep{Bel1971, Ala1971}, followed by two-dimensional models for the solar corona \citep{Bra1997, Rud1998, Usm2000}. The interaction between forward-propagation and reflected Alfv\'{e}n waves, leading to a non-linear turbulent cascade and hence, coronal heating were first discussed in models described by \citet{Vel1989, Zan1996, Mat1999, Suz2006, Ver2007, Cra2010, Cha2011, Mat2012}. Recently, 3D models simulating the solar corona have been developed \citep{Lio2009, Dow2010, Van2010}. 

Another aspect vital to coronal modeling is energy partitioning among particle species. It is now known that the electron and ion temperatures are quite different beyond 2 \Rs, as the plasma becomes collisionless \citep{Har1968}. The simplest description is a single fluid approach with separate temperatures for electrons and protons, which was developed by \citet{Tu1995, Lai2003, Vai2003} by including Alfv\'{e}n waves accounting for the heating and acceleration of the solar wind plasma. Using remote observations from Ultraviolet Coronagraph Spectrometer (UVCS), \citet{Koh1998} and \citet{Li1998} showed the proton temperature anisotropy in the coronal holes. The perpendicular (to the local magnetic field direction) ion temperature was found to be much larger than the parallel ion temperature in the solar corona (SC) as well as the inner heliosphere (IH), as seen in Helios observations \citep{Mar1982}. This temperature anisotropy appeared in various 1D numerical models, for example, \citealp{Lee1972,Cha2011} as well as in 2D models, \citet{Vas2003, Li2004}.

Our coronal and solar wind model, the Alfv\'{e}n Wave Solar atmosphere Model (AWSoM) is a component within the Space weather Modeling framework (SWMF; \citealp{Tot2012}) and follows similar lines of development to provide a self-consistent physics-based global description of coronal heating and solar wind acceleration \citep{Sok2013, Van2014}. AWSoM inherits many aspects of the model of \citet{Van2010}, including a description of low-frequency forward and counter-propagating Alfv\'{e}n waves that non-linearly interact resulting in a turbulent cascade and dissipative heating. In addition, there are separate temperatures for electrons and protons with collisional heat conduction applied only to electrons and radiative losses based on the Chianti model \citep{Der1997}. AWSoM is significantly advanced by extending the model to the base of the transition region and balanced turbulence \citep{Sok2013}. Later model advances \citep{Van2014, Men2015} include a self-consistent treatment of Alfv\'{e}n wave reflection and a stochastic heating model by \citet{Cha2011} as well as a description of proton parallel and perpendicular temperatures and kinetic instabilities based on temperature anisotropy and plasma beta. 

%\citet{Eva2012} enhanced the model by including surface Alfv\'{e}n waves. (This aspect is controversial ans was not retained in AWSoM) 

%%%%%%%% GONG and ADAPT MAPS
\begin{figure*}[ht!]
\gridline{\fig{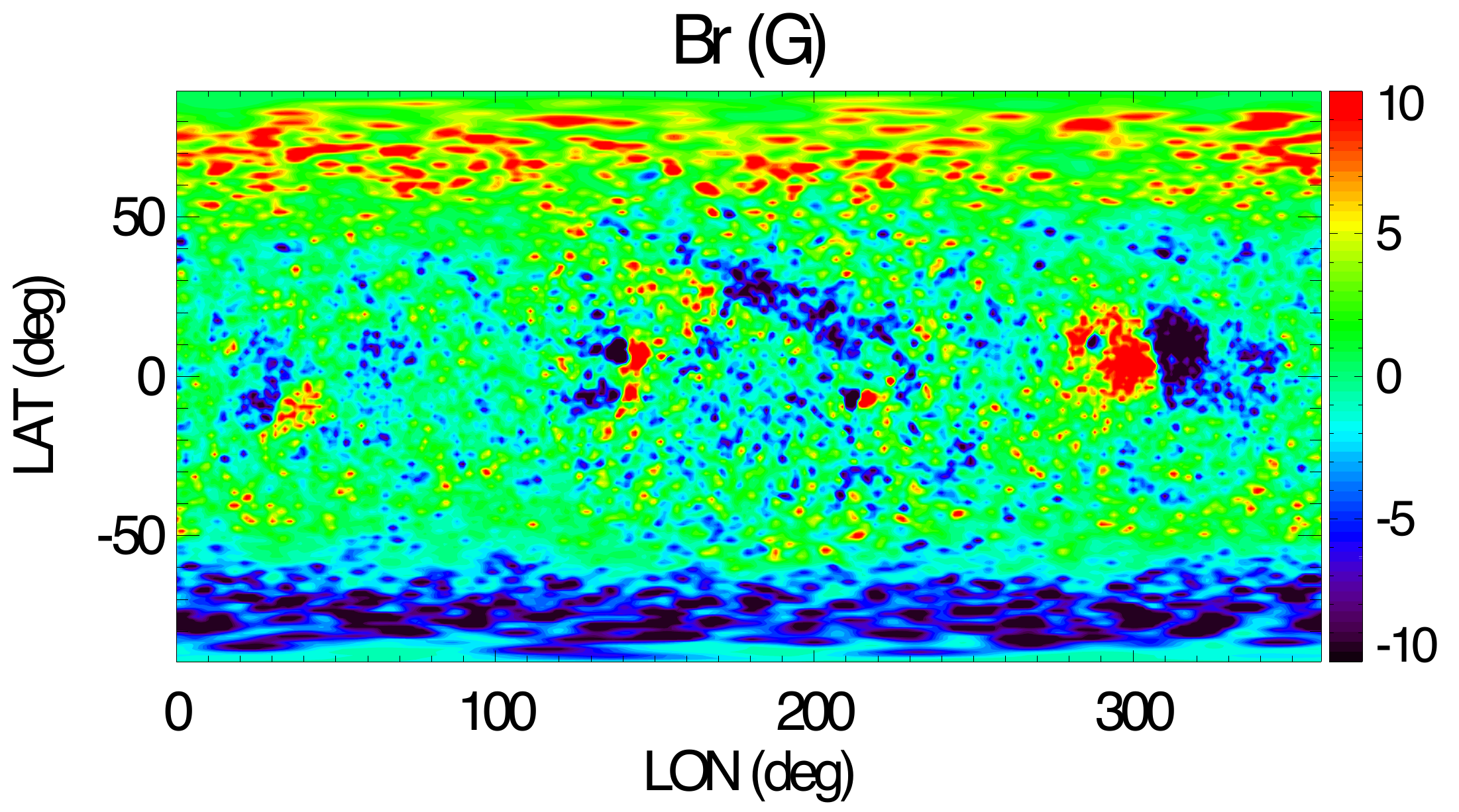}{0.34\textheight}{(a) ADAPT-GONG map}
           \fig{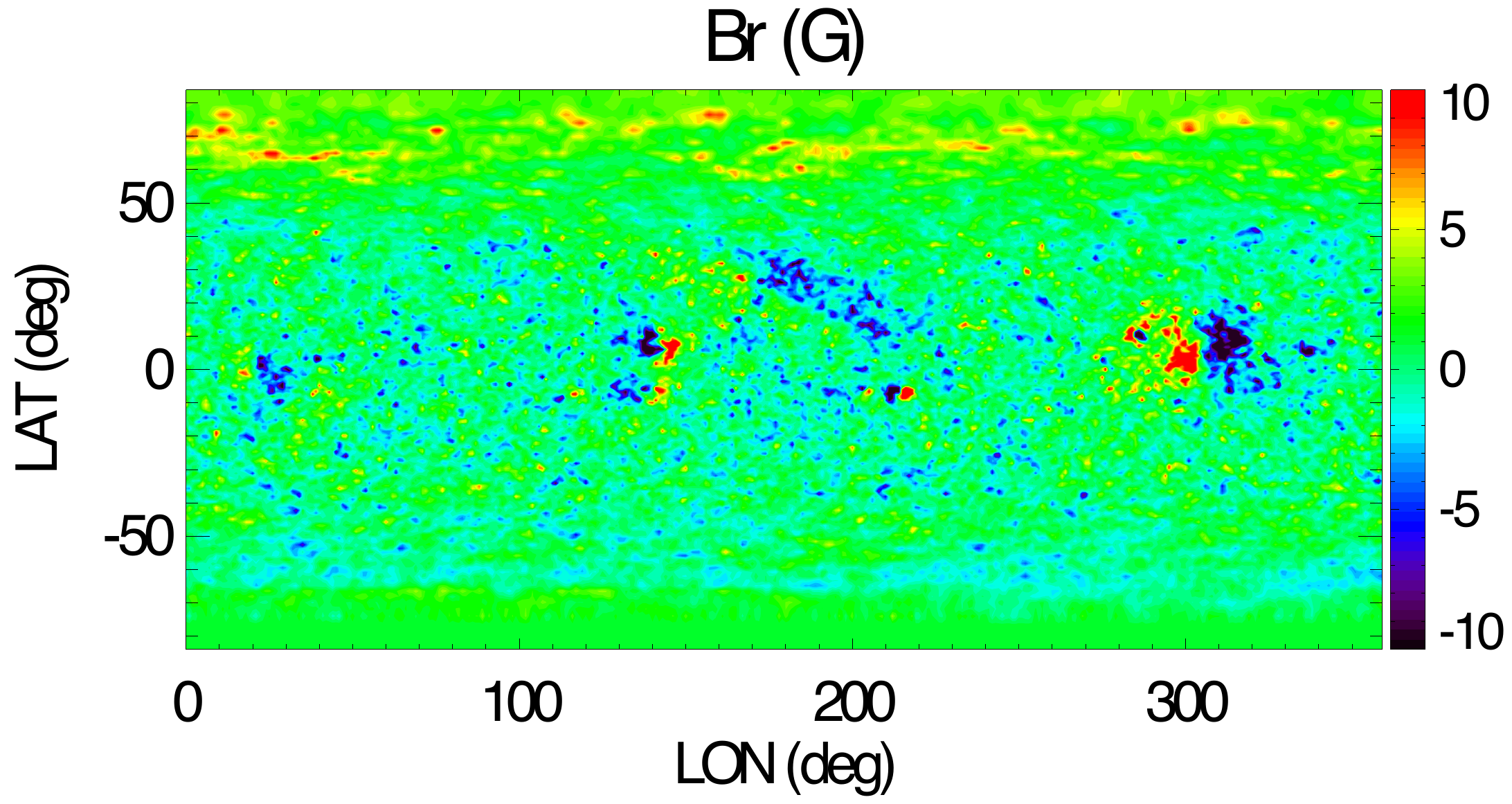}{0.36\textheight}{(b) GONG map}
         }
\caption{The radial magnetic fields for CR2208 using one realization of the ADAPT-GONG ensemble of synchronic maps (left) and the GONG synoptic map (right) provided by National Solar Observatory (NSO). The magnetic fields in this plot are saturated at $\pm$ 10 G.\label{fig:two_maps}}
\end{figure*}

AWSoM is a data-driven model capable of simulating the detailed 3D structure of the corona with boundary conditions supplied by GONG or Helioseismic and Magnetic Imager (HMI) synoptic magnetic maps. These data, combined with the physical processes of wave dissipation, heat conduction and radiative cooling, give AWSoM the capability of capturing the temperature and mass density structure of the corona. As a result, synthetic EUV images can be made with AWSoM, which reproduce multi-wavelength observations including features such as coronal hole morphology and active region brightness \citep{Sok2013, Van2014}, similar to those first produced by \citet{Dow2010}. The model results have been compared to {\it in situ} observations from ACE, Wind and STEREO data at 1 AU \citep{Men2015, Van2019} and to observations from {\it Ulysses} \citep{Ora2013, Jia2016}.
In addition to steady state conditions, our solar wind models have been applied to study coronal mass ejection (CMEs). \citet{Man2012, Jin2013} applied the model of \cite{Van2010} to show that the two-temperature model accurately reproduced the CME shock structure without unphysical heat precursors ahead of CMEs, which can appear due electron heat conduction  applied to ions. \citet{Man2014} and \citet{Jin2017} also simulated observed fast CME events with the Gibson-Low (GL) flux rope model \citep{Gib1998} and demonstrate the ability to reproduce many observed features near the Sun and at 1 AU by comparing with observations from SDO, SOHO, and STEREO A/B.

In this paper, we follow the work of \citet{Jin2012} and perform a comprehensive validation of the coronal model.  We describe the SC-IH simulation results for solar minimum conditions using the latest version of the AWSoM model within the SWMF. The input is obtained from ADAPT-GONG global magnetic maps for Carrington rotations, CR2208 (2018-09-02 to 2018-09-29) and CR2209 (2018-09-29 to 2018-10-26). We compare the model predicted results with an extensive suite of observations ranging from near the Sun up to 1 AU. The observations include STEREO-A EUVI and AIA images, tomographic reconstructions of electron density and temperatures from AIA data between 1.025 and 1.225\,\Rs~and reconstruction of the electron density from LASCO-C2 data between 2.55 to 6\,\Rs. We also include model comparisons with InterPlanetary Scintillation (IPS) data at 20\,\Rs~, 100\,\Rs~and 1 AU. Finally, comparisons with OMNI data at 1 AU are shown.
The paper is organized as follows, Section \ref{sec:Model} details the AWSoM model characteristics, input global photospheric magnetic field maps and simulation parameters. In Section \ref{sec:Results}, we validate the results of the solar wind model for CR2208 and CR2209 with observations. We conclude with a summary and discussion in Section \ref{sec:Summary}.

\section{Computational Model and Simulation}\label{sec:Model}
\subsection{Alfv\'{e}n Wave Solar atmosphere Model (AWSoM) description}\label{sec:Awsom}
We describe here the main characteristics of the 3D global MHD Alfv\'{e}n Wave Solar atmosphere Model (AWSoM) model included within the Space Weather Modeling Framework (SWMF; \citealp{Tot2012}). This model uses the Block-Adaptive-Tree-Solarwind-Roe-Upwind-Scheme (BATS-R-US; \citealp{Pow1999}) numerical scheme to solve the MHD equations. AWSoM extends from the upper chromosphere, through the transition region, into the solar corona (SC) and the inner heliosphere (IH; up to 1 AU and beyond). 
%Due to high density and frequent collisions between electrons and protons near the Sun, the temperatures for both species equilibrate. Farther out, as the medium rarifies and becomes collisionless, the electron and proton temperatures decouple. Different perpendicular and parallel ion temperatures have been observed via remote sensing near the Sun \citep{Koh1998,Li1998} as well as in the IH \citep{Mar1982}. 

%%%%%%% USED INPUT MAPS ADAPT-GONG
\begin{figure*}[ht!]
\gridline{\fig{CR2208_adapt05_new.pdf}{0.35\textheight}{(a) CR2208}
           \fig{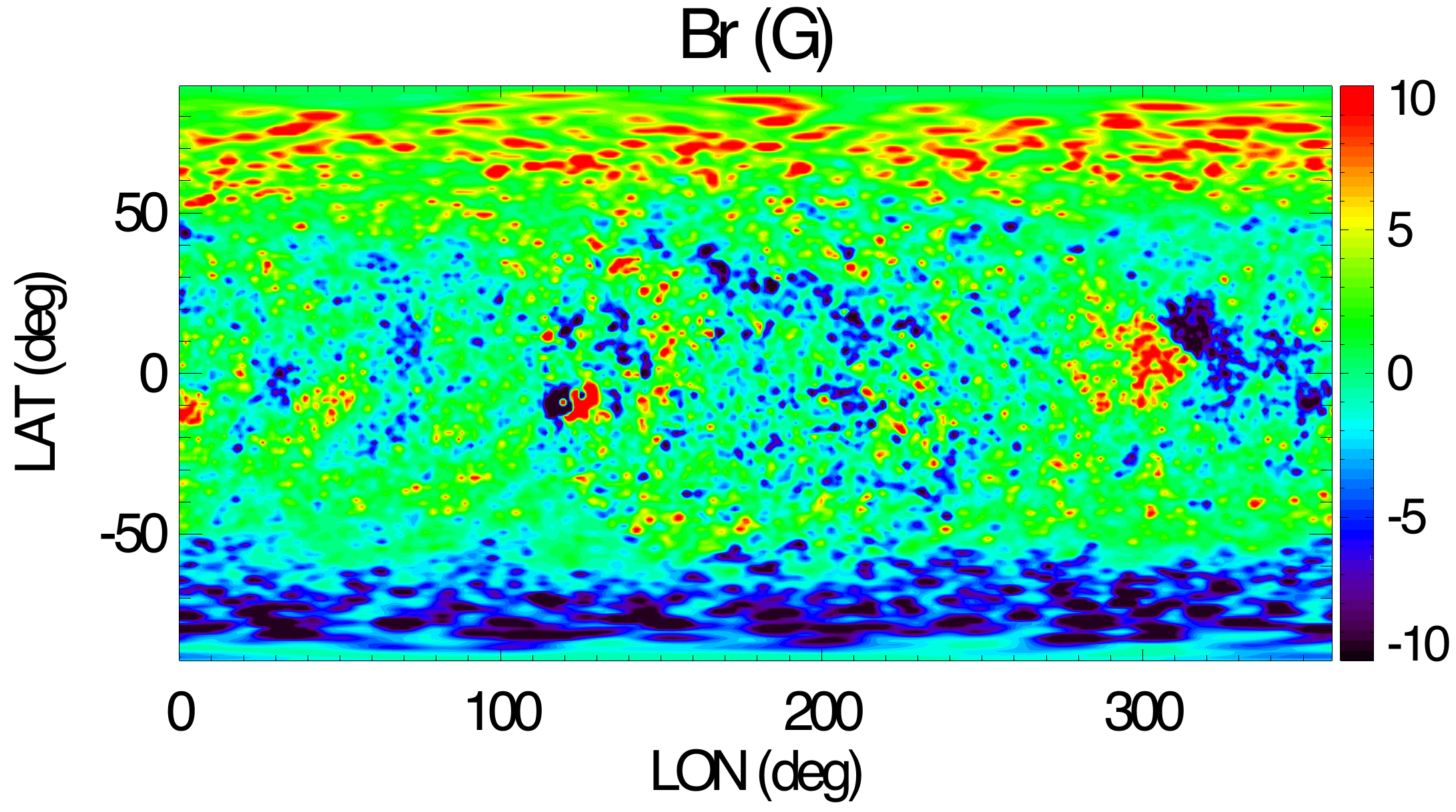}{0.35\textheight}{(b) CR2209}
         }
\caption{The input radial magnetic field maps for CR2208 (left) and CR2209 (right) using ADAPT-GONG global maps. The magnetic fields in this plot are saturated at $\pm$ 10 G.\label{fig:maps}}
\end{figure*}
AWSoM includes isotropic electron temperature as well as anisotropic (distinct perpendicular and parallel) proton temperatures. It addresses the coronal heating and solar wind acceleration with low-frequency Alfv\'{e}n wave turbulence. The wave pressure gradient accelerates the plasma and wave dissipation heats it. The model includes non-linear interaction between outward propagating and counter-propagating (reflected) Alfv\'{e}n waves that gives rise to a transverse turbulent cascade from the outer scale to smaller perpendicular scales where dissipation and coronal heating takes place. To distribute the coronal heating among three temperatures, AWSoM uses the physics-based theories of linear wave damping and stochastic heating. At the proton gyro-radius scale the kinetic Alfv\'{e}n wave turbulence has a range of parallel wave numbers, but for the damping rates we need to assign a single wave number. This wave number is determined by the critical balance condition in which we set the Alfv\'en wave frequency equal to the inverse of the cascade time of the minor wave \citep{Lit2007}. This is an improvement with respect to the energy partitioning used in \citep{Cha2011, Van2014}, where the cascade time of the major wave was used. This change leads to more electron heating and less solar wind acceleration, resulting in significantly improved model-data comparisons. Details of the changes in the energy partitioning will be reported in \citet{Van2019b}. No ad hoc heating functions are used. The model also includes the electron heat conduction both for the collisonal and collisonless regimes. MHD equations included in the AWSoM model are described in detail in \citet{Van2014}.

\subsection{Input Global Magnetic Maps}\label{sec:Mag}
The primary data input to solar MHD models is the synoptic magnetogram which provides estimates of the photospheric magnetic field of the Sun. These synoptic maps are essential for modeling the solar corona and the solar wind accurately for the purpose of prediction. Therefore, it is important that the magnetic field estimates of the Sun are reliable. The Global Oscillation Network Group (GONG) provides such standard synoptic magnetograms. These are full disk surface maps of the radial component of the photospheric magnetic field. 
To create a synoptic map, first the full disk line-of-sight images are merged and mapped to heliographic coordinates. It is assumed that the photospheric magnetic field is radial and that the Sun rotates as a solid body with a 27.27 days rotation rate. The remapped images are then merged together for a Carrington rotation with parts of the overlapping coordinates merged. In addition, as the polar fields are not well observed from the ecliptic, the processing in GONG maps estimates them by polynomial fits to the observed fields from neighbouring latitudes leading to uncertainties. These uncertainties in the polar magnetic flux distribution propagate into the solar wind simulations in the coronal models \citep{Ber2014}.

\citet{Wor2000} developed a model to create synchronic synoptic maps which evolve the magnetic flux on the Sun based on super-granulation, diffusion, differential rotation, meridional circulation, flux-emergence and data merging. These processes are used in the model to provide missing data where observations are not available. The Air Force Data Assimilation Photospheric Flux Transport (ADAPT; \citealp{Arg2010,Arg2013,Hen2012}) model incorporates this \citet{Wor2000} model and the Los Alamos National Lab (LANL) data assimilation code \citep{Hic2015} to create synchronic maps based on observations and dynamic physical processes. The data assimilation technique produces multiple realizations of the magnetic field maps to account for different parameters and their uncertainties in the photospheric flux-transport model. ADAPT maps using observations from different instruments are available at \url{https://www.nso.edu/data/nisp-data/adapt-maps/}. 
%The 12 realizations evolve smoothly over time, independent of each other, without abrupt changes. The changes for any given realization (from one rotation to another) are driven smoothly by different supergranulation flow patterns.

Figure \ref{fig:two_maps} shows the ADAPT-GONG and GONG global maps for CR2208. The two maps show significant differences, especially in the polar regions. We find that using ADAPT-GONG maps as input to the AWSoM model produces significantly better results in comparison to using GONG maps. Therefore, in this work we use ADAPT-GONG global magnetic maps for both CR2208 and CR2209. These are shown in Figure \ref{fig:maps}.
%%%%%%%%%%% EUV AND AIA COMPARISON

\begin{figure*}
\gridline{\fig{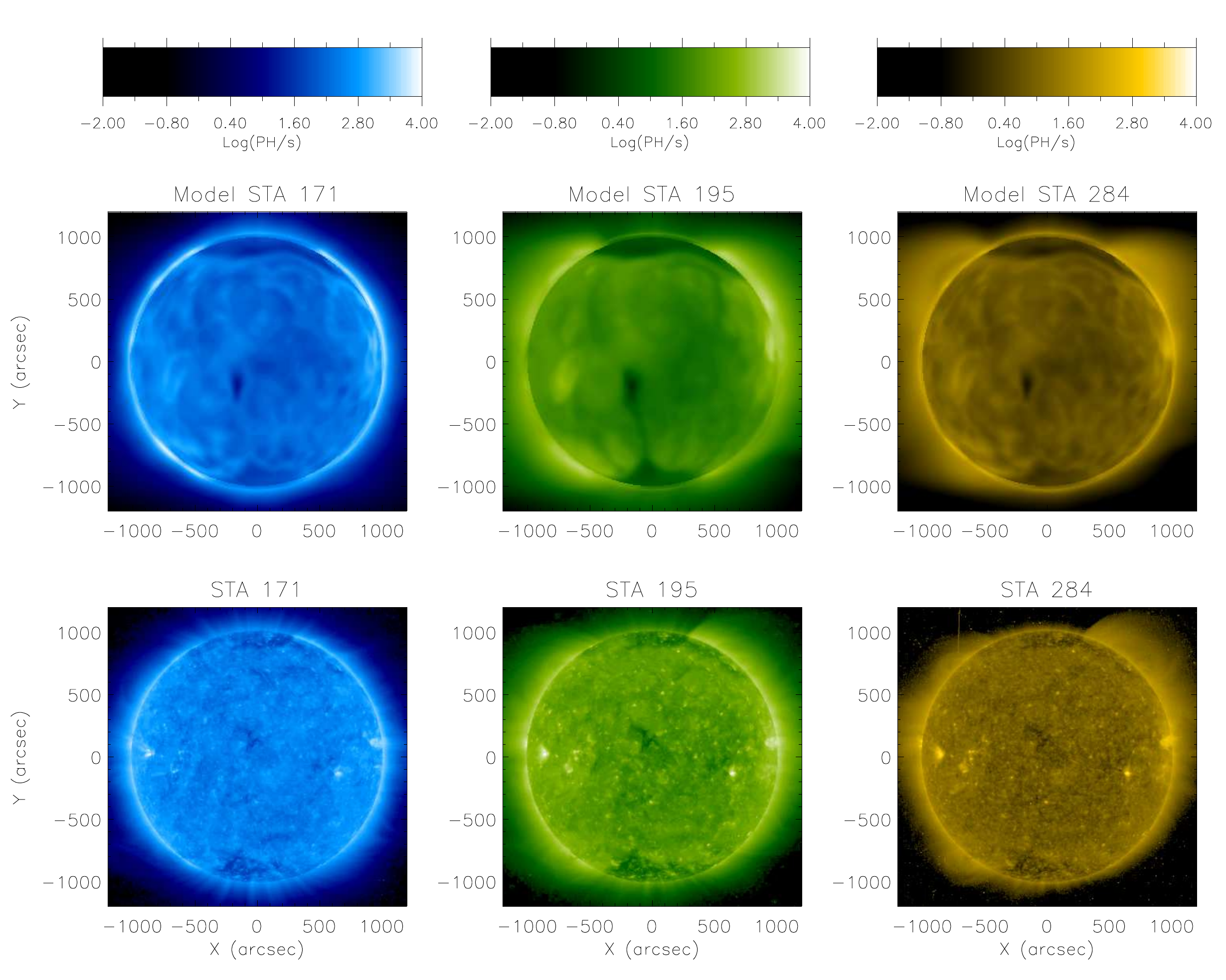}
{0.5\textwidth}{}
          \fig{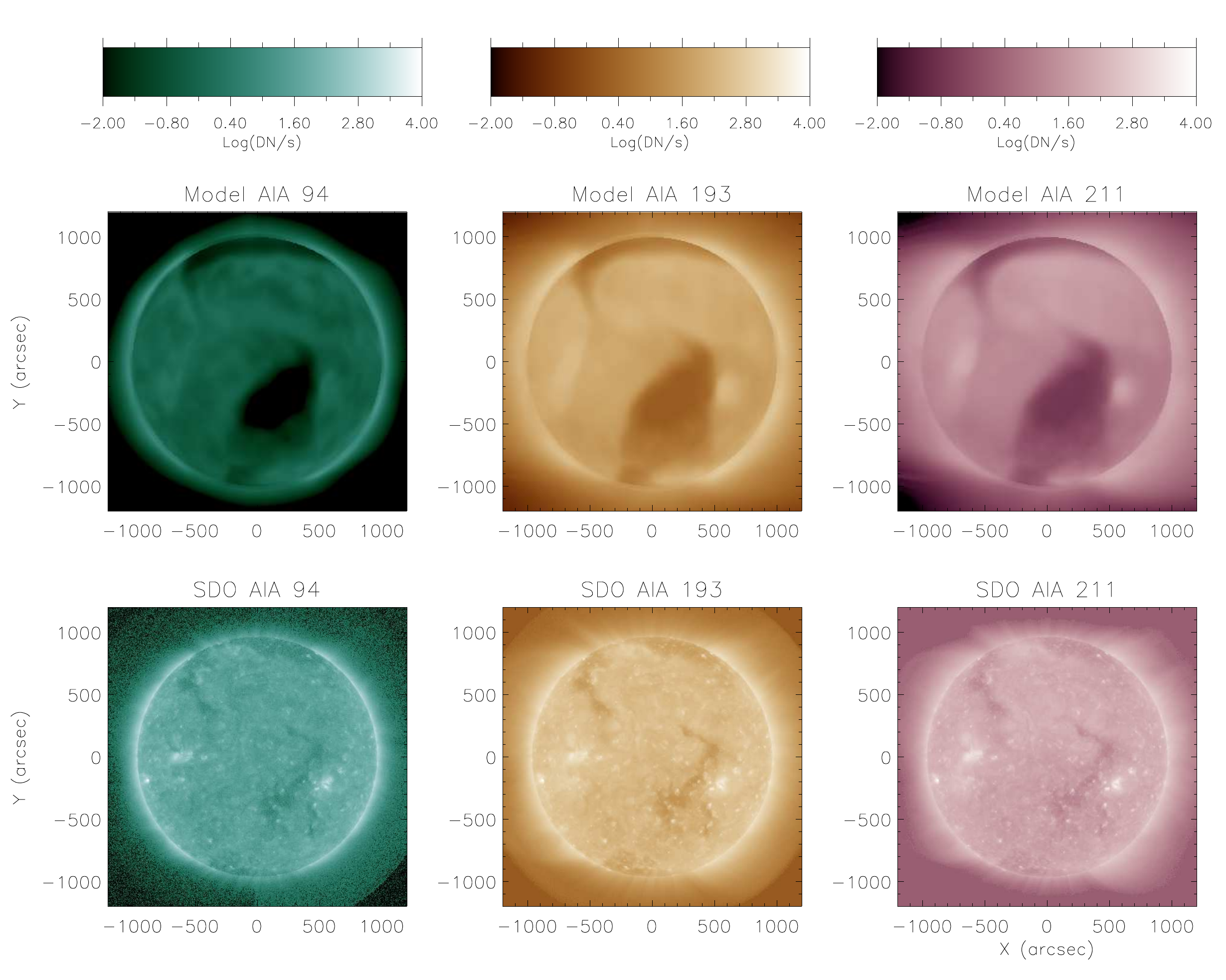}{0.5\textwidth}{}
          }
          \caption{Comparison of model synthesized LOS EUV images with STEREO-A/EUVI (left) for 171, 195 and 284\,\AA~and SDO/AIA extreme ultraviolet images (right) for 94, 193 and 211\,\AA~for CR2208. The top panels show the LOS images from the AWSoM model and the bottom panels show the observations. \label{fig:euv08}}
\end{figure*}

\begin{figure*}
\gridline{\fig{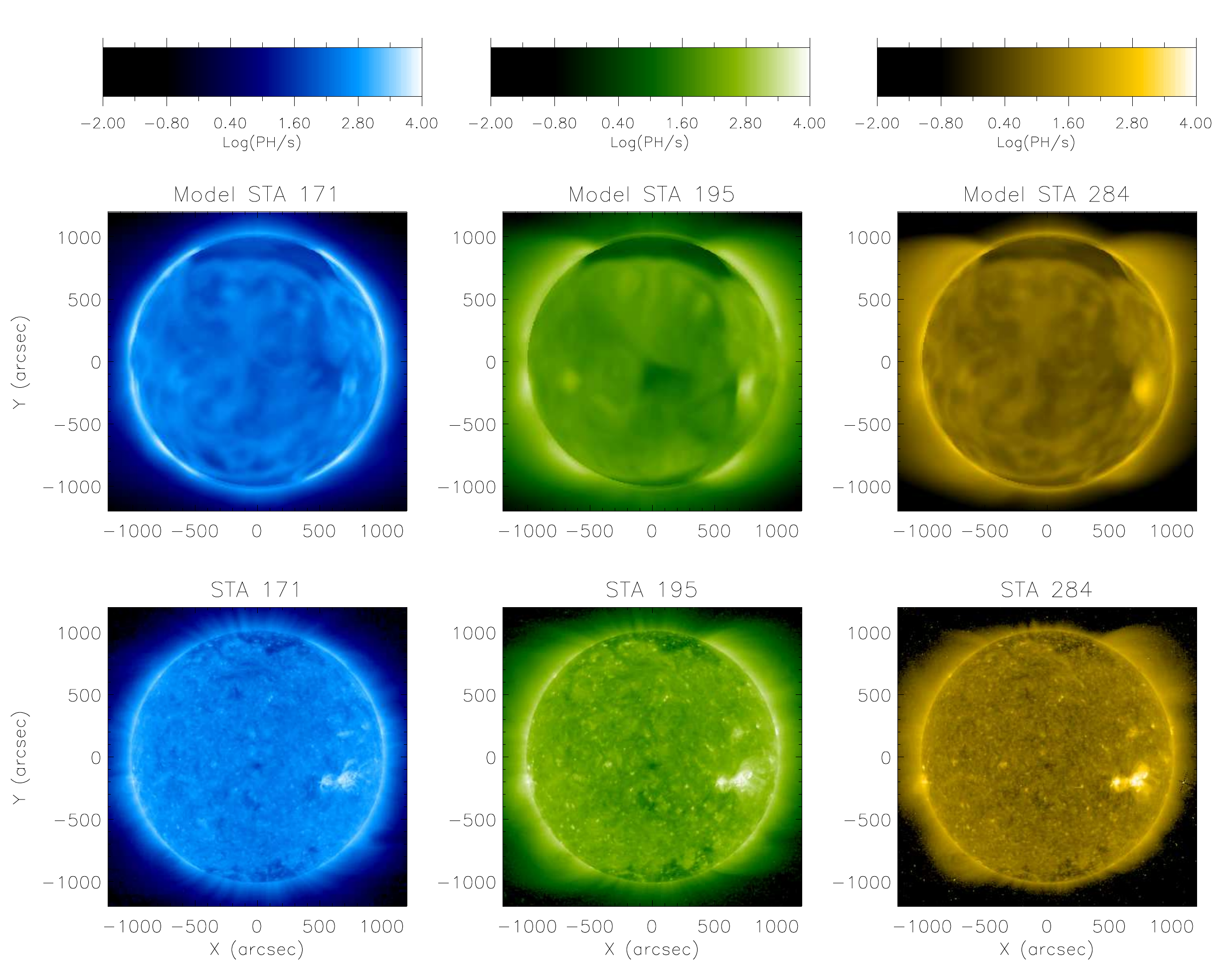}
{0.5\textwidth}{}
          \fig{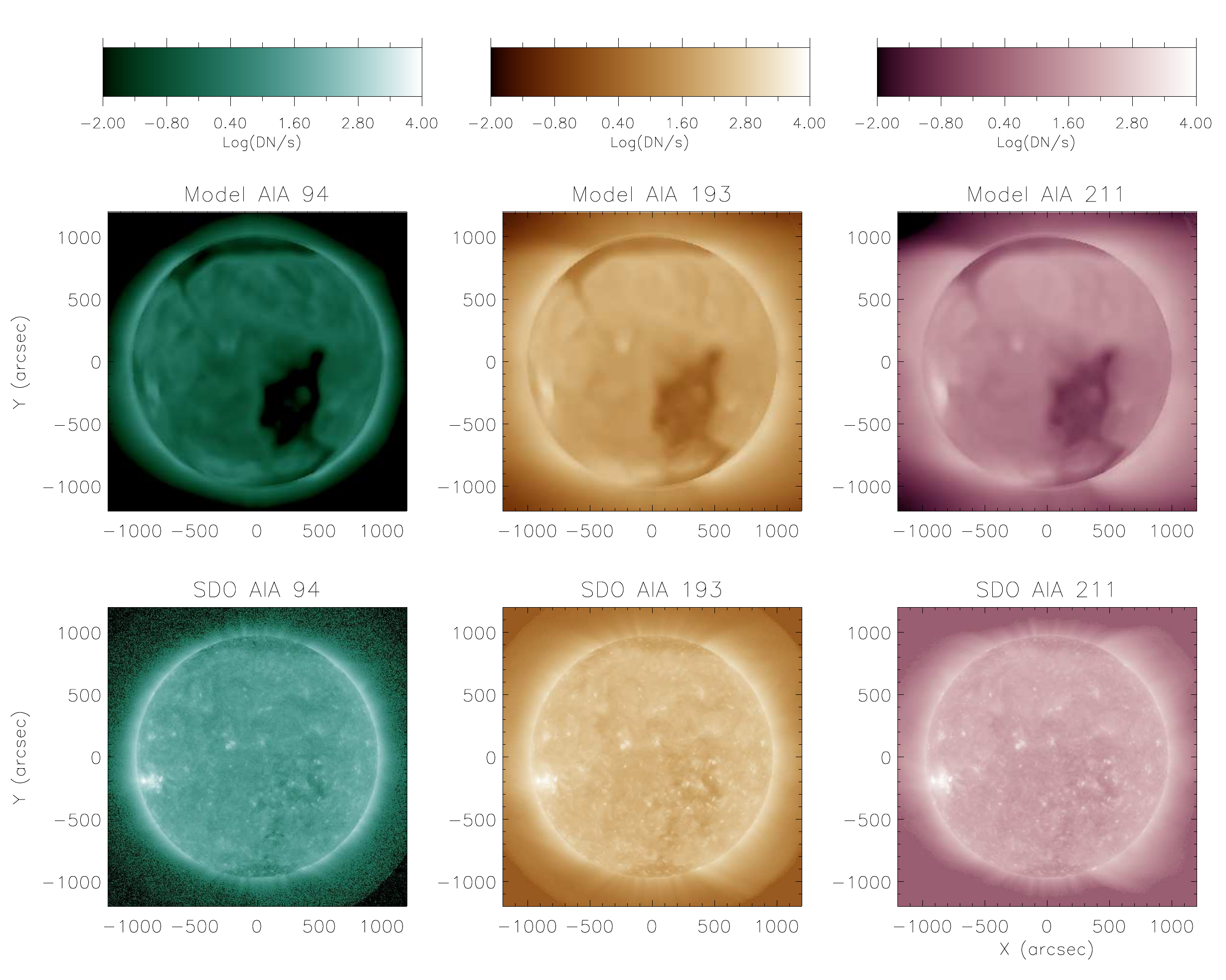}{0.5\textwidth}{}
          }
          \caption{Comparison of model synthesized LOS EUV images with STEREO-A/EUVI (left) for 171, 195 and 284\,\AA~and SDO/AIA extreme ultraviolet images (right) for 94, 193 and 211\,\AA~for CR2209. The top panels show the LOS images from the AWSoM model and the bottom panels show the observations. \label{fig:euv09}}
\end{figure*}

\subsection{Simulation Parameters and Setup}\label{sec:Sim}
In this section, we set up the solar wind model. The SWMF facilitates the simultaneous execution and coupling of different components of the space environment covering various physics models. Besides space weather applications for the Sun-Earth system, the SWMF has been used for many planetary, comet and moon applications \citep{Tot2005}. Tools for SWMF and numerical schemes of BATS-R-US MHD solver are described in \citet{Tot2012}. We use the Solar Corona (SC) and Inner Heliosphere (IH) components of SWMF in this paper. The SC model uses a 3D spherical grid and the IH model uses a Cartesian grid, with an overlapping buffer grid which couples the solutions from SC over to IH. 
The computational domain for SC model lies within the radial coordinate ranging from 1\,\Rs~to 24\,\Rs~using a radially stretched grid and the z-axis aligned with the rotation axis. The stretched grid, with a radial resolution of 0.001\,\Rs~close to the Sun provides a high numerical resolution for the steep density gradients in the upper transition region. The Adaptive Mesh Refinement (AMR) for SC, between 1.0\,\Rs~and 1.7\,\Rs~refines the angular cell size to $1.4^{\circ}$. Outside this radial range, the grid is one level coarser, with an angular resolution of 2.8$^{\circ}$. The MHD equations described in \citet{Van2014} are solved in the heliographic rotating (HGR) frame including contributions from the Coriolis and centrifugal forces. The heliospheric current sheet (HCS) is resolved with two extra levels of refinement with $1.4^{\circ}$ cell size in the longitude and latitude directions. We decompose the SC domain into $6 \times 8 \times 8$ grid blocks. The number of cells used in the SC component is of the order of 3 million and local time stepping is used for speeding up the convergence of the simulation to a steady state solar wind solution. 

The initial as well as the boundary condition for the magnetic field is specified by the synchronic ADAPT-GONG maps provided by NSO. We use a Potential Field Source Surface Model (PFSSM) to extrapolate the 3D magnetic field (from the 2D photospheric magnetic field maps), which we represent as spherical harmonics. The source surface is taken to be at $r=2.5$\,\Rs. Beyond the source surface the magnetic field is purely radial. The initial condition is specified by all the components of the magnetic field while the radial component of the magnetic field specifies the boundary condition. At the inner boundary, the radial component of the magnetic field is held fixed (according to the PFSSM solution) and the latitudinal and longitudinal components of the magnetic field are allowed to adjust freely in response to the interior dynamics.

The inner boundary of the model is at the base of the transition region ($\approx$1.0\,\Rs) which is artificially broadened to obtain higher resolution near the Sun \citep{Lio2009,Sok2013}. The density at the inner boundary is taken to be an overestimate, $N_{e}=N_{i}=N_{\odot}=2\times10^{17}$\,m$^{-3}$ corresponding to the isotropized temperature values, $T_{e}=T_{i}=T_{i\parallel}=T_{\odot}=50,000$\,K. 
This ensures that the base is not affected by chromospheric evaporation and the upper chromosphere extends for the density to fall rapidly to correct (lower) values \citep{Lio2009}. To account for the energy partitioning between electrons and protons, the stochastic heating exponent and amplitude are set to 0.21 and 0.18 respectively \citep{Cha2011}.
The Poynting flux of the outgoing wave sets the empirical boundary condition for the Alfv\'{e}n wave energy density (\rm{w}). As, $S_{A} \propto V_{A}\,\rm{w} \propto B_{\odot}$, the proportionality constant is estimated as, $(\frac{S_{A}}{B})_{\odot}=1.0\times 10^{6}$\,Wm$^{-2}$T$^{-1}$, where, S$_{A}$ is the Poynting flux, V$_{A}$ is the Alfv\'{e}n wave velocity and $B_{\odot}$ is the field strength at the inner boundary \citep{Sok2013}. The correlation length (L$_{\perp}$) of the Alfv\'{e}n waves (transverse to the magnetic field direction) is proportional to B$^{-1/2}$. The proportionality constant, L$_{\perp}\sqrt{B}$ is an adjustable input parameter in the model and is set to $1.5\times10^{5}$ m\,$\sqrt{T}$.
To synthesize high resolution line of sight (LOS) EUV images from the model, we use the fifth order numerical scheme with MP5 limiter \citep{Sur1997,Che2016} within 1.5\,\Rs, and the standard second-order shock-capturing schemes in the remainder of the SC region \citep{Tot2012}.
%Grid blocks used for fifth order scheme remain the same ($6\times8\times8$), for resolving in the longitude and latitude directions.

The computational domain for the IH component is a cube surrounding the spherical domain of SC extending 
%with a grid decomposition of $8\times 8\times 8$. The box range is given by,
-250\,\Rs~$\leq (x,y,z) \leq $~250\,\Rs. The adaptive Cartesian grid ranges from cell size less than 0.5\,\Rs~to $\approx$ 8\,\Rs. Total number of cells for the IH component are of the order of 8 million.

The SC component runs for 60000 stpdf to reach a steady state. The SC and IH components are then coupled once. Following which, SC is switched off and IH runs for 5000 stpdf until it converges.
In this paper, we show simulation results for Carrington rotations CR2208 and CR2209. The two Carrington rotations represent the near solar minimum conditions during the end of the decaying phase of solar cycle 24, close to the beginning of solar cycle 25. The ADAPT-GONG global magnetic maps used as input for these rotations are shown in Figure \ref{fig:maps}.
%We use spherical harmonics to extrapolate the radial magnetic field component from the global magnetic field maps to a 3D Potential Field Source Surface (PFSS). The source surface is taken to be at $r=2.5$\,\Rs. This PFSS solution is used for both the initial and boundary conditions. 
The following section describes the results of the SC-IH simulations when compared with an extensive set of observations ranging from the lower corona up to 1 AU.

\section{Comparisons with observations}\label{sec:Results}
In this section, we present the results of steady-state solar wind simulations for both CR2208 and CR2209 representing solar minimum conditions. The results shown here are using one of the 12 realizations of the ADAPT-GONG maps. We compare the steady state AWSoM model simulations with observations at various radial distance ranges. Beginning from close to the Sun (Extreme Ultraviolet images, STEREO-A/EUVI and SDO/AIA), followed by tomographic reconstructions of plasma parameters using AIA (Atmospheric Imaging Assembly) data, SOHO/LASCO-C2 data, and InterPlanetary Scintillation (IPS) data. Finally, we compare the model results with 1 AU observations (OMNI data).

\subsection{Extreme UltraViolet Images (EUVI)}\label{sec:EUVI}
%For validation of the steady state solution using AWSoM model, we compare the results with observations near the Sun. 
The model simulated electron density and temperature are used to synthesize extreme ultraviolet (EUV) line-of-sight (LOS) images. These are compared to the multi-wavelength EUV observations from STEREO-A/EUVI and SDO/AIA. Figures \ref{fig:euv08} and \ref{fig:euv09} show these comparisons for CR2208 and CR2209 respectively. Synthetic images are shown corresponding to STEREO-A/EUVI, 171\,\AA, 195\,\AA~and 284\,\AA~bands and SDO/AIA, 94\,\AA, 193\,\AA~and 211\,\AA~bands, corresponding to Fe emission lines. The observation time for CR2208 is $\approx 22:00:00\,$UT on 2018 September, 15 and for CR2209 it is $\approx 06:00:00\,$ UT on 2018 October, 13. These times coincide with the central meridian times of the ADAPT-GONG map used for the respective simulations. No STEREO-B images are available for comparison, as the spacecraft ceased to operate before these rotations.

For each rotation, the top row shows the model simulated LOS EUV image while the bottom row shows the observation. The corresponding wavelengths are indicated at the top of each panel. As mentioned before, this model accounts for the partial reflection of outward propagating waves and their interaction with the counter-propagating (reflected) waves. This leads to turbulent cascade dissipation and hence, coronal heating. As a result, in regions of strong magnetic fields, such as, active regions, stronger reflection and therefore, more dissipation occurs, which results in an intensified EUV emission. 

The LOS images are produced under the assumption that for all wavelengths considered here, the plasma is optically thin. In general, there tends to be a dominant stray light component in EUV images caused by long-range scatter. \citet{She2012} showed that 70$\%$ of the emission in coronal holes on the solar disk is made up of this stray light in EUVI. The STEREO-A/EUVI observations shown in Figures \ref{fig:euv08} and \ref{fig:euv09} are stray-light corrected. We see the extended coronal hole in the north reproduced in the model results. The narrow southern coronal hole is also visible in the model simulations in all wavelengths in the STEREO-A/EUVI images. The average brightness of the EUV images is captured quantitatively by the model simulations for both STEREO-A/EUVI and SDO/AIA images. However, our model results show coronal holes that are darker in comparison to the AIA observations, which is at least partially due to the neglected scattering in the synthetic EUV images. 

%\textcolor{red}{I'm not sure how to fit this text in, maybe it is not needed.} The portion of the corona facing STEREO-A faces away from HMI and must rely on older data that has been forward evolved by ADAPT.

With the exception that our model shows far less brightness in coronal holes, specially in comparison to AIA observations, we find that the coronal hole locations are pretty-well captured in our analyses. As expected, the small scale structure is partially captured, with larger active regions clearly reproduced. We note that the steady state simulation is performed for a synchronic magnetic field map over a complete Carrington rotation whereas the observations are for particular time stamps, thus, the model cannot reproduce time dependent activity during the rotation.

%%%%%%%%%%%%%% DEMT FIGURES %%%%%%%%%%%%%%
%CR2208 DEMT DENSITY
\begin{figure*}
\gridline{\fig{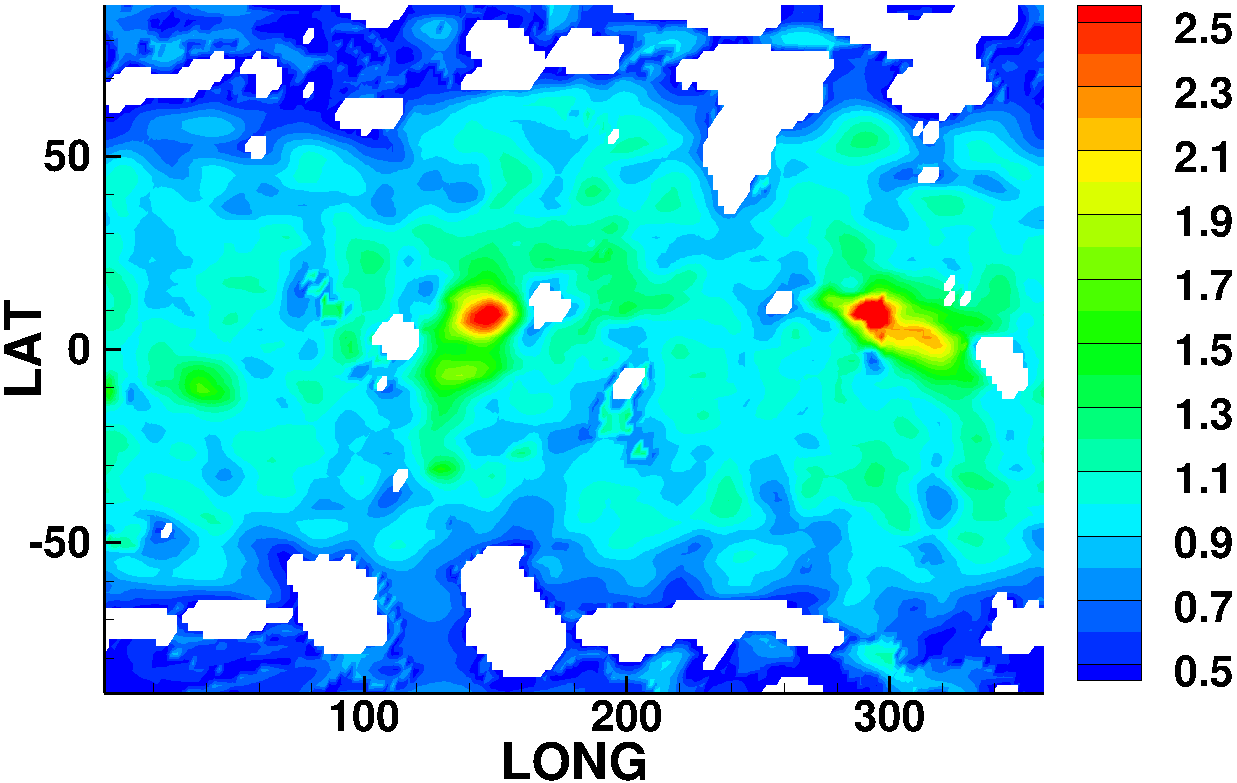}{0.33\textwidth}{Ne DEMT (10$^{8}$ cm$^{-3}$)}
          \fig{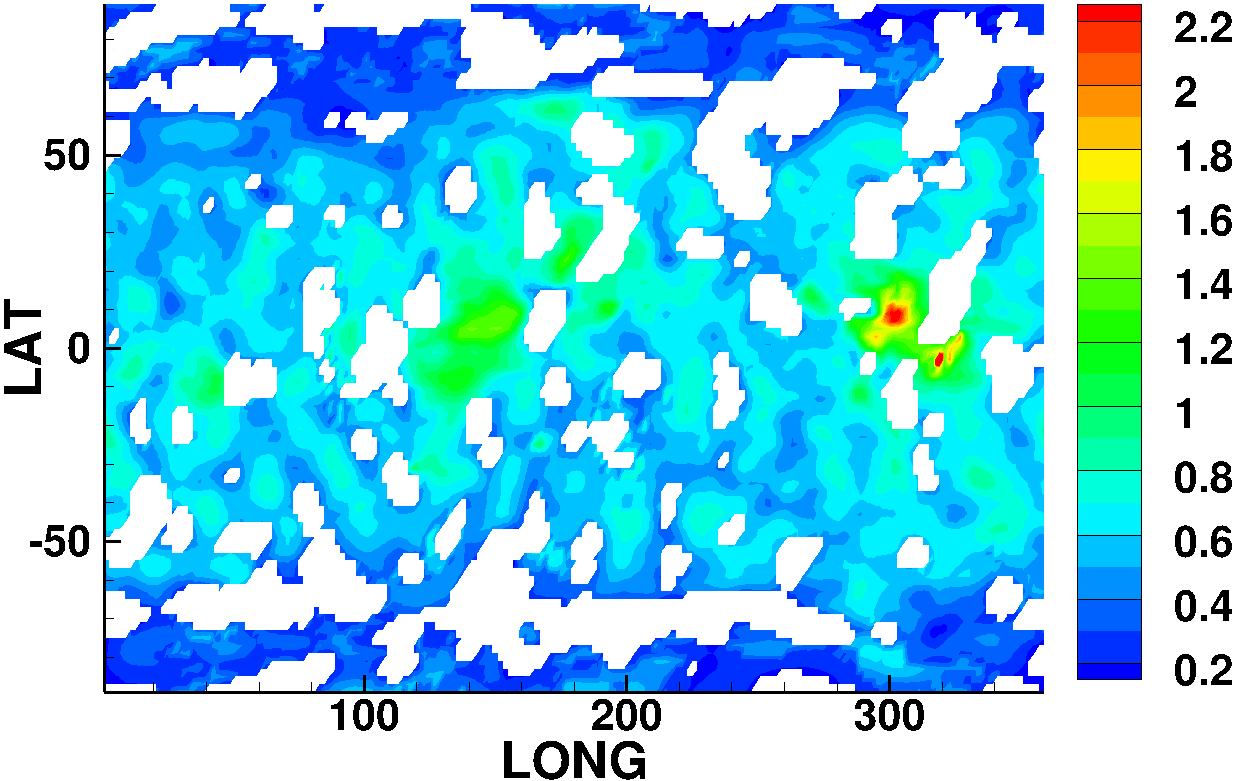}{0.33\textwidth}{Ne DEMT (10$^{8}$ cm$^{-3}$)}
          \fig{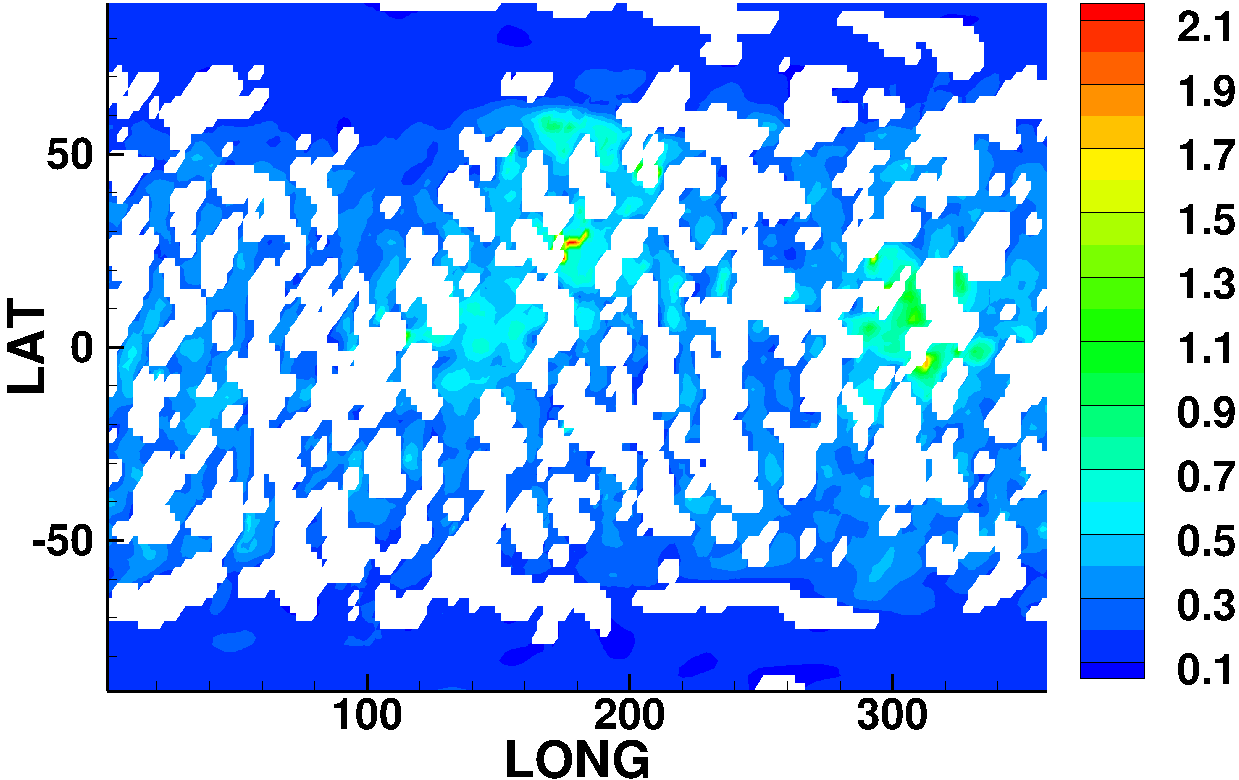}{0.33\textwidth}{Ne DEMT (10$^{8}$ cm$^{-3}$)}
          }
\gridline{\fig{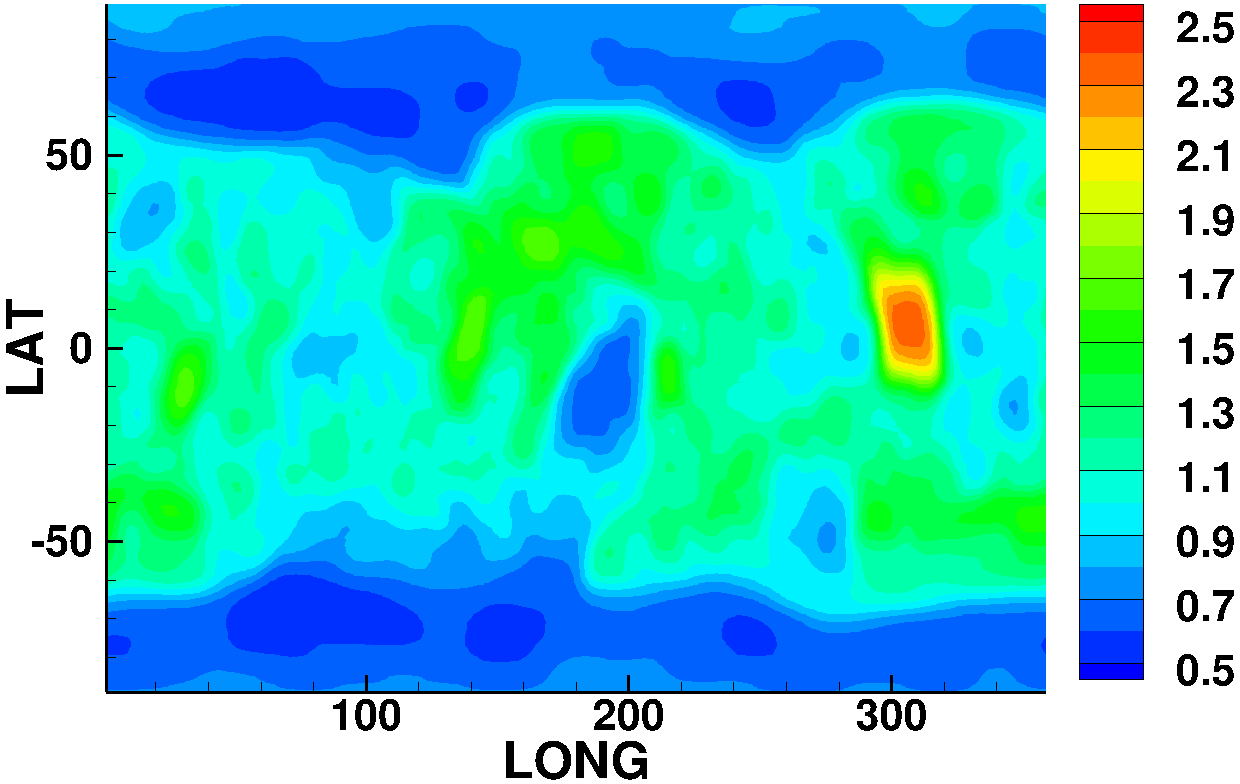}{0.33\textwidth}{Ne AWSoM (10$^{8}$ cm$^{-3}$)}
          \fig{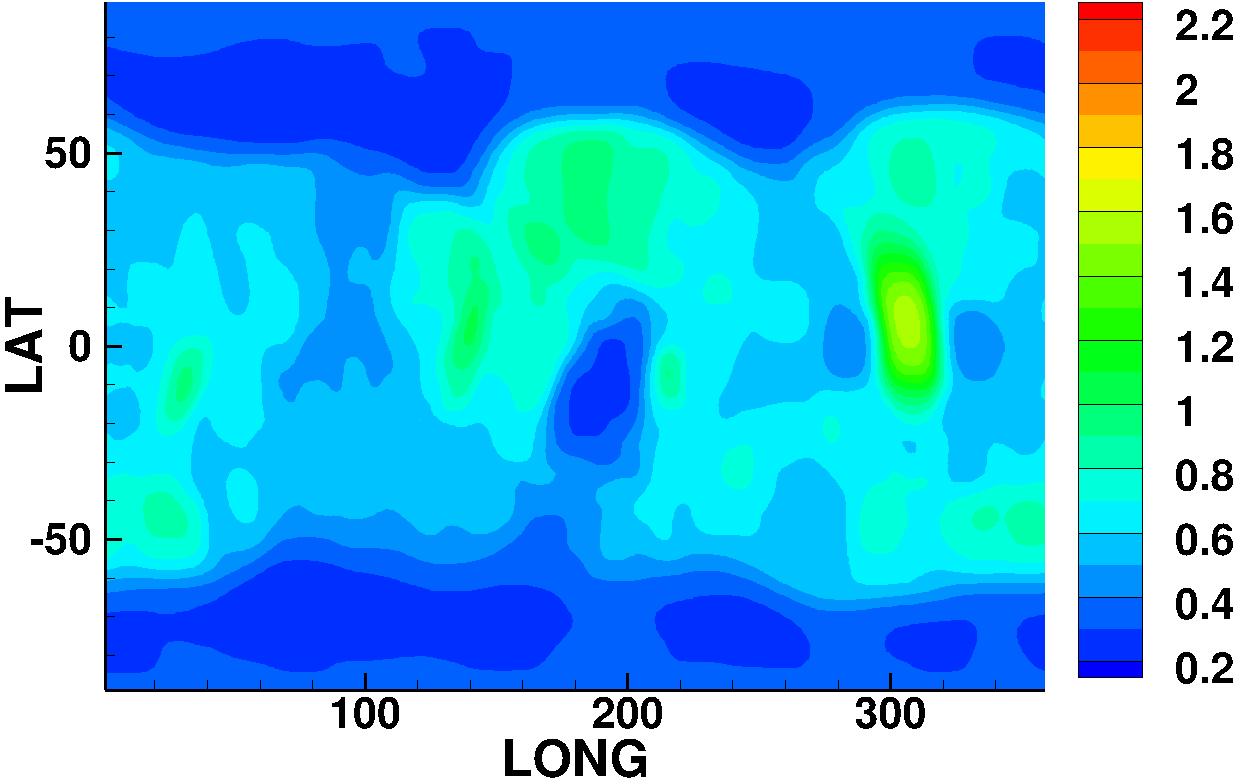}{0.33\textwidth}{Ne AWSoM(10$^{8}$ cm$^{-3}$)}
          \fig{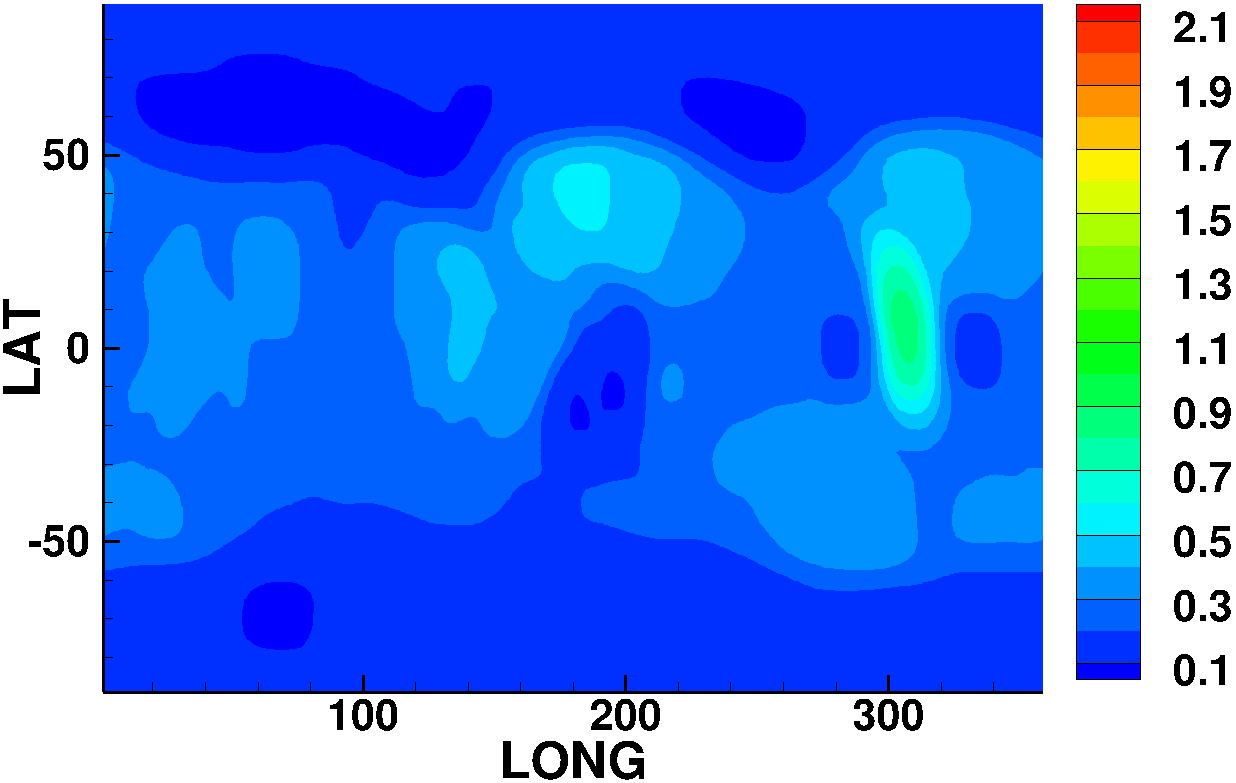}{0.33\textwidth}{Ne AWSoM (10$^{8}$ cm$^{-3}$)}
          }
\gridline{\fig{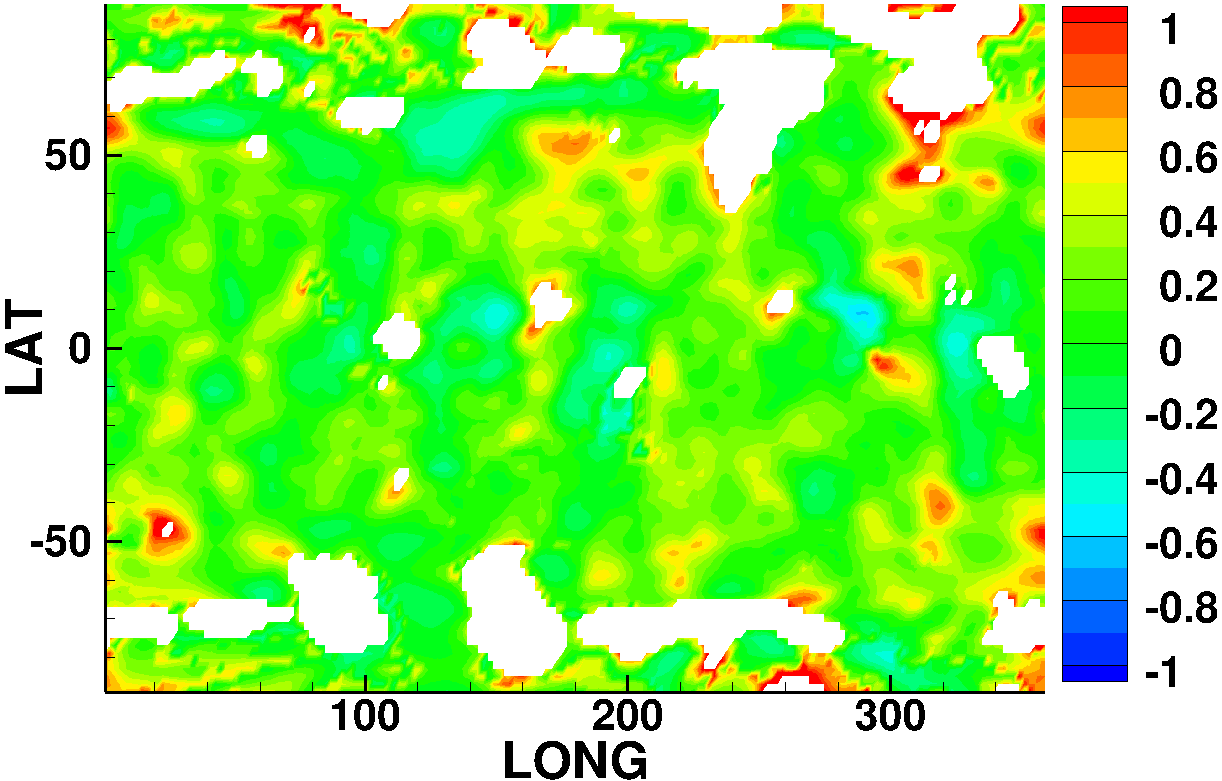}{0.33\textwidth}{Ne Rel Diff}
          \fig{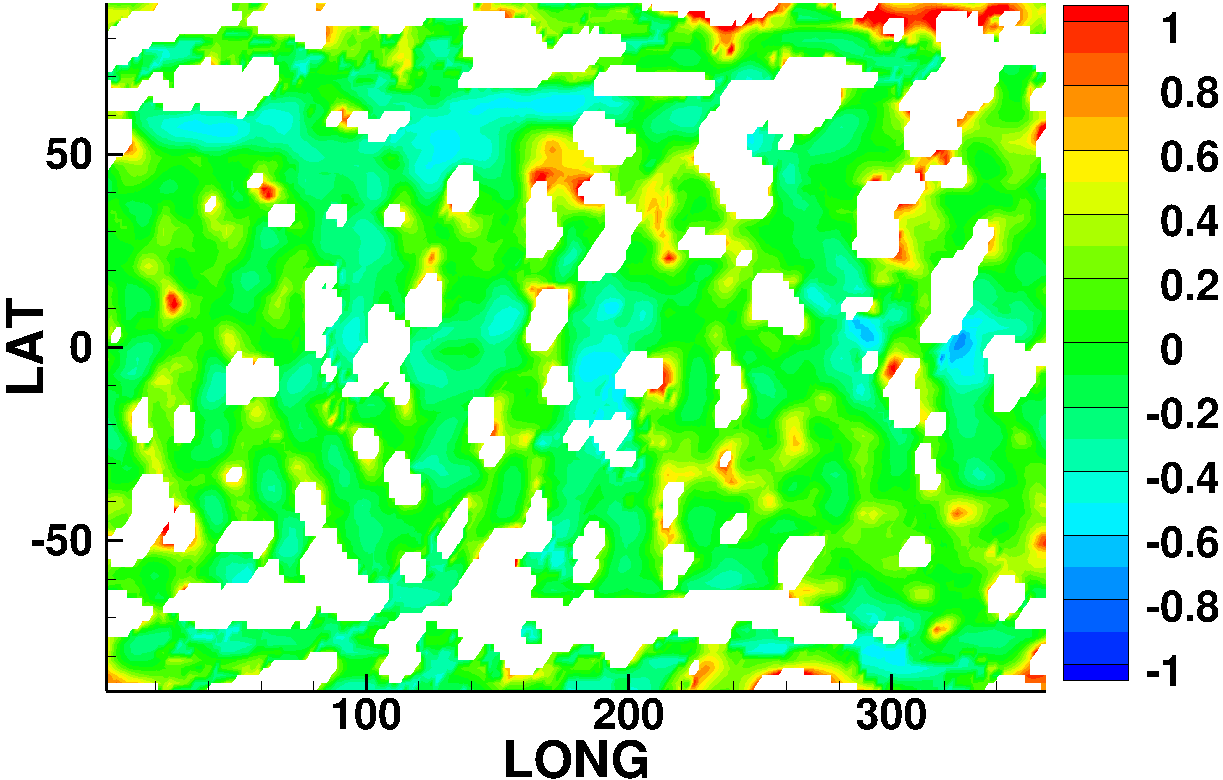}{0.33\textwidth}{Ne Rel Diff}
          \fig{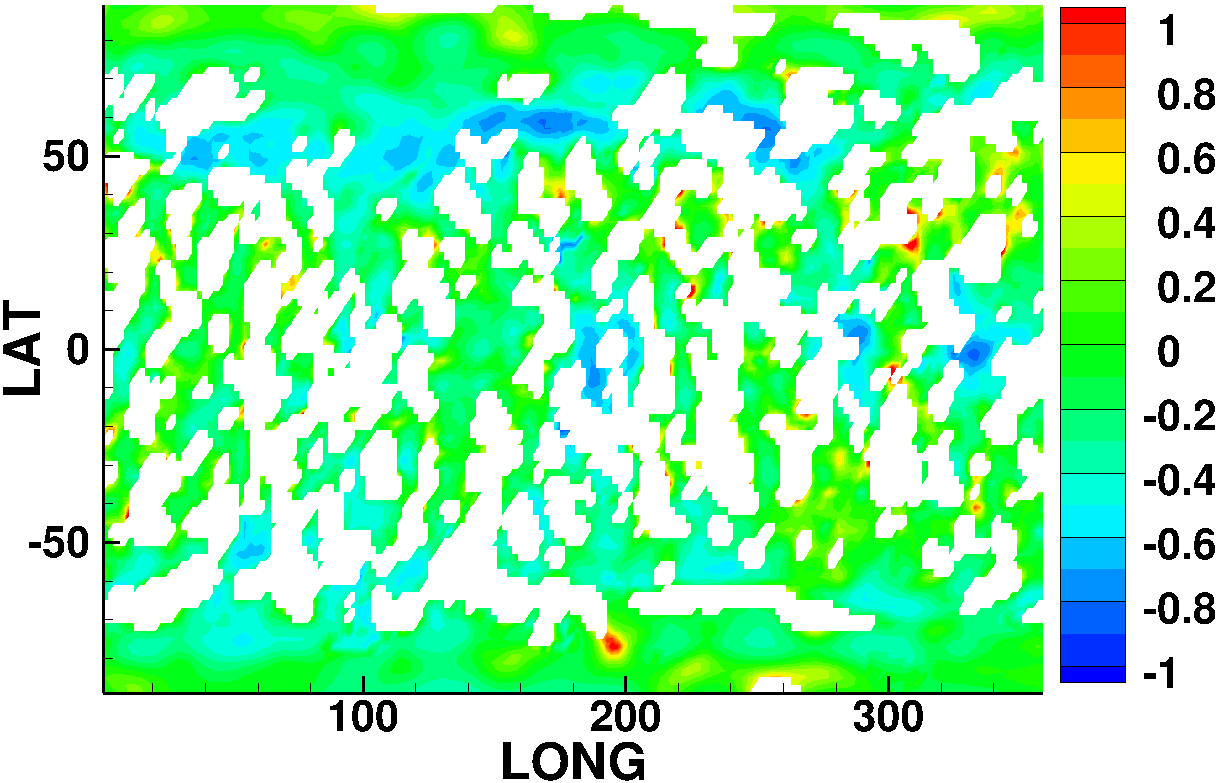}{0.33\textwidth}{Ne Rel Diff}
          }
\gridline{\fig{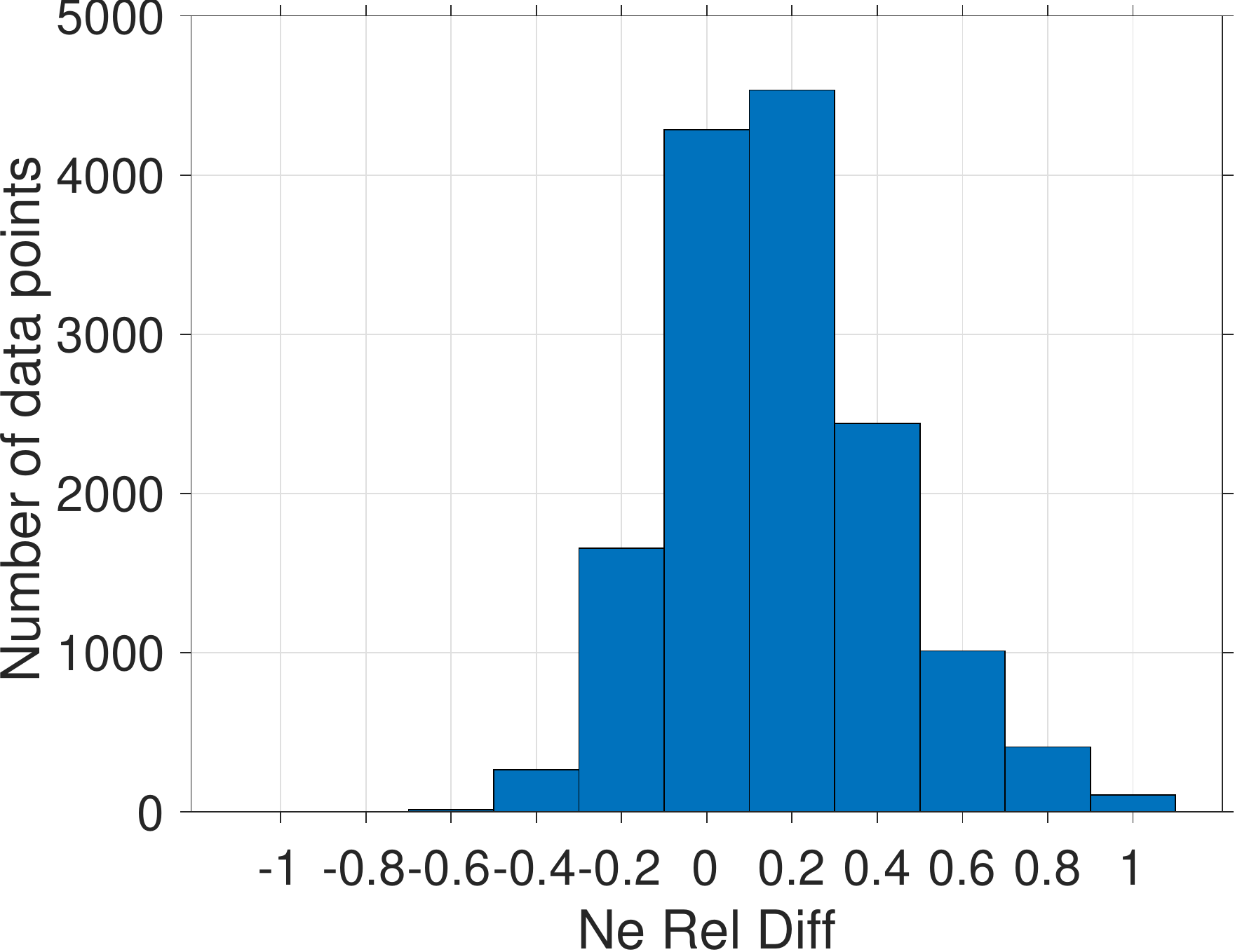}{0.3\textwidth}{(a) r=1.055 \Rs}
          \fig{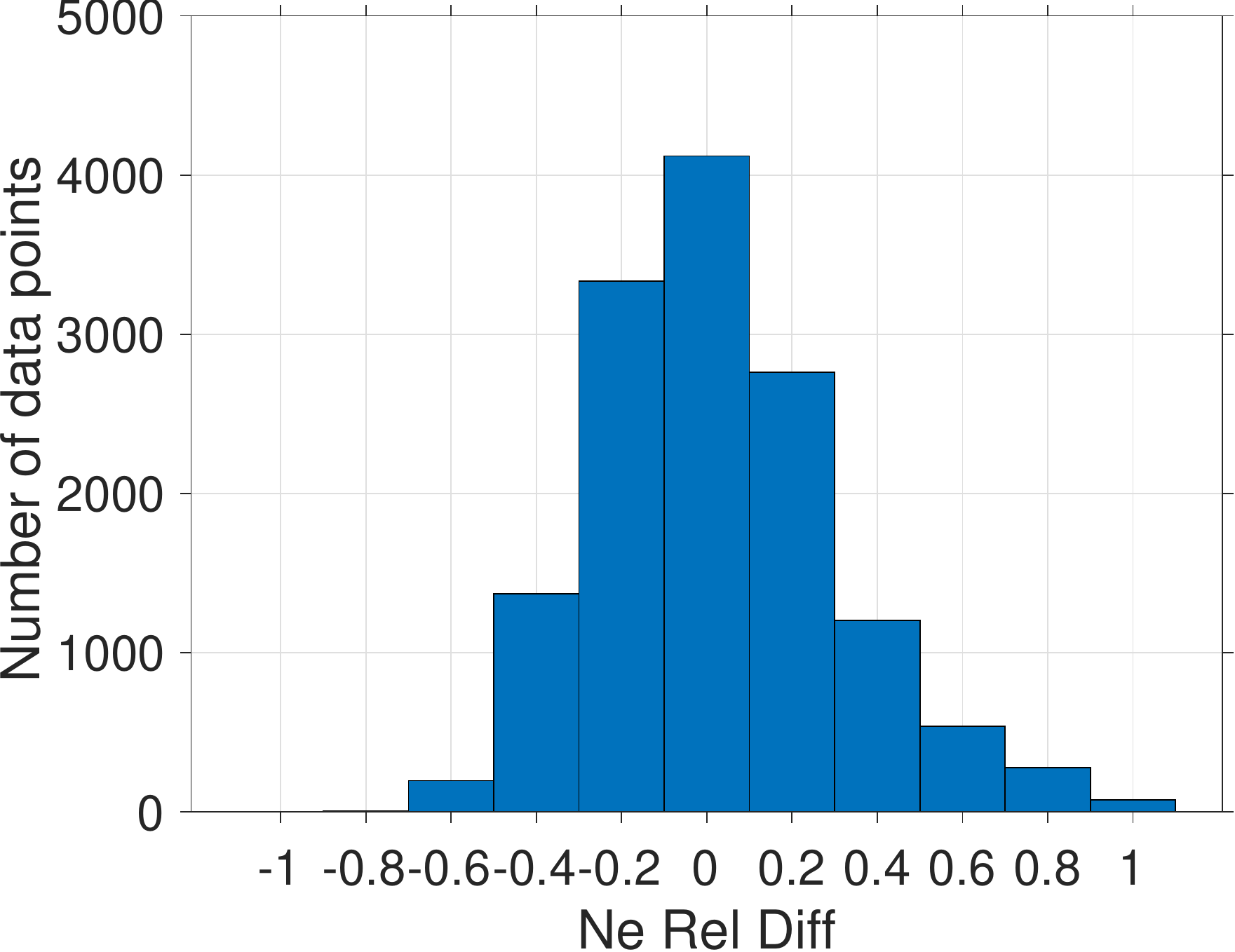}{0.3\textwidth}{(b) r=1.105 \Rs}
          \fig{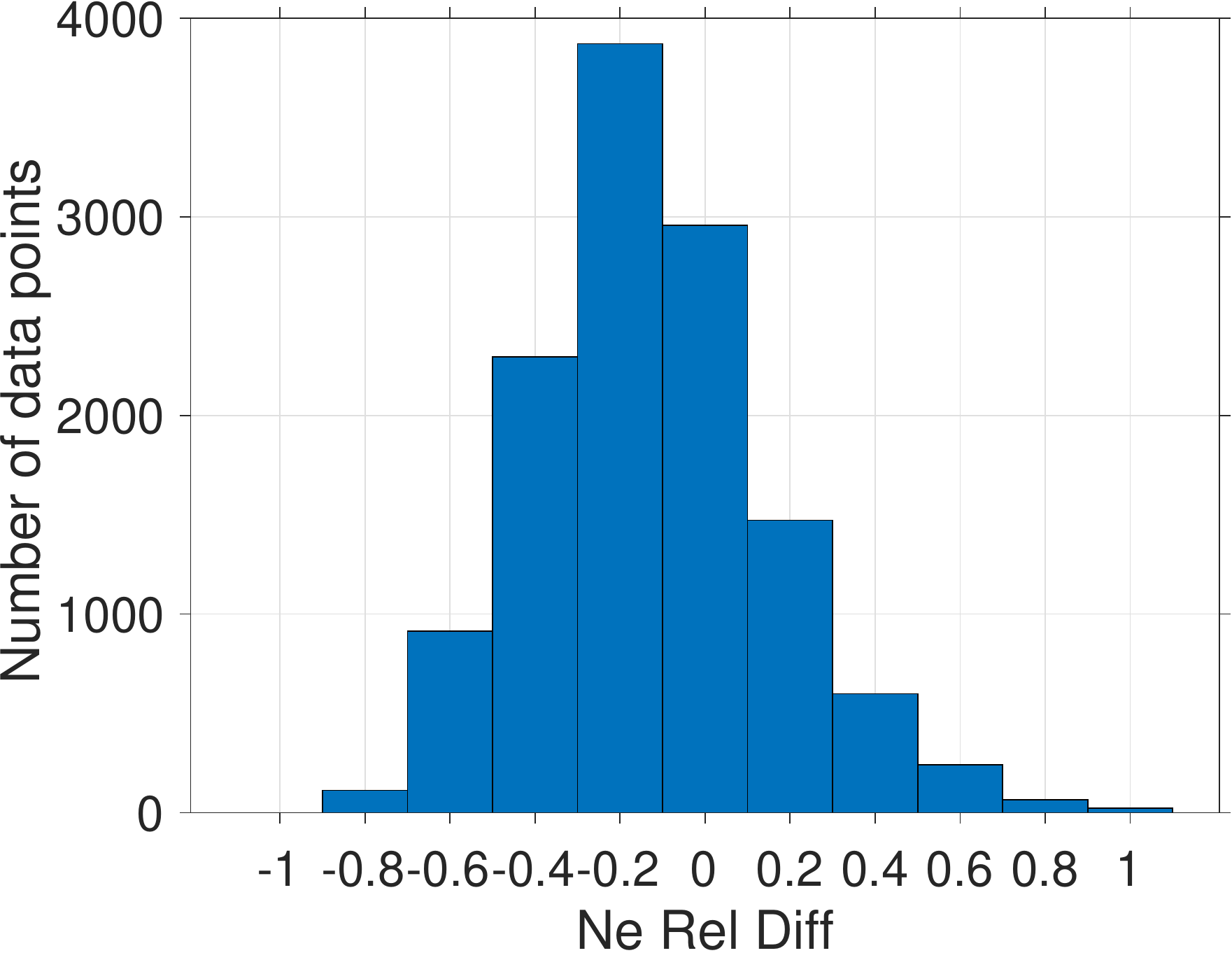}{0.3\textwidth}{(c) r=1.205 \Rs}
          }
\caption{Comparison of tomographic reconstructions of electron density from EUV observations and AWSoM model simulation results for CR2208 at (a) 1.055\,\Rs, (b) 1.105\,\Rs~and (c) 1.205\,\Rs. First and second rows show the 3D reconstructed density from SDO/AIA observations using DEMT (Ne DEMT) and the model predicted density (Ne AWSoM), respectively in units of 10$^{8}$ cm$^{-3}$. The third row depicts the relative difference between the observations and model results. The quantity shown is \rm{Ne Rel Diff}=\big($\frac{Ne_{AWSoM}}{Ne_{DEMT}}-1 \big)$. Bottom row shows the histogram distribution for \rm{Ne Rel Diff}.}
\label{fig:demt1_08}
\end{figure*}
%%%%CR2208 DEMT  TEMPERATURE
\begin{figure*}[ht!]
\gridline{\fig{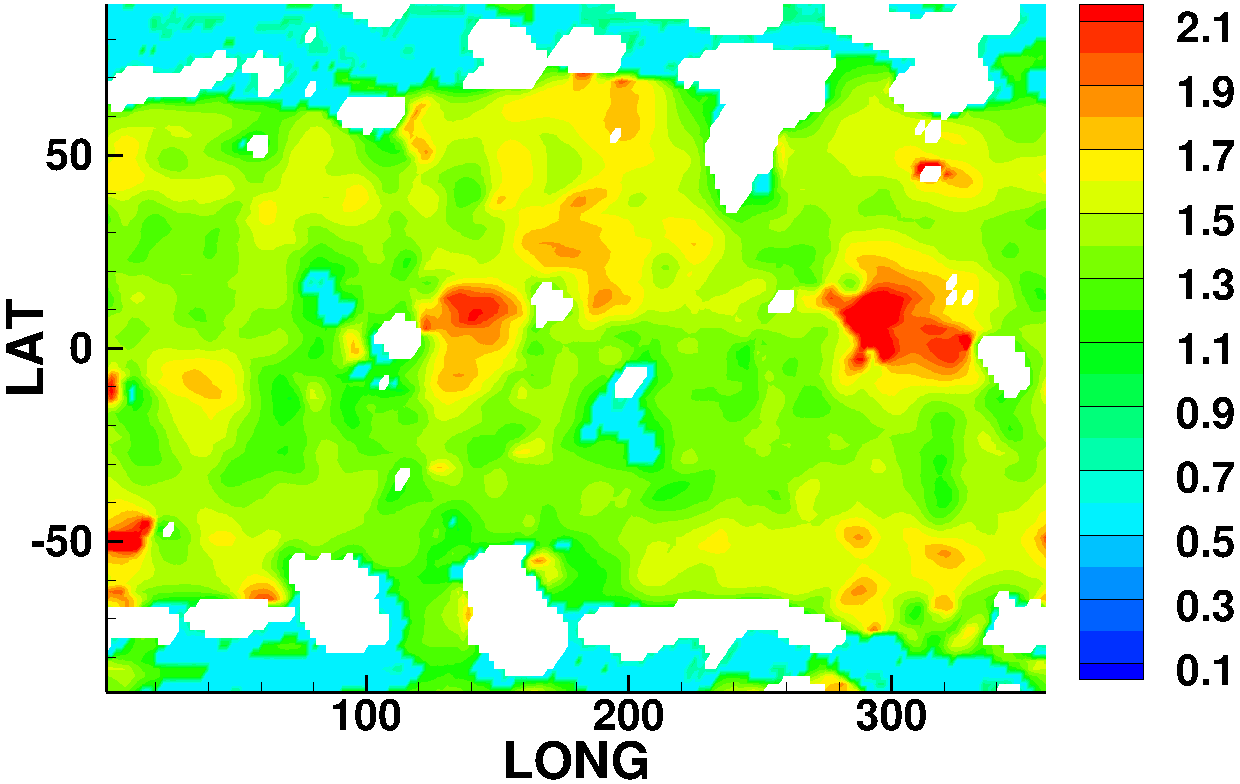}{0.33\textwidth}{Te DEMT (MK)}
          \fig{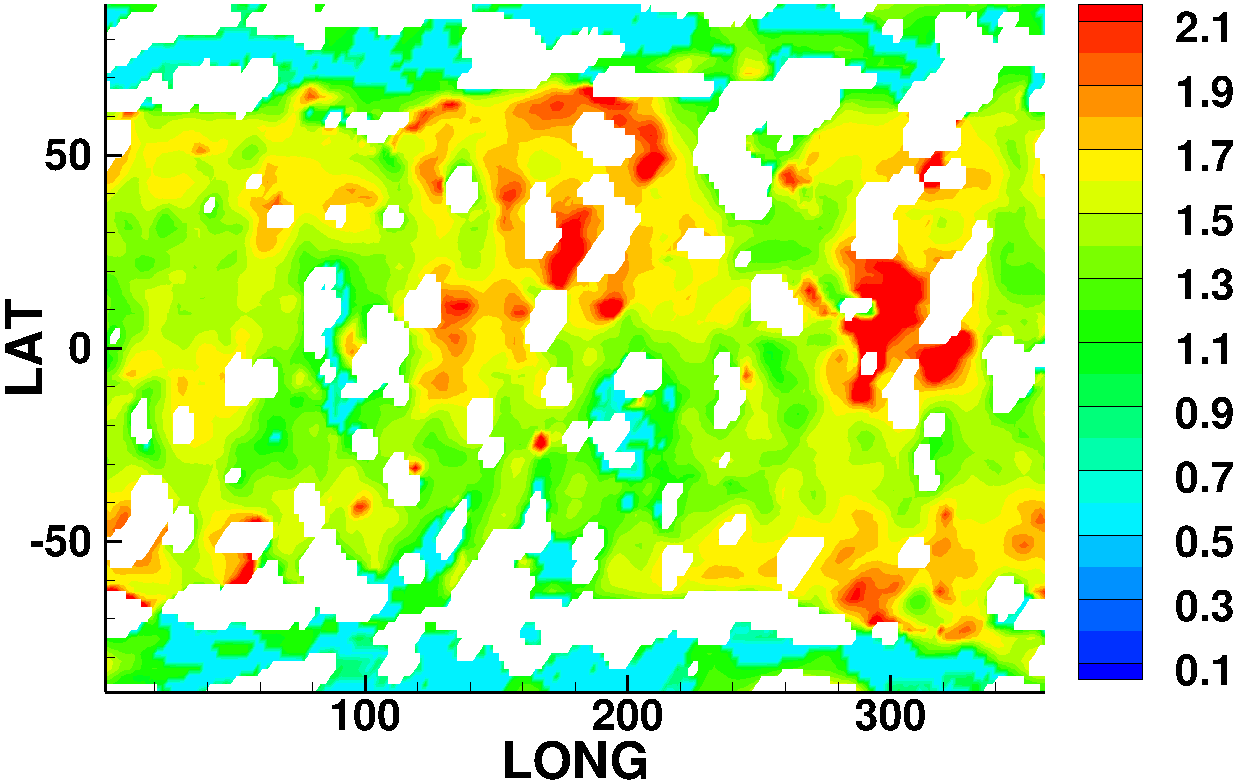}{0.33\textwidth}{Te DEMT (MK)}
          \fig{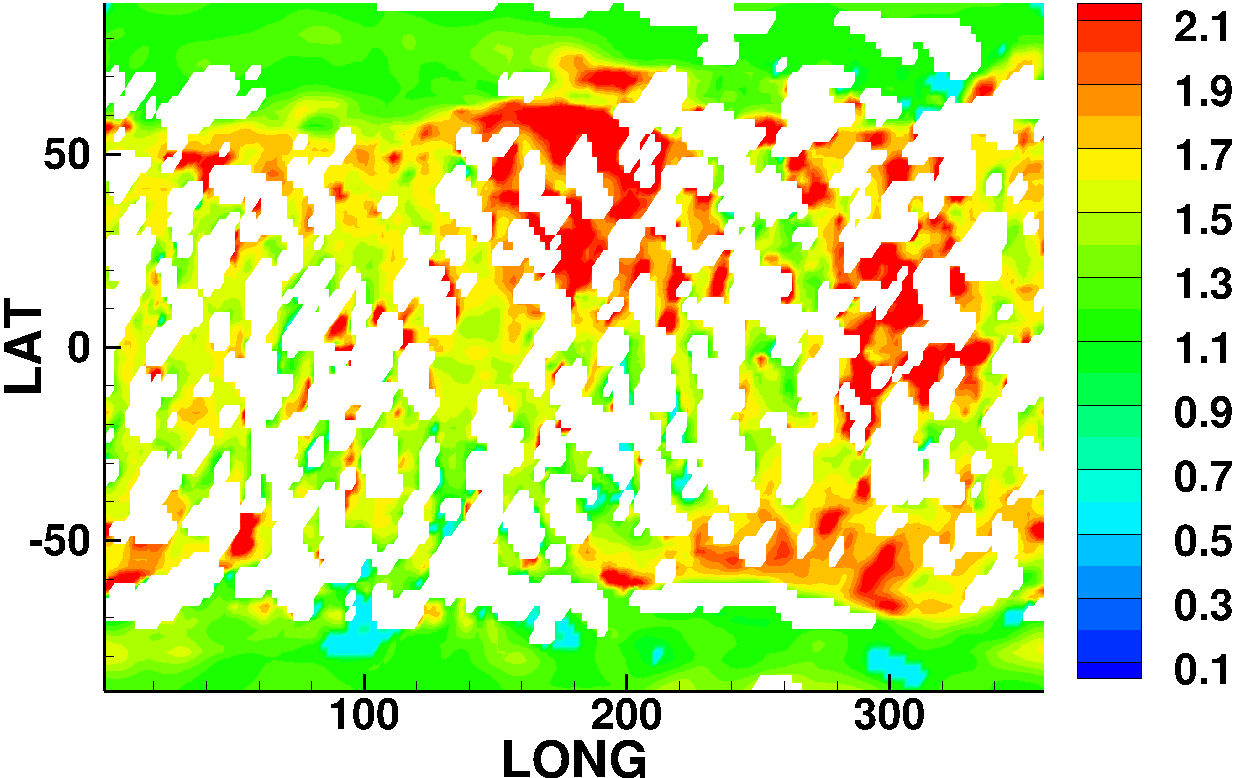}{0.33\textwidth}{Te DEMT (MK)}
          }
\gridline{\fig{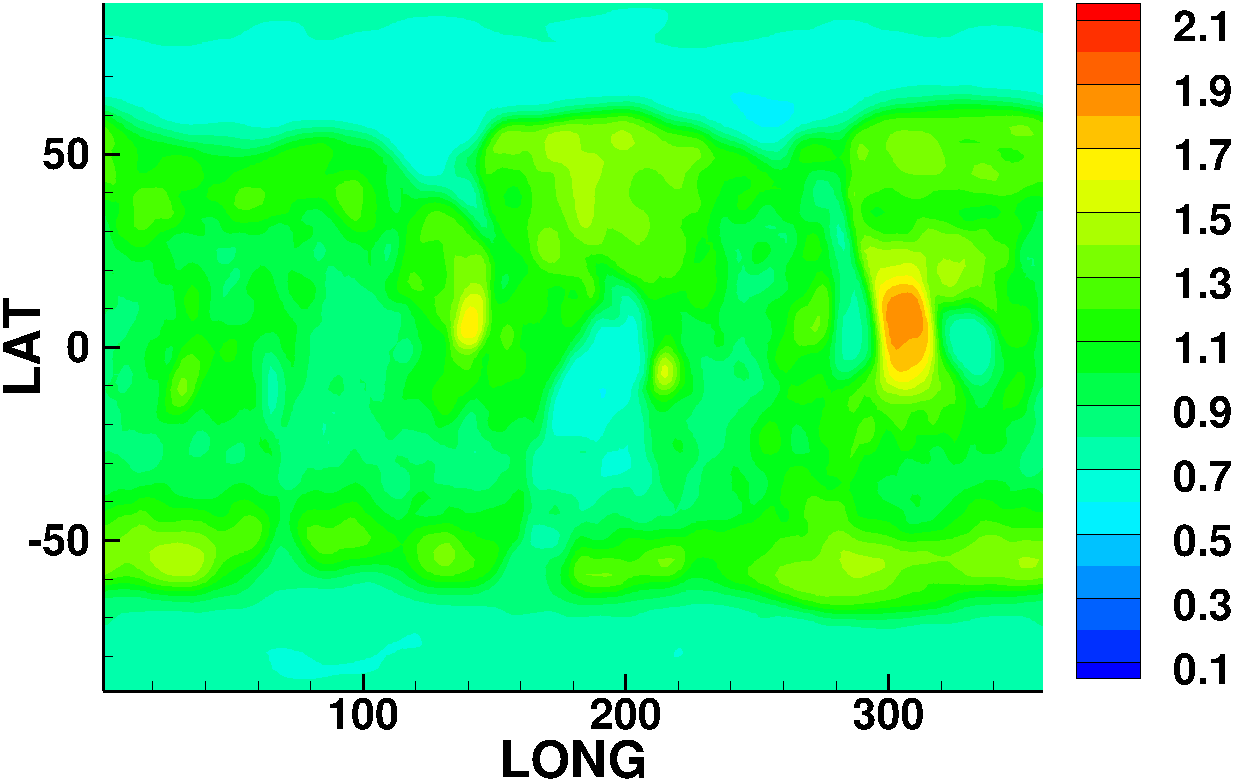}{0.33\textwidth}{Te AWSoM (MK)}
          \fig{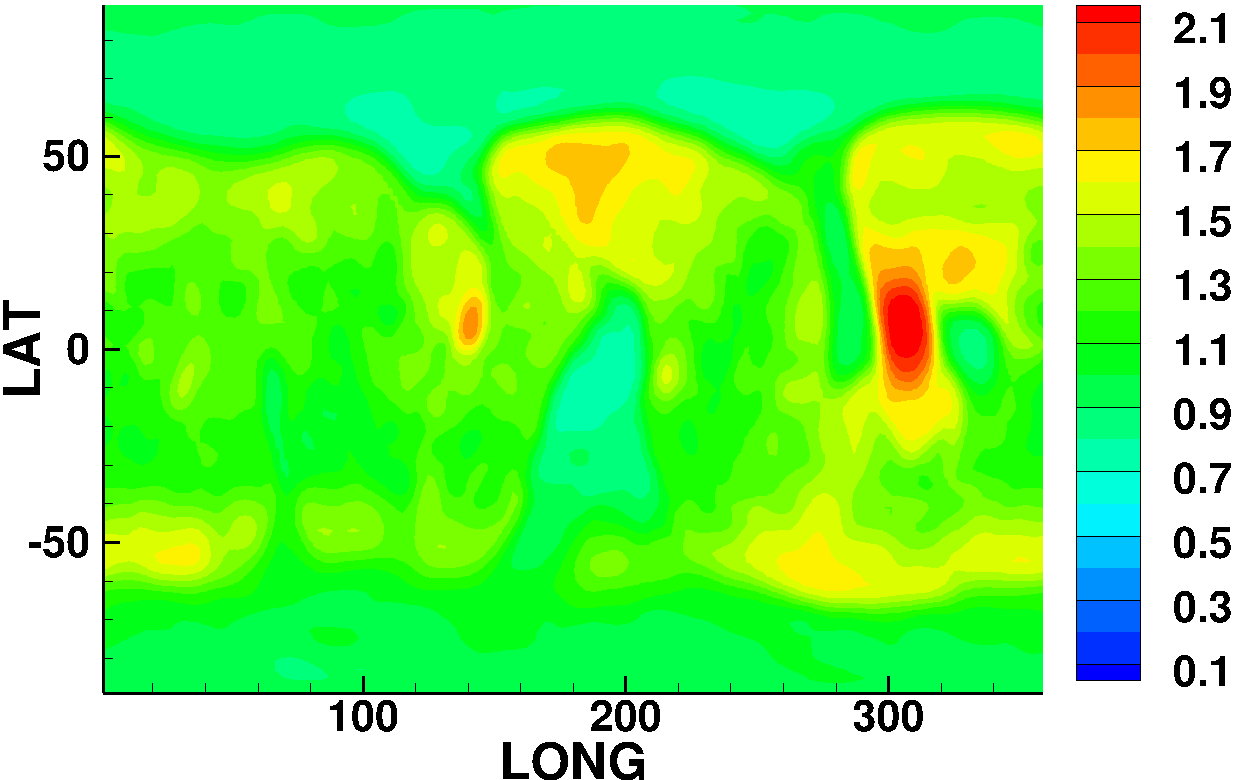}{0.33\textwidth}{Te AWSoM (MK)}
          \fig{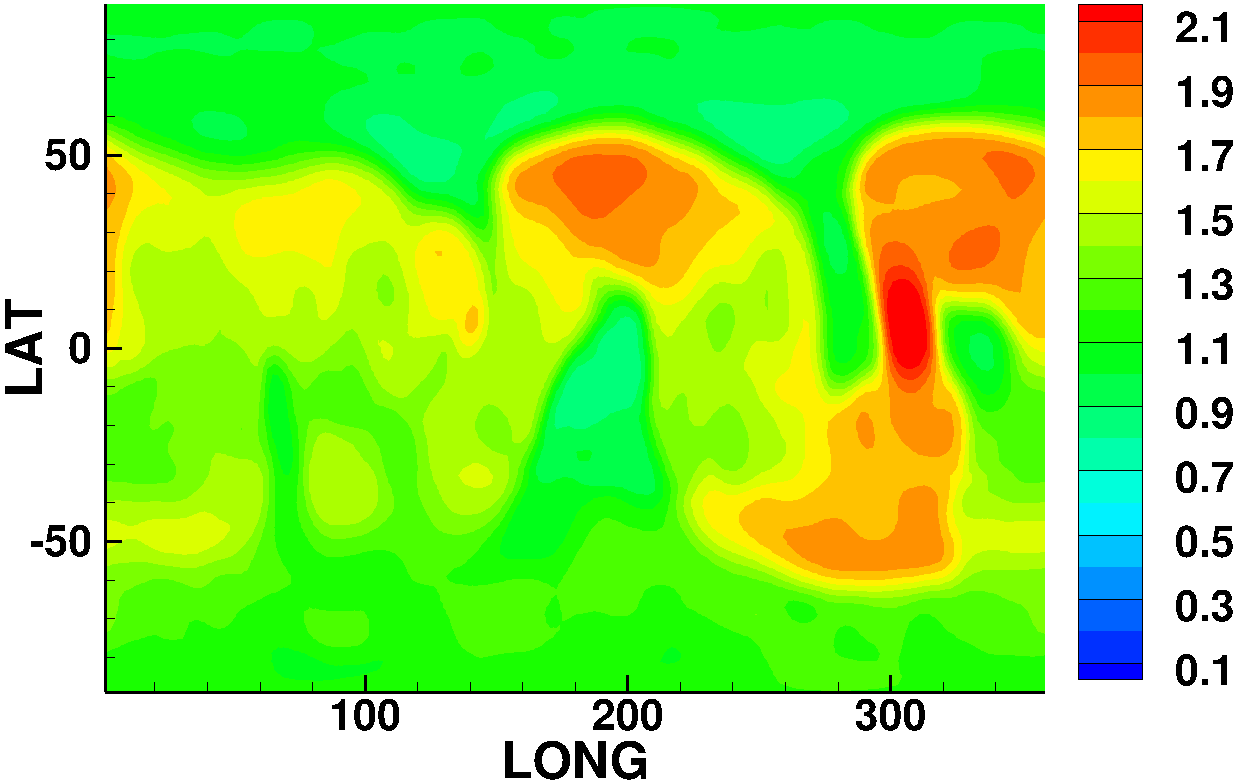}{0.33\textwidth}{Te AWSoM (MK)}
          }
\gridline{\fig{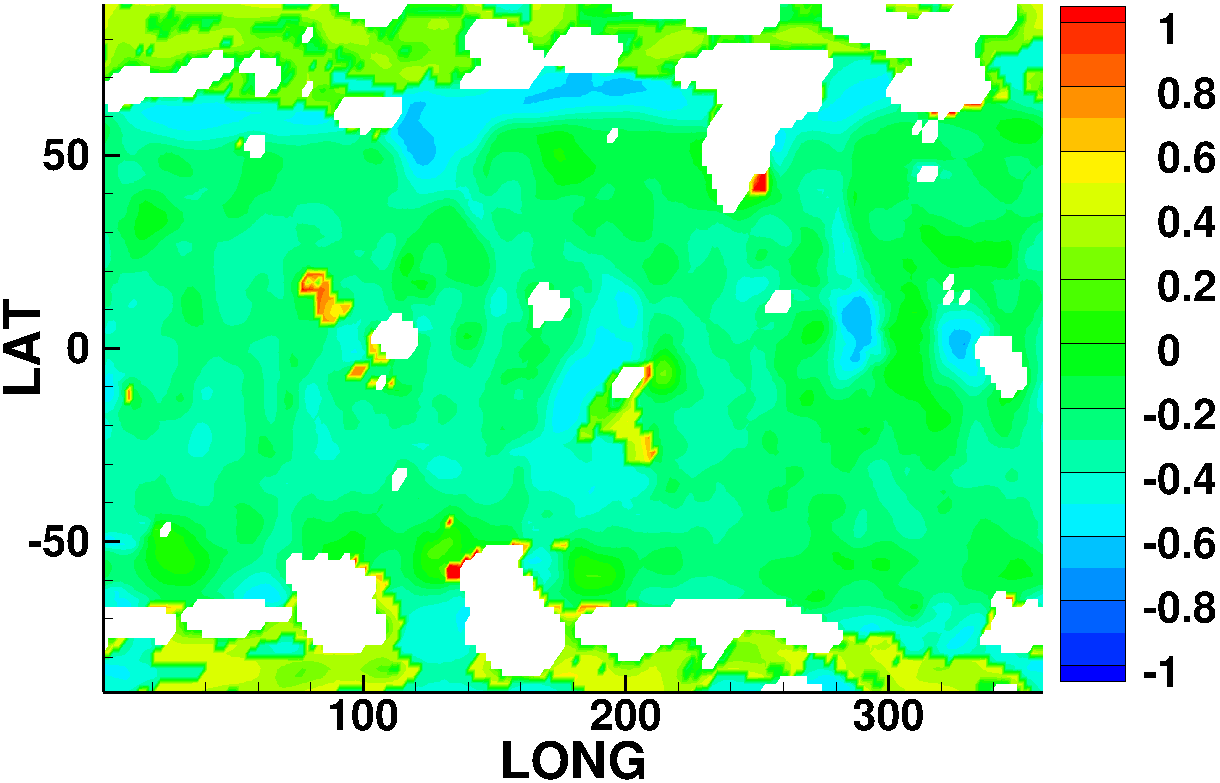}{0.33\textwidth}{Te Rel Diff}
          \fig{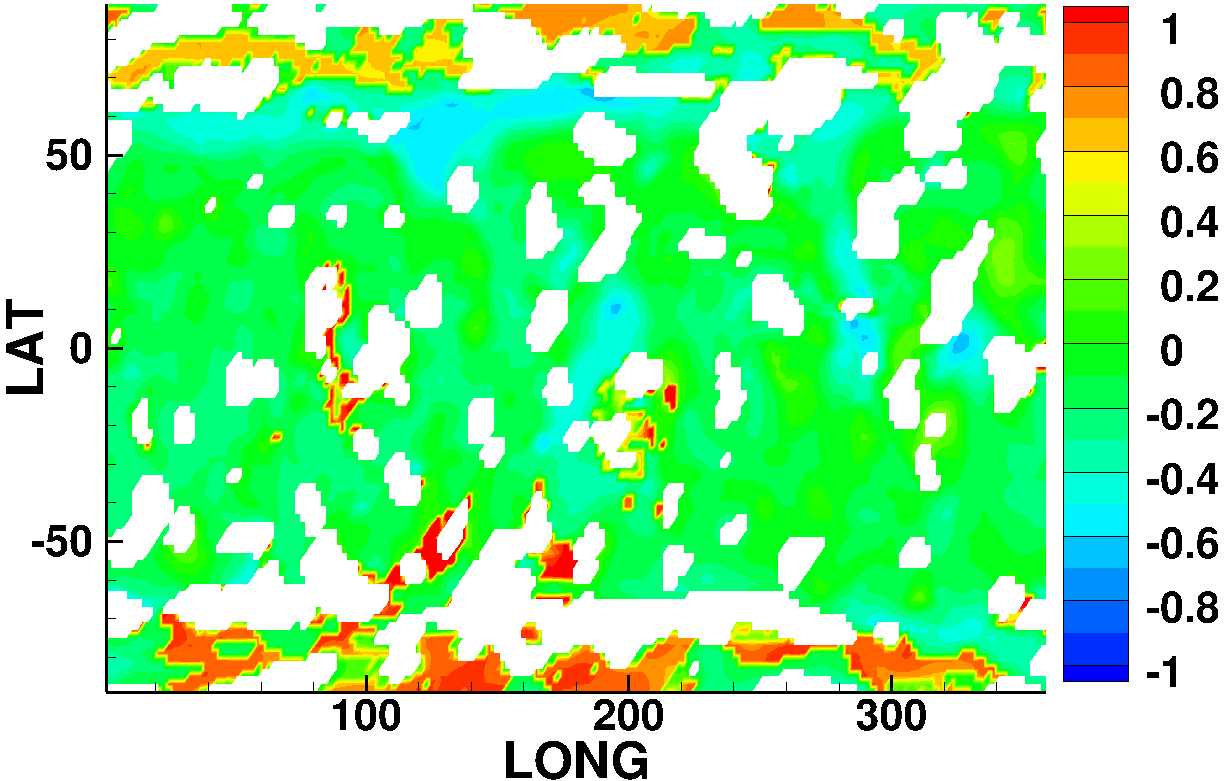}{0.33\textwidth}{Te Rel Diff}
          \fig{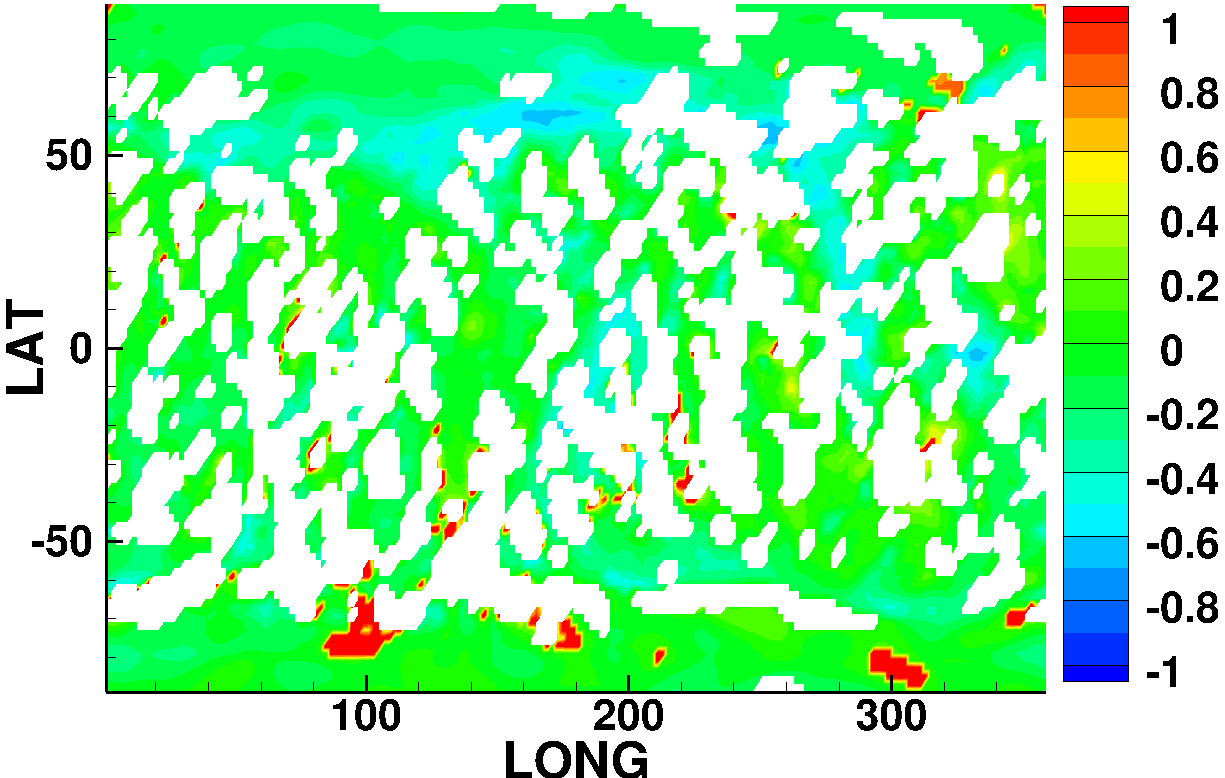}{0.33\textwidth}{Te Rel Diff}
          }
\gridline{\fig{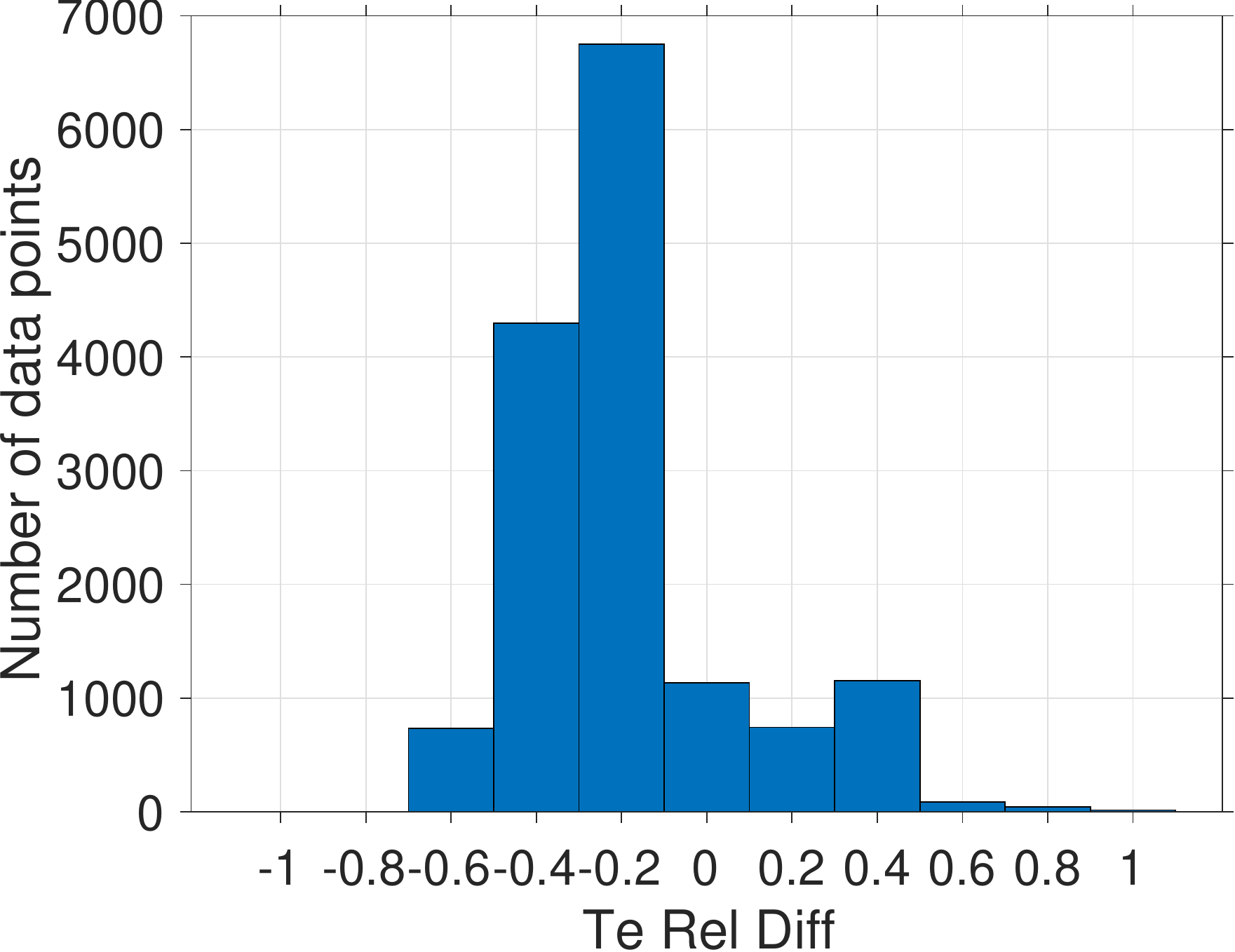}{0.3\textwidth}{(a) r=1.055 \Rs}
          \fig{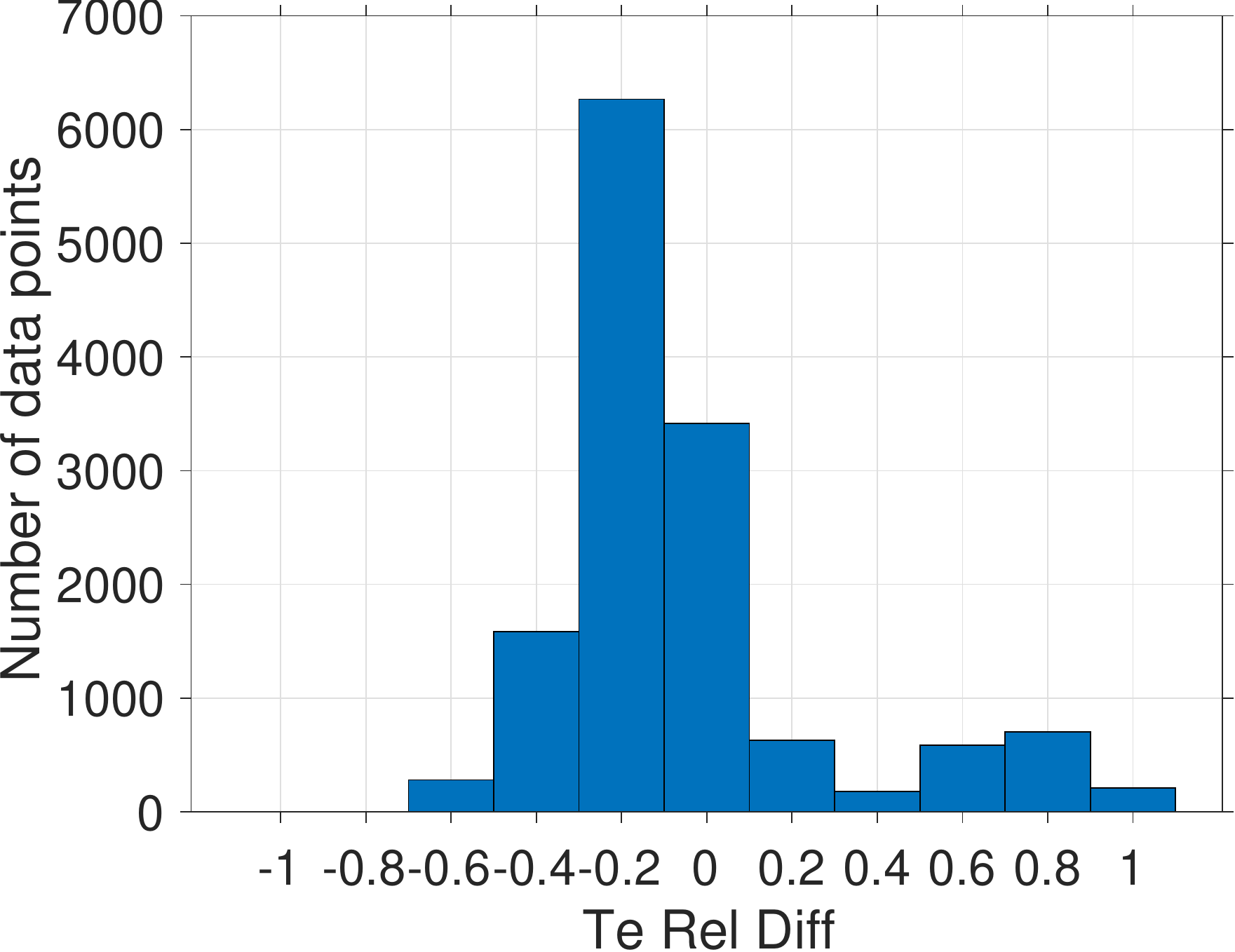}{0.3\textwidth}{(b) r=1.105 \Rs}
          \fig{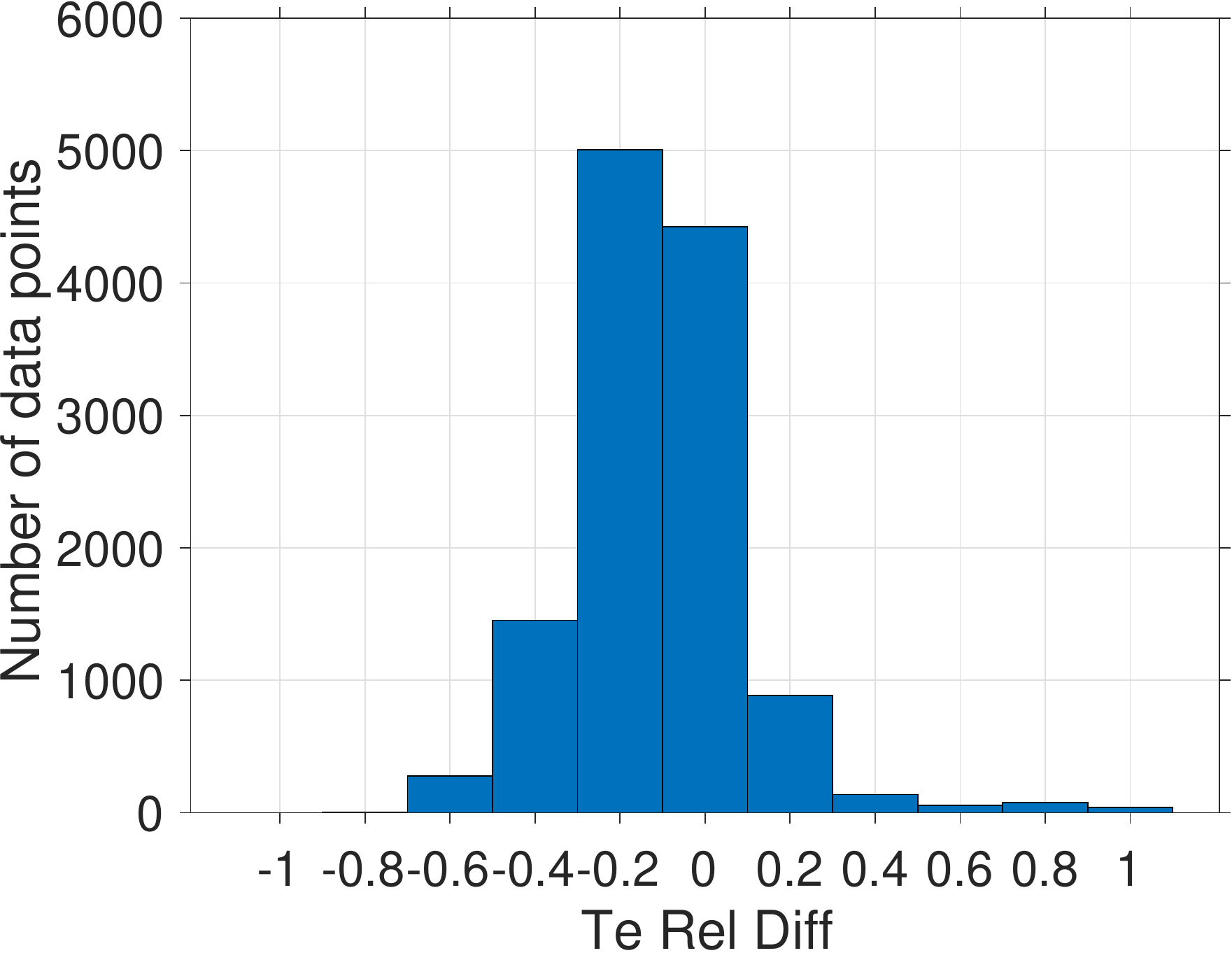}{0.3\textwidth}{(c) r=1.205 \Rs}
          }
\caption{Comparison of tomographic reconstructions of electron temperature from EUV observations and AWSoM model simulation results for CR2208 at (a) 1.055\,\Rs, (b) 1.105\,\Rs~and (c) 1.205\,\Rs. First and second rows show the 3D reconstructed temperature from SDO/AIA observations using DEMT (Te DEMT) and the model predicted temperature (Te AWSoM), respectively in units of $10^{6}$ K (MK). The third row depicts the relative difference between the observations and model results. The quantity shown is \rm{Te Rel Diff} $=\big(\frac{Te_{AWSoM}}{Te_{DEMT}}-1 \big)$. Bottom row shows the histogram distribution for \rm{Te Rel Diff}.}
\label{fig:demt2_08}
\end{figure*}
%CR2209 DEMT DENSITY
\begin{figure*}
\gridline{\fig{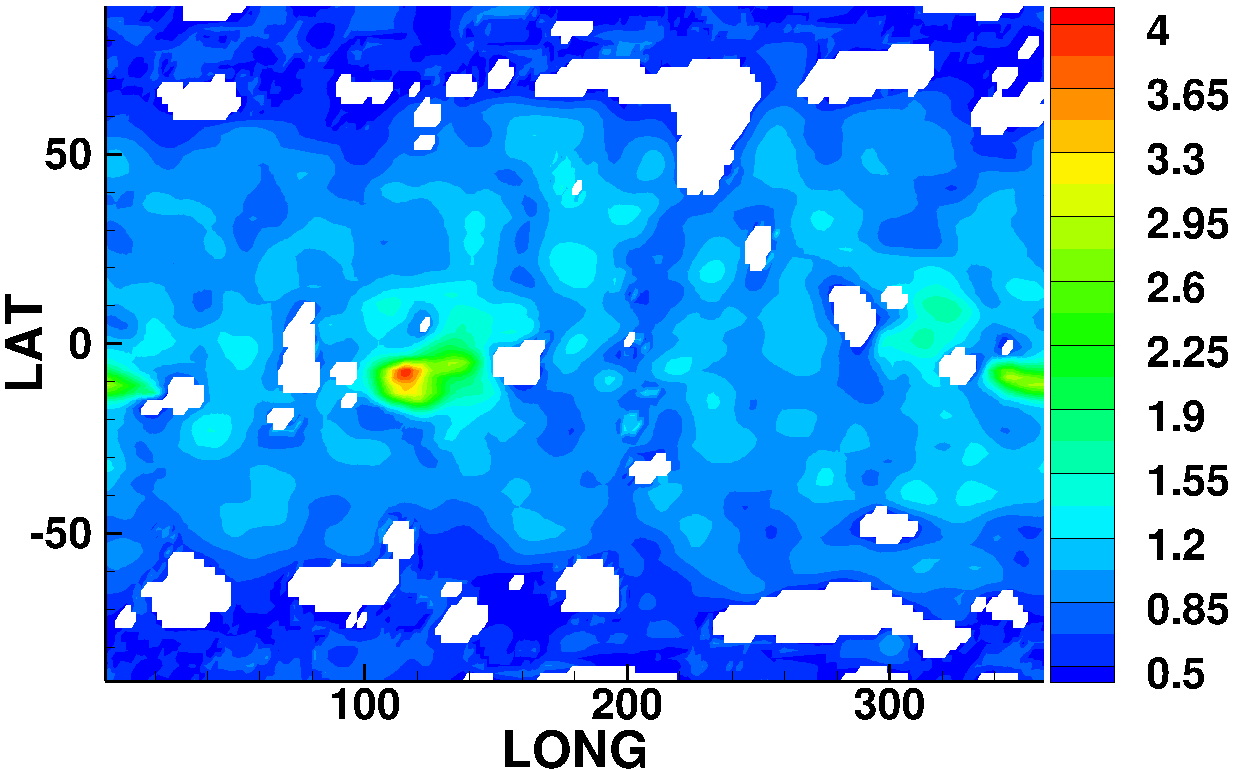}{0.33\textwidth}{Ne DEMT (10$^{8}$ cm$^{-3}$)}
          \fig{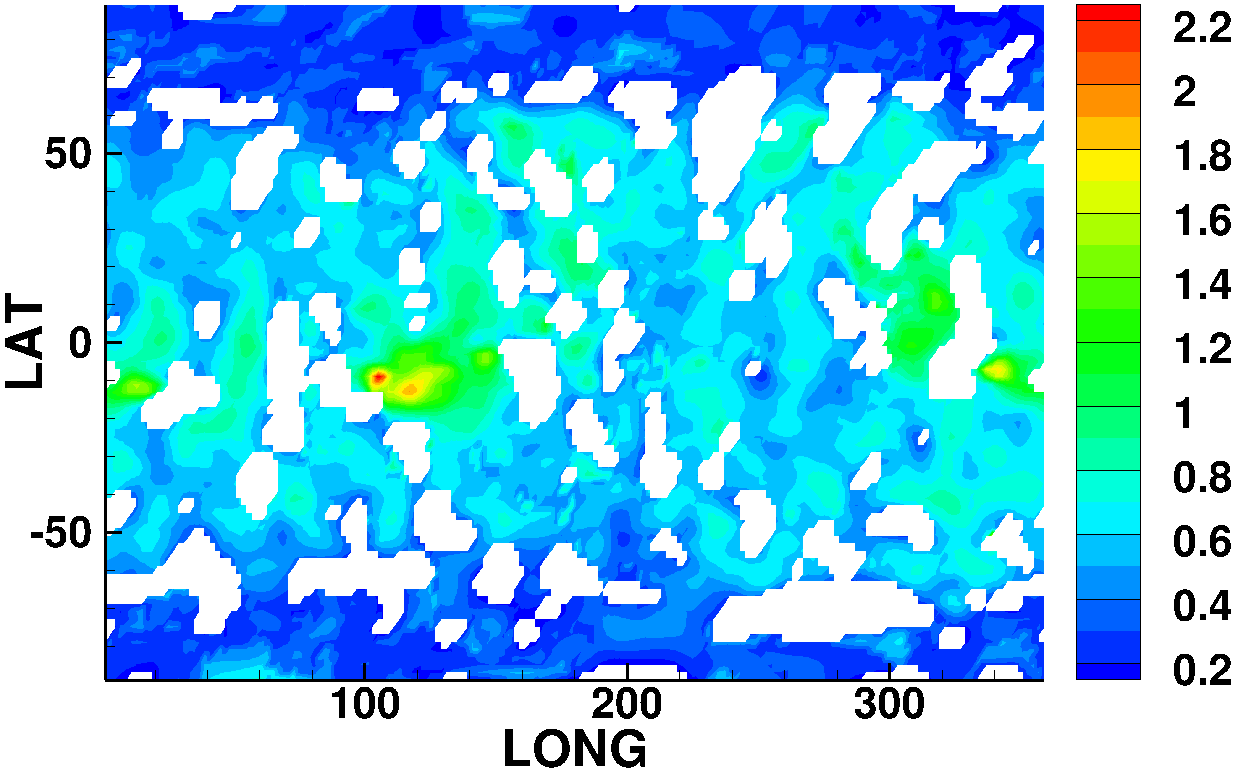}{0.33\textwidth}{Ne DEMT (10$^{8}$ cm$^{-3}$)}
          \fig{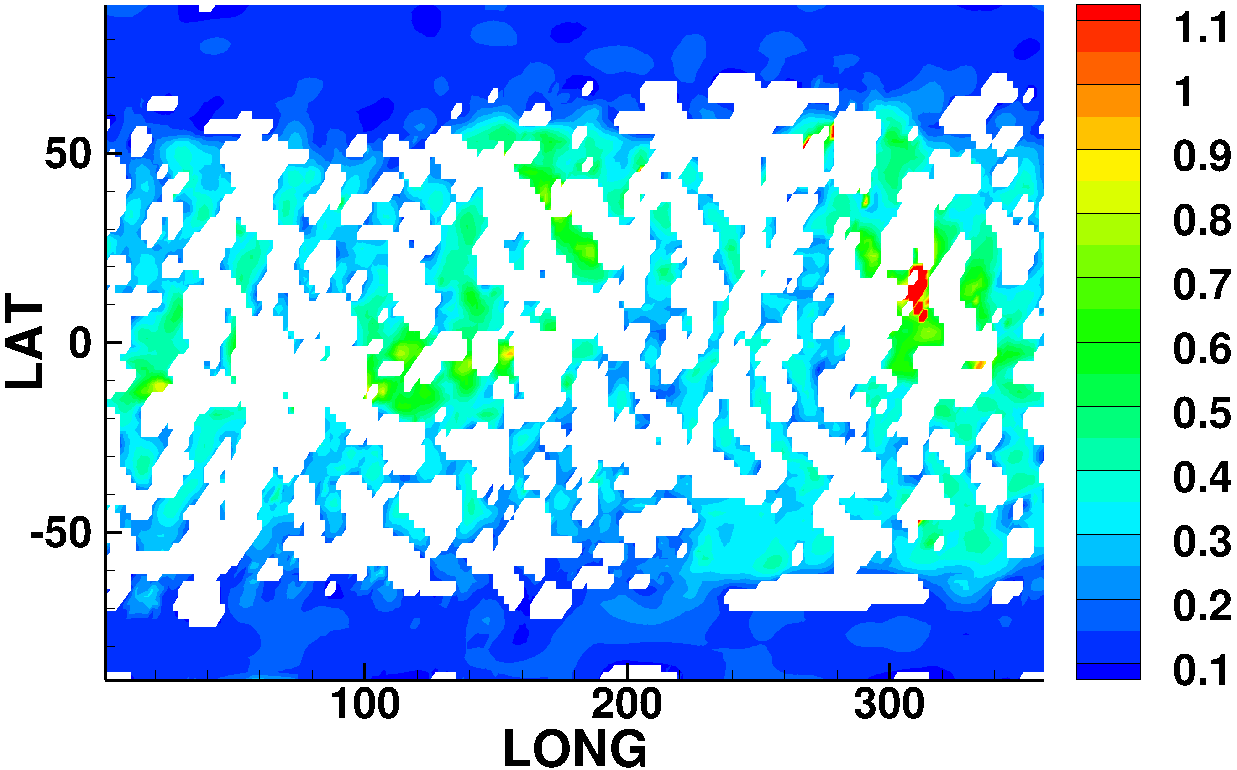}{0.33\textwidth}{Ne DEMT (10$^{8}$ cm$^{-3}$)}
          }
\gridline{\fig{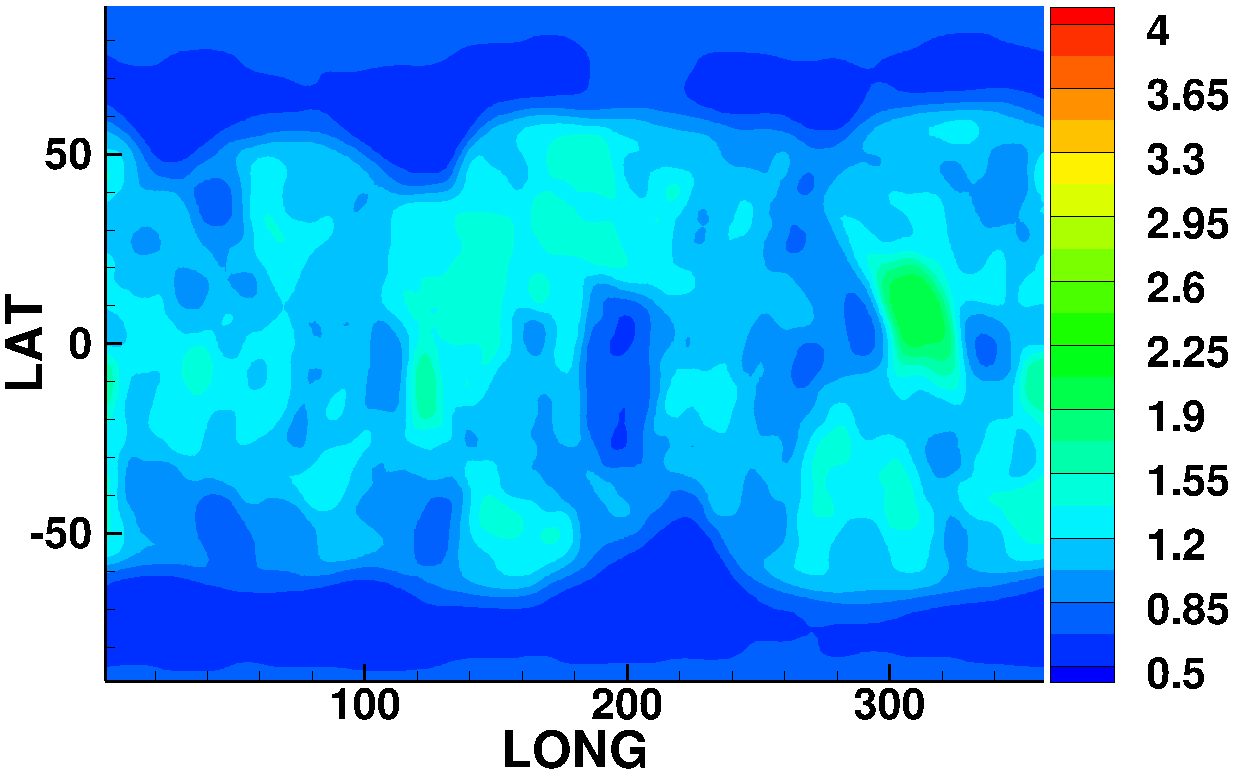}{0.33\textwidth}{Ne AWSoM (10$^{8}$ cm$^{-3}$)}
          \fig{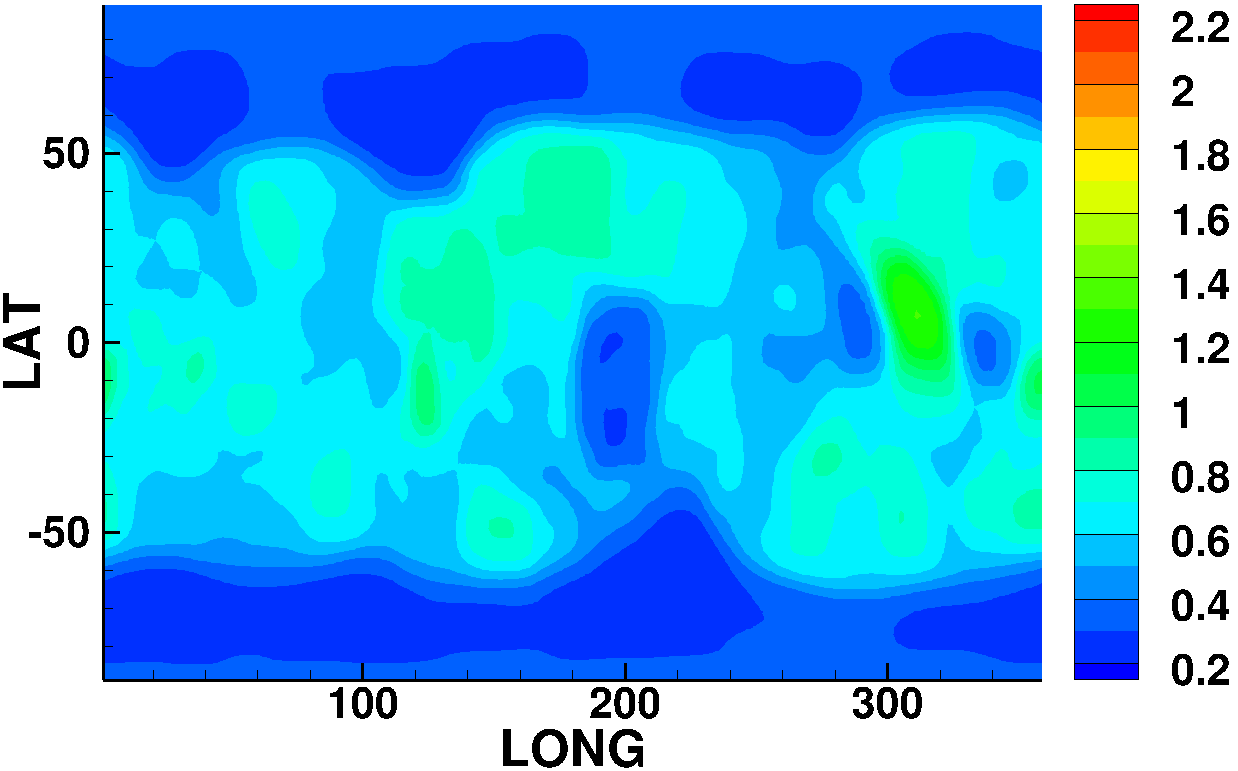}{0.33\textwidth}{Ne AWSoM (10$^{8}$ cm$^{-3}$)}
          \fig{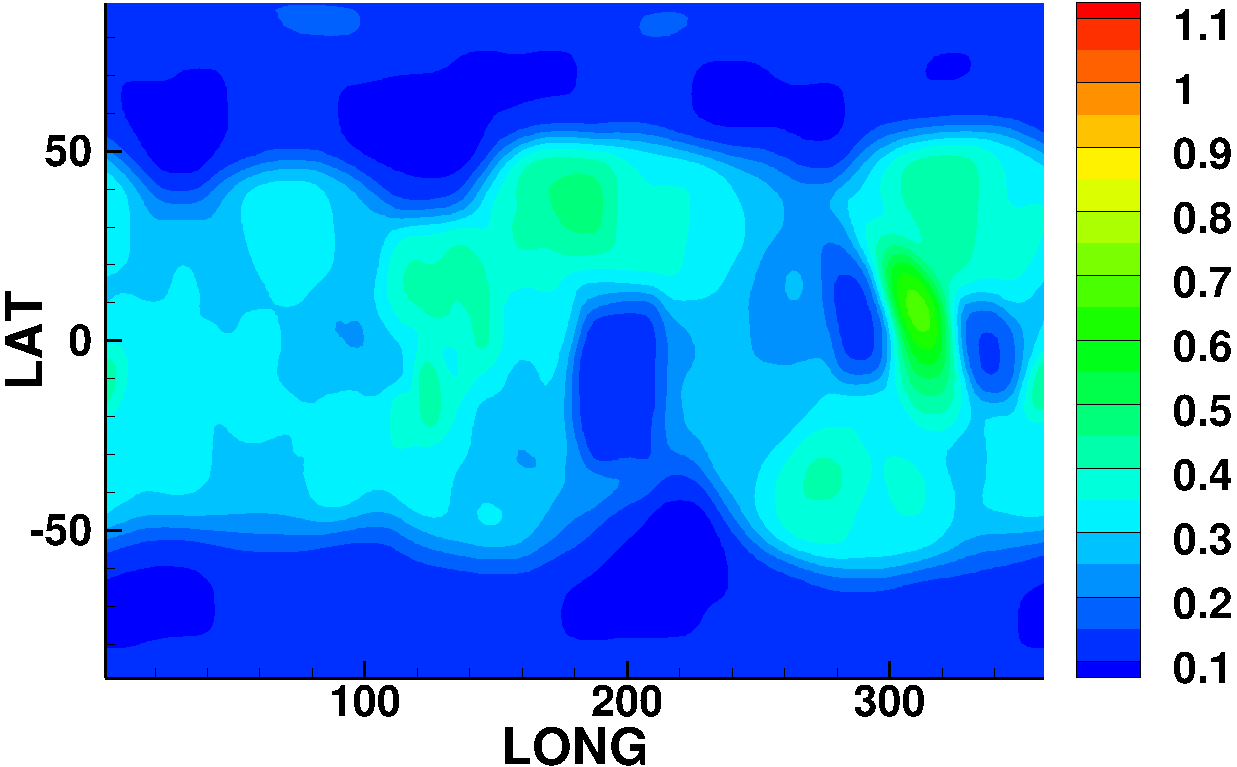}{0.33\textwidth}{Ne AWSoM (10$^{8}$ cm$^{-3}$)}
          }
\gridline{\fig{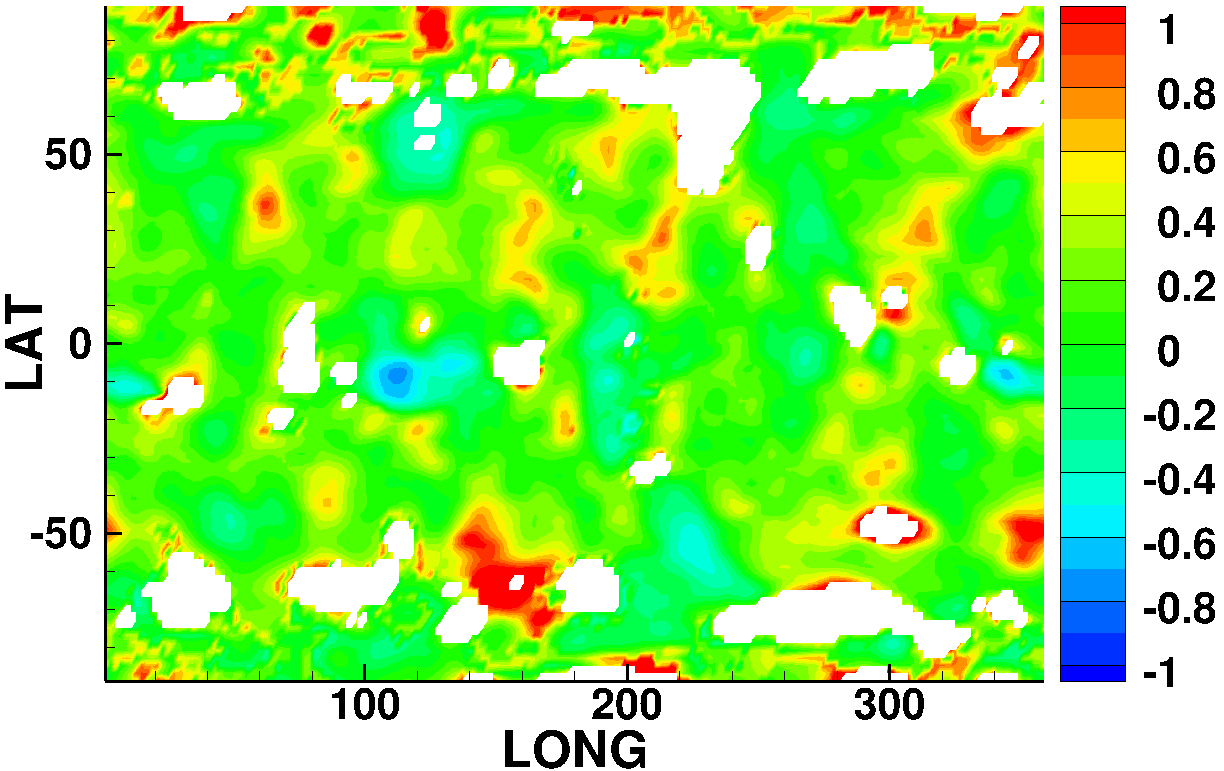}{0.33\textwidth}{Ne Rel Diff}
          \fig{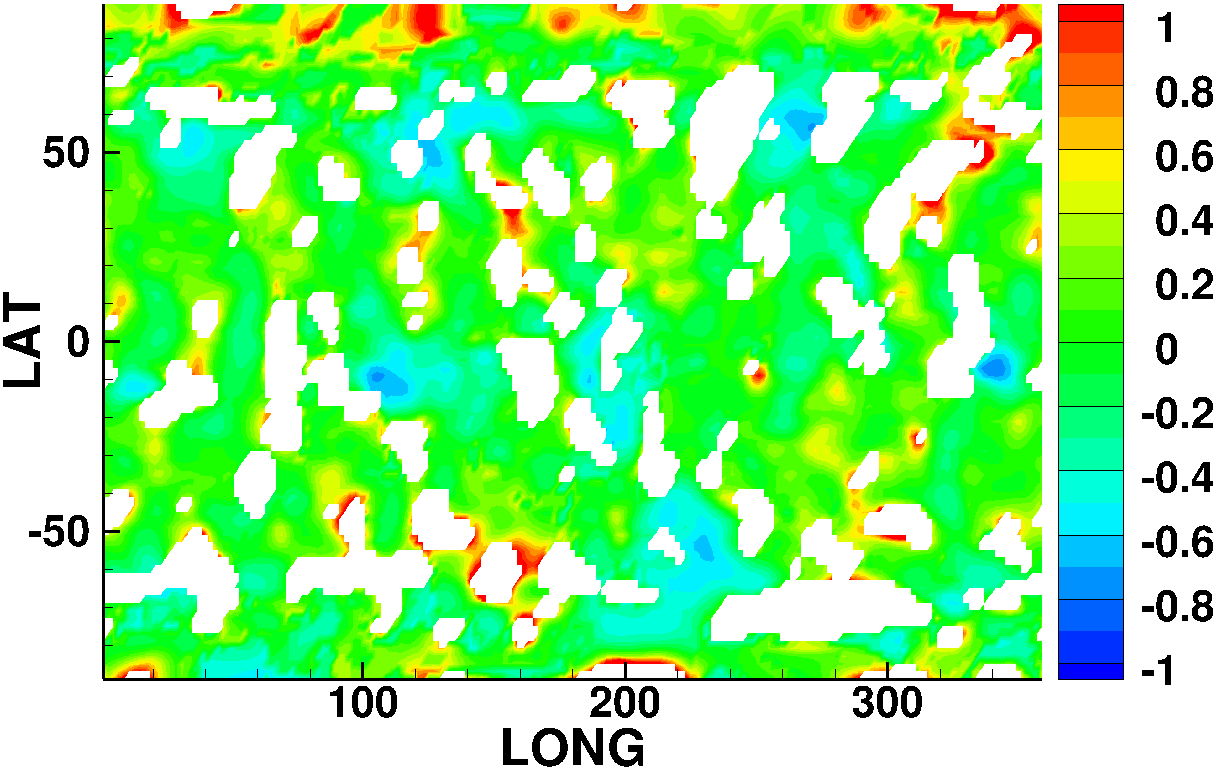}{0.33\textwidth}{Ne Rel Diff}
          \fig{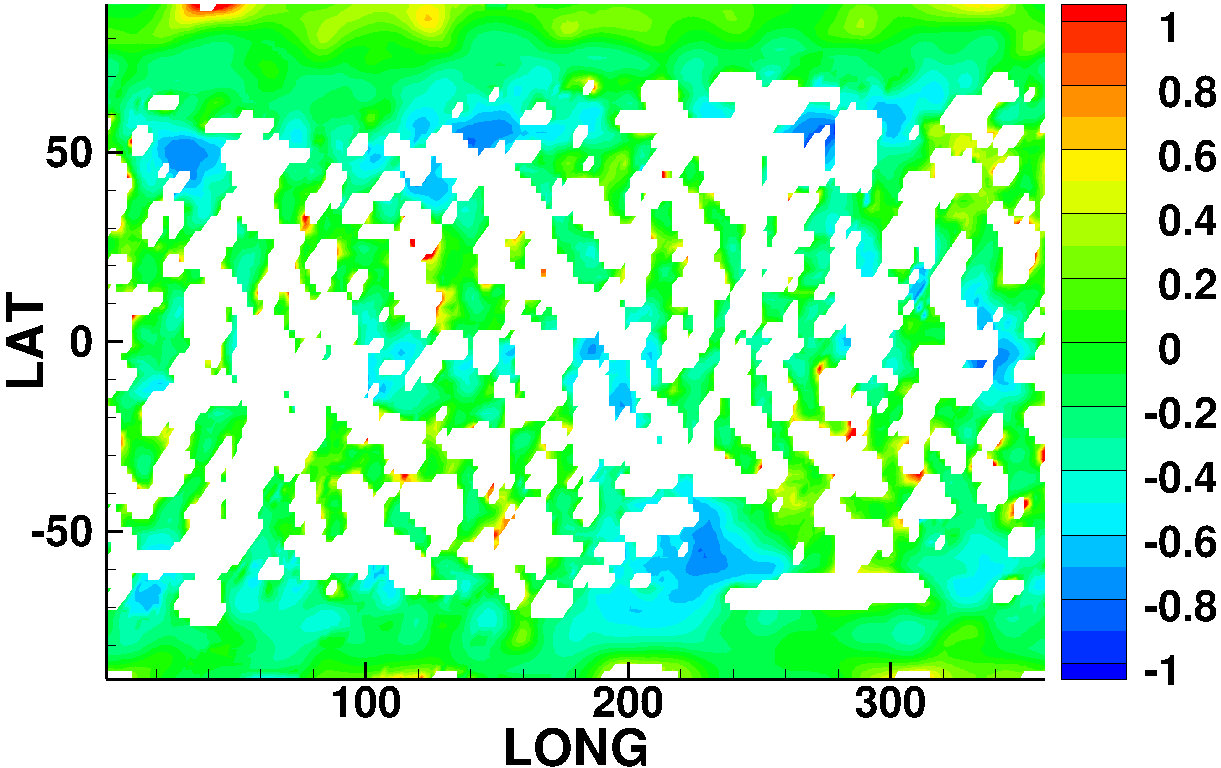}{0.33\textwidth}{Ne Rel Diff}
          }
\gridline{\fig{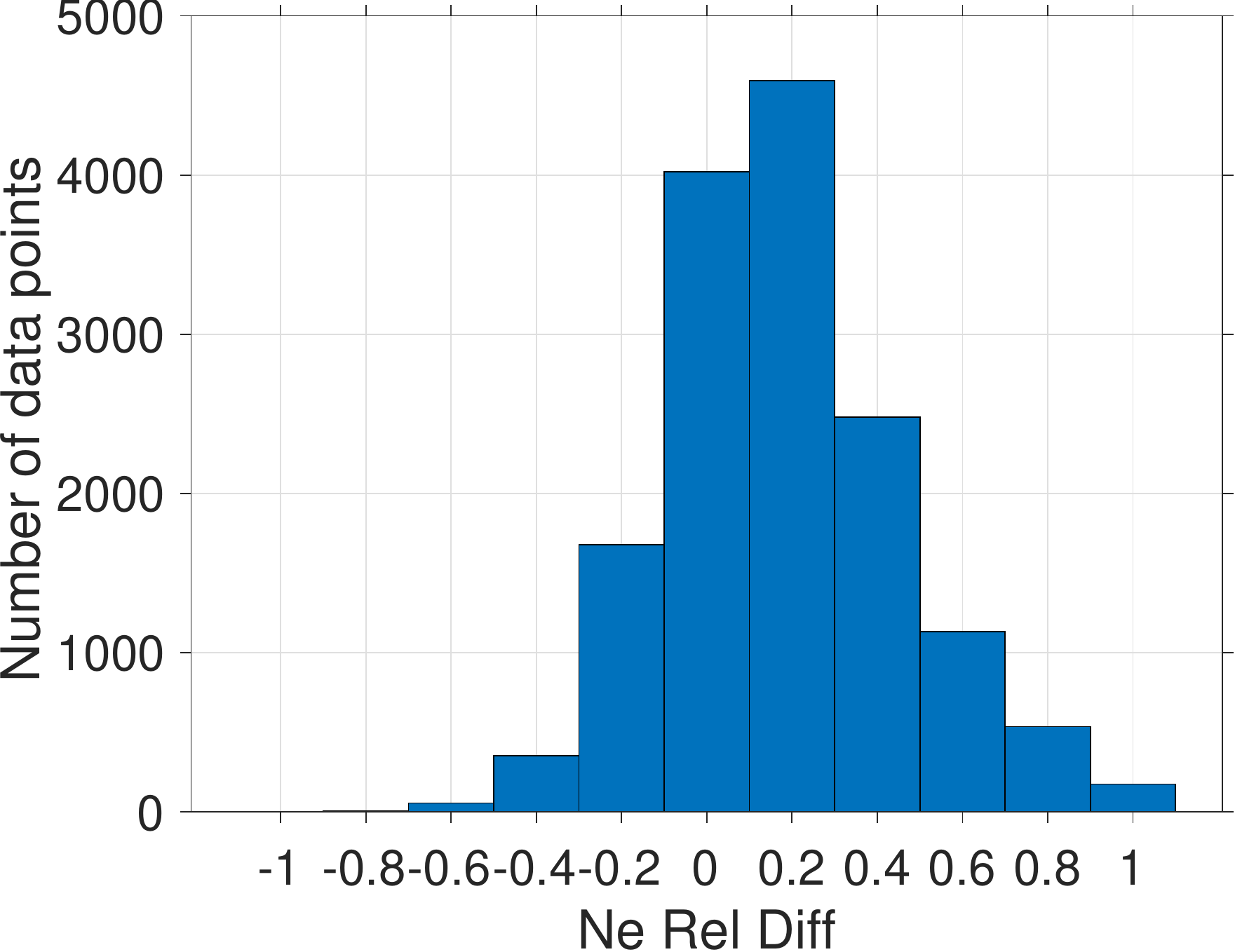}{0.3\textwidth}{(a) r=1.055 \Rs}
          \fig{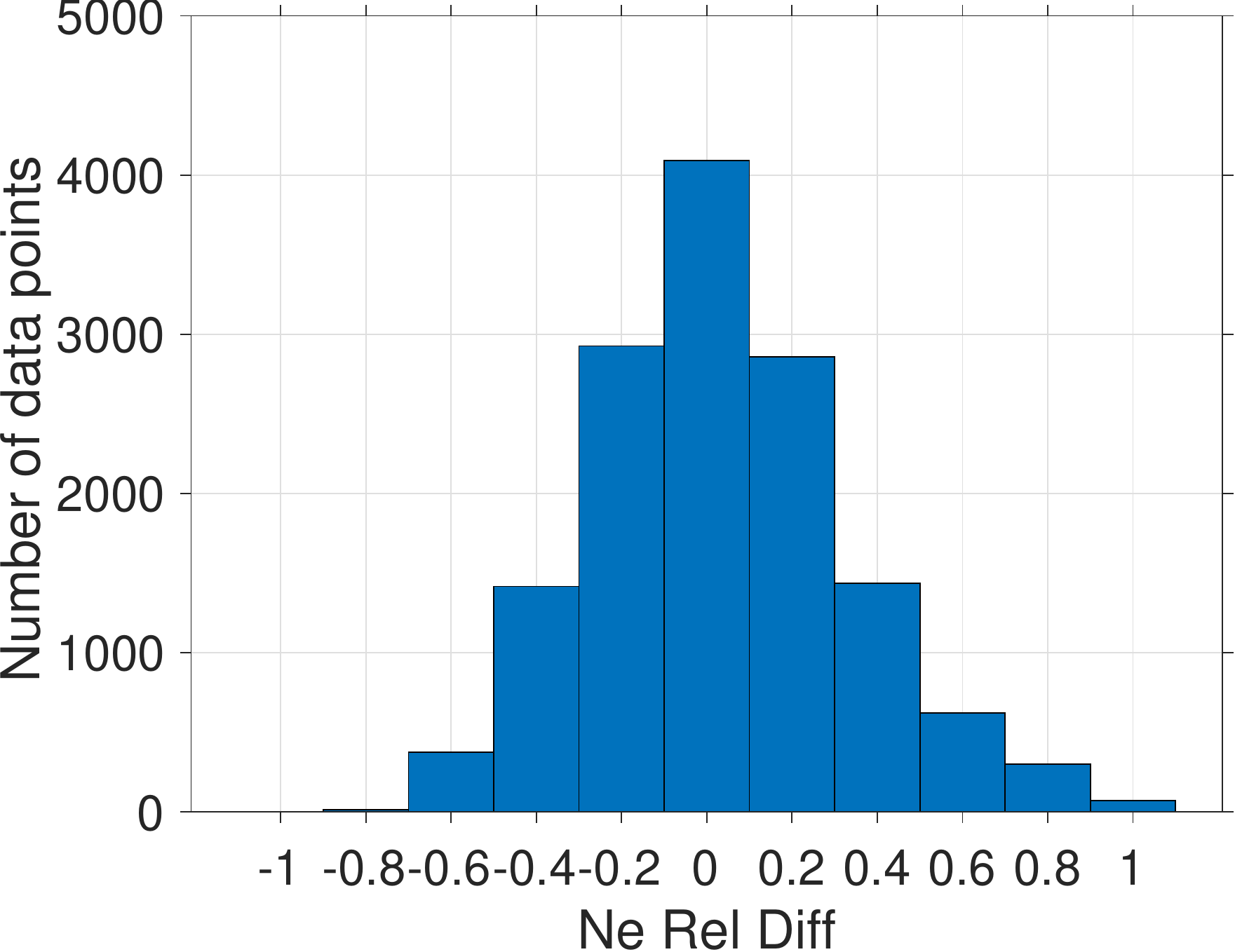}{0.3\textwidth}{(b) r=1.105 \Rs}
          \fig{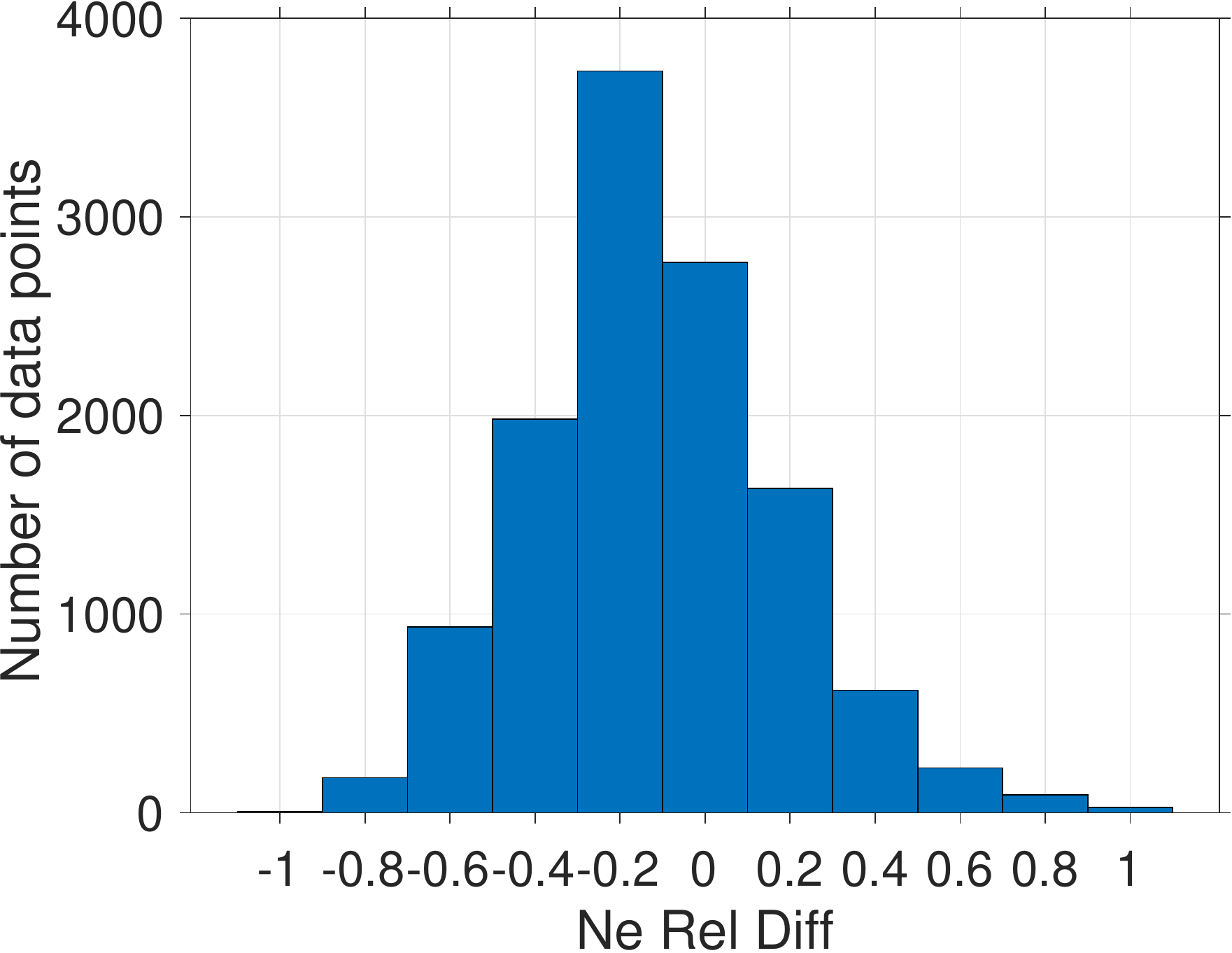}{0.3\textwidth}{(c) r=1.205 \Rs}
          }
\caption{Comparison of tomographic reconstructions of electron density from EUV observations and AWSoM model simulation results for CR2209 at (a) 1.055\,\Rs, (b) 1.105\,\Rs~and (c) 1.205\,\Rs. First and second rows show the 3D reconstructed density from SDO/AIA observations using DEMT (Ne DEMT) and the model predicted density (Ne AWSoM), respectively in units of 10$^{8}$ cm$^{-3}$. The third row depicts the relative difference between the observations and model results. The quantity shown is \rm{Ne Rel Diff}=\big($\frac{Ne_{AWSoM}}{Ne_{DEMT}}-1 \big)$. Bottom row shows the histogram distribution for \rm{Ne Rel Diff}.}
\label{fig:demt3_09}
\end{figure*}
%%%%CR2209 DEMT  TEMPERATURE
\begin{figure*}[ht!]
\gridline{\fig{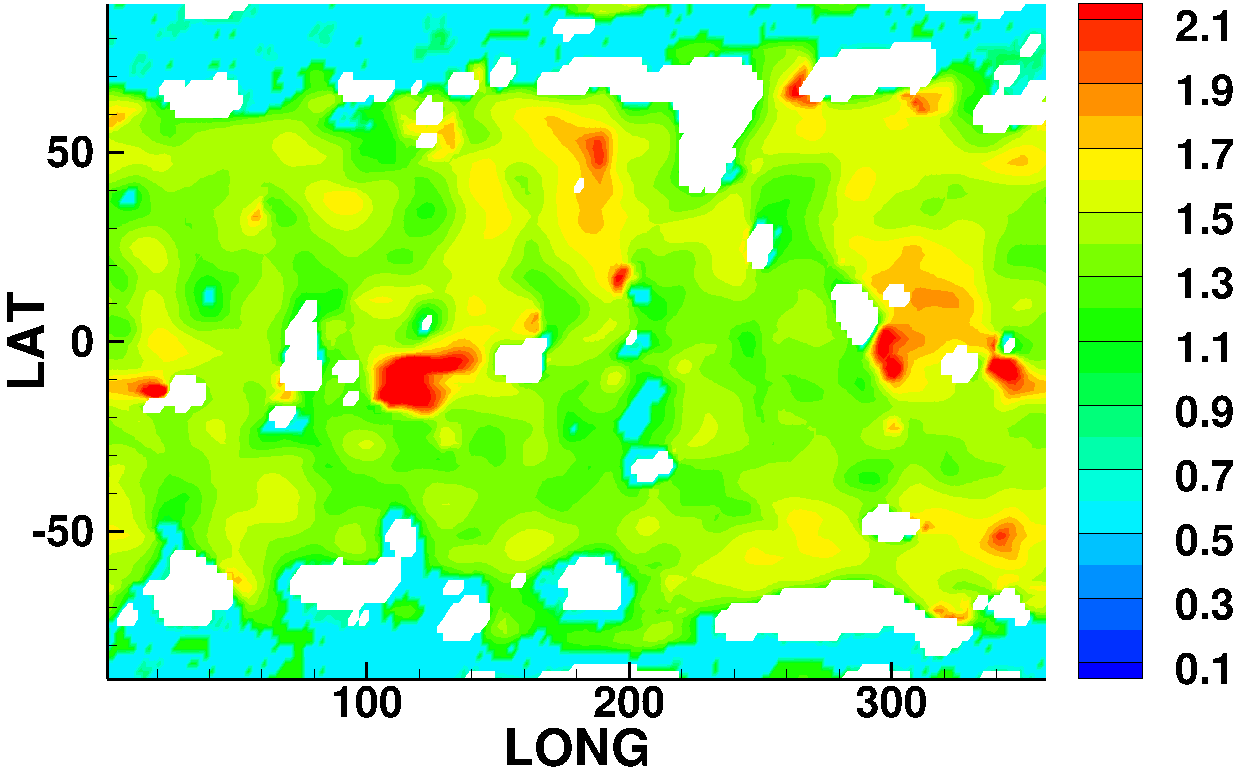}{0.33\textwidth}{Te DEMT (MK)}
          \fig{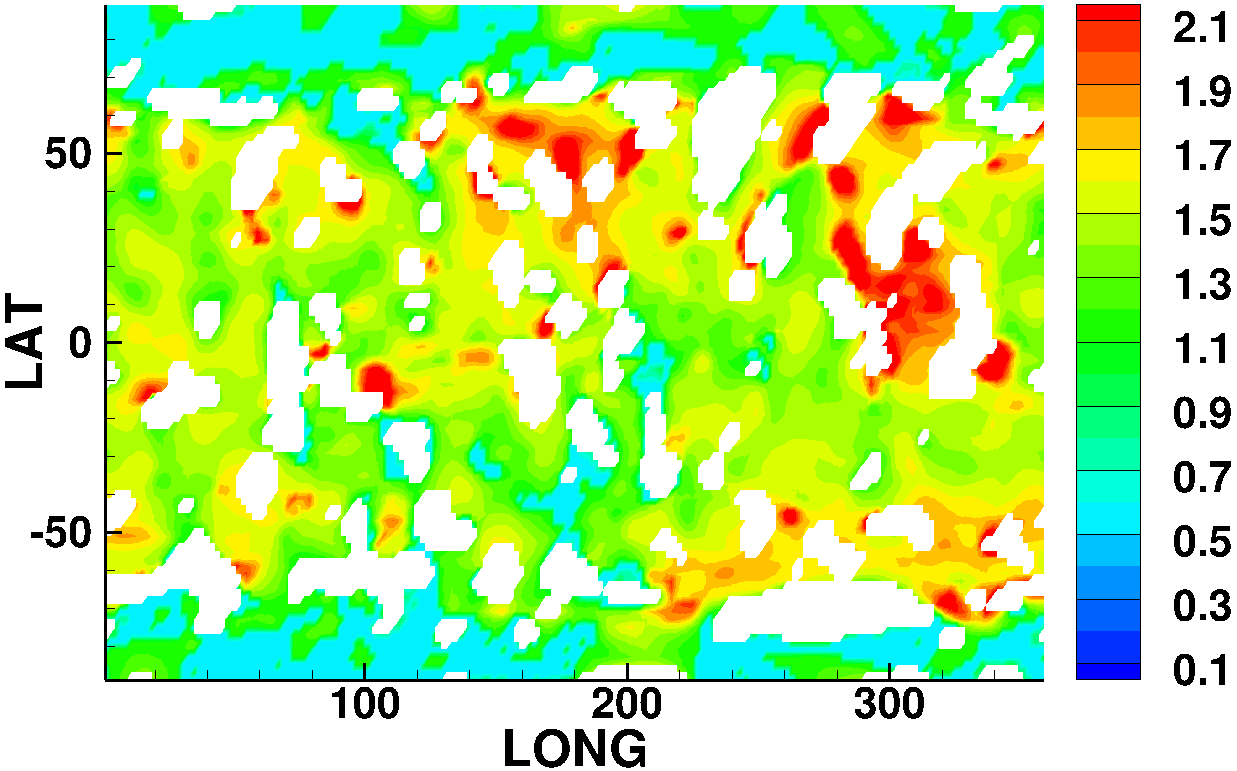}{0.33\textwidth}{Te DEMT (MK)}
          \fig{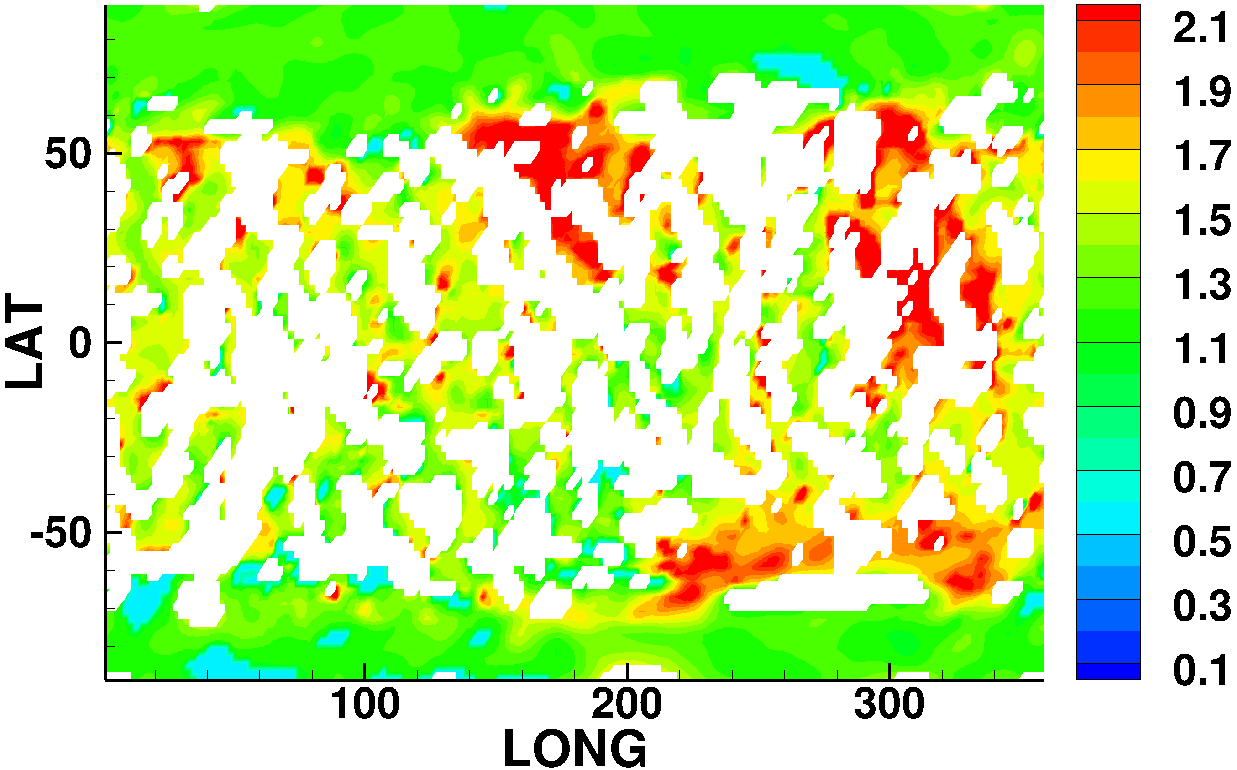}{0.33\textwidth}{Te DEMT (MK)}
          }
\gridline{\fig{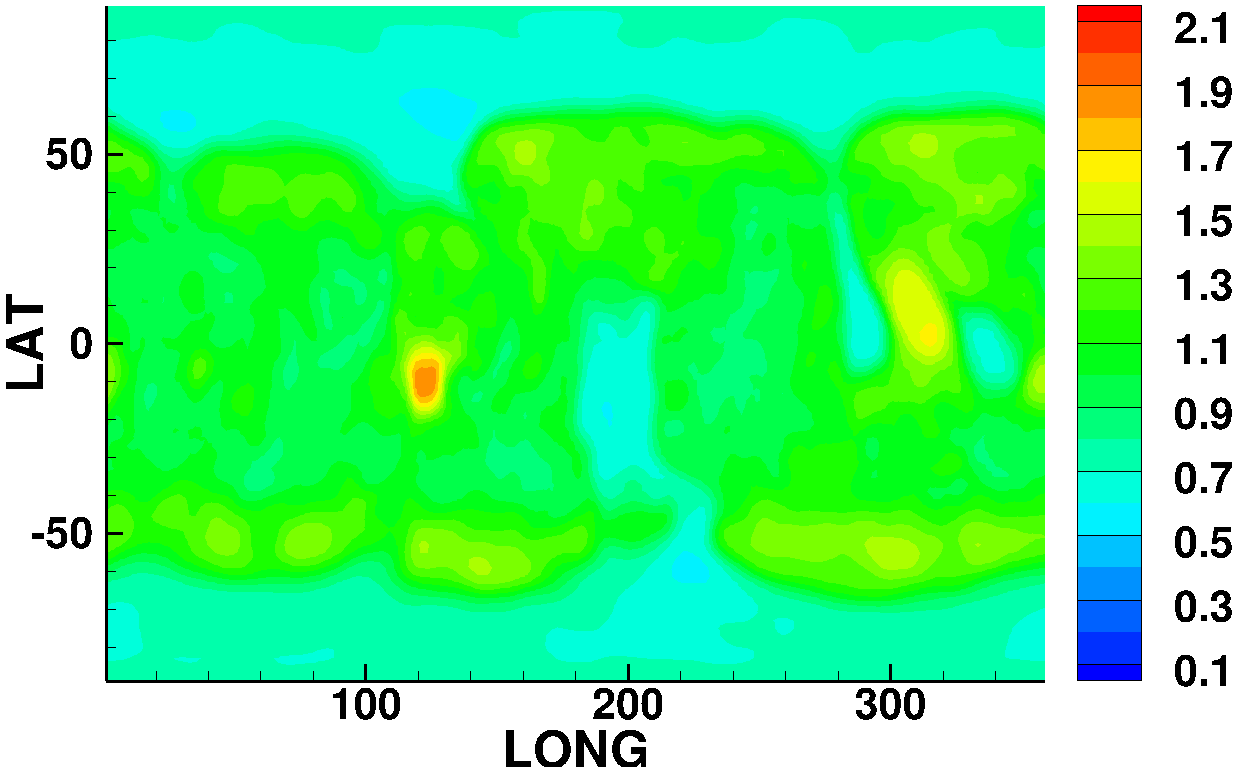}{0.33\textwidth}{Te AWSoM (MK)}
          \fig{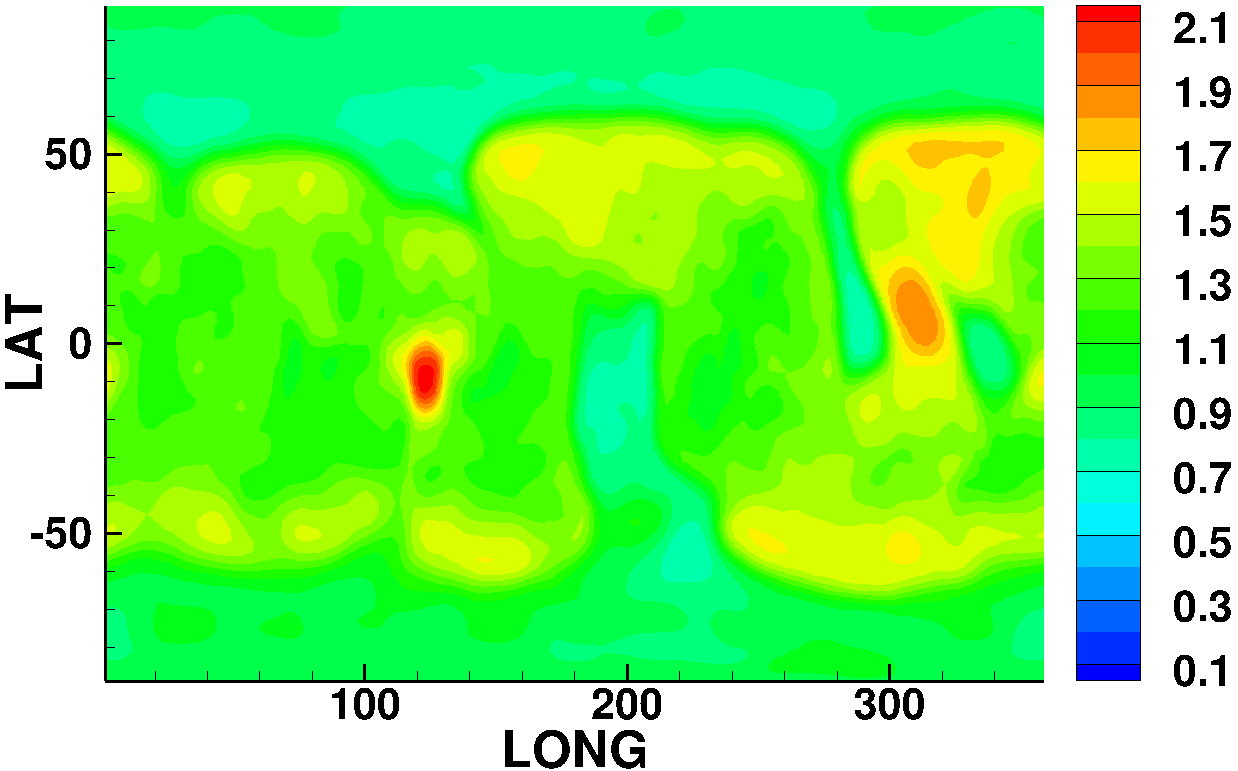}{0.33\textwidth}{Te AWSoM (MK)}
        \fig{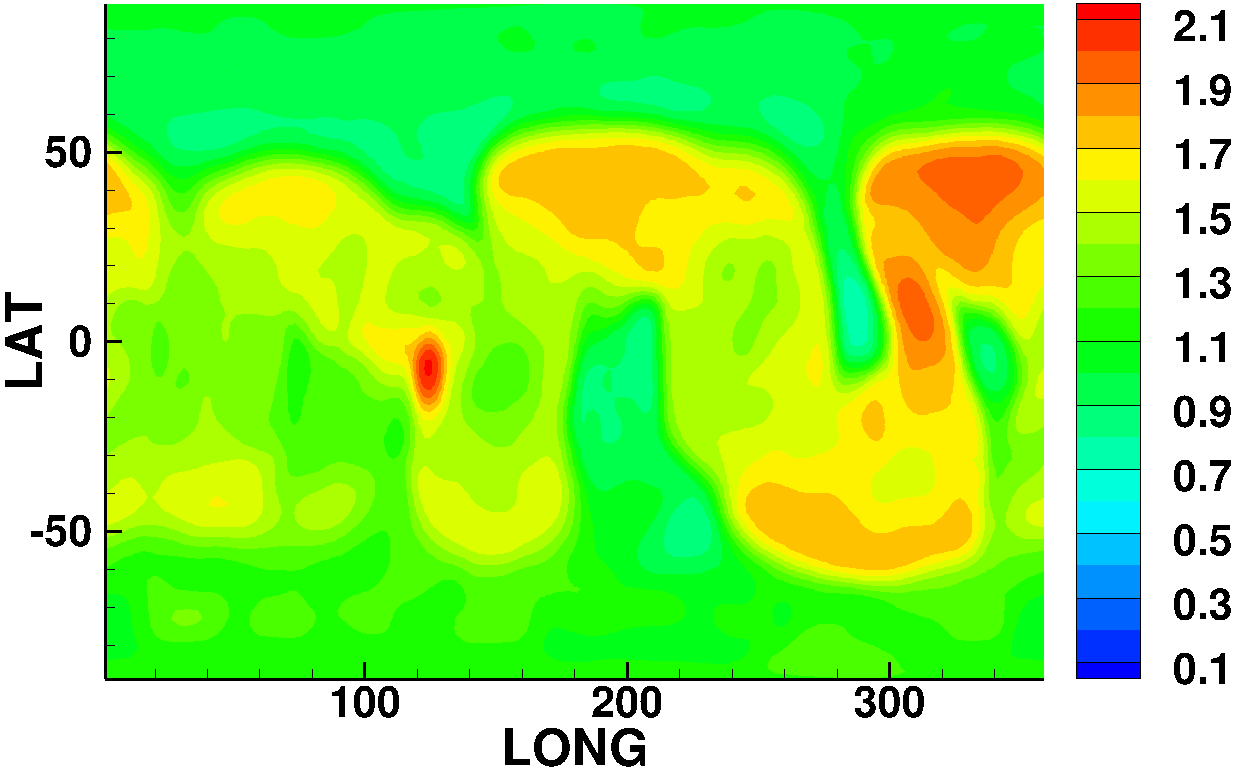}{0.33\textwidth}{Te AWSoM (MK)}
          }
\gridline{\fig{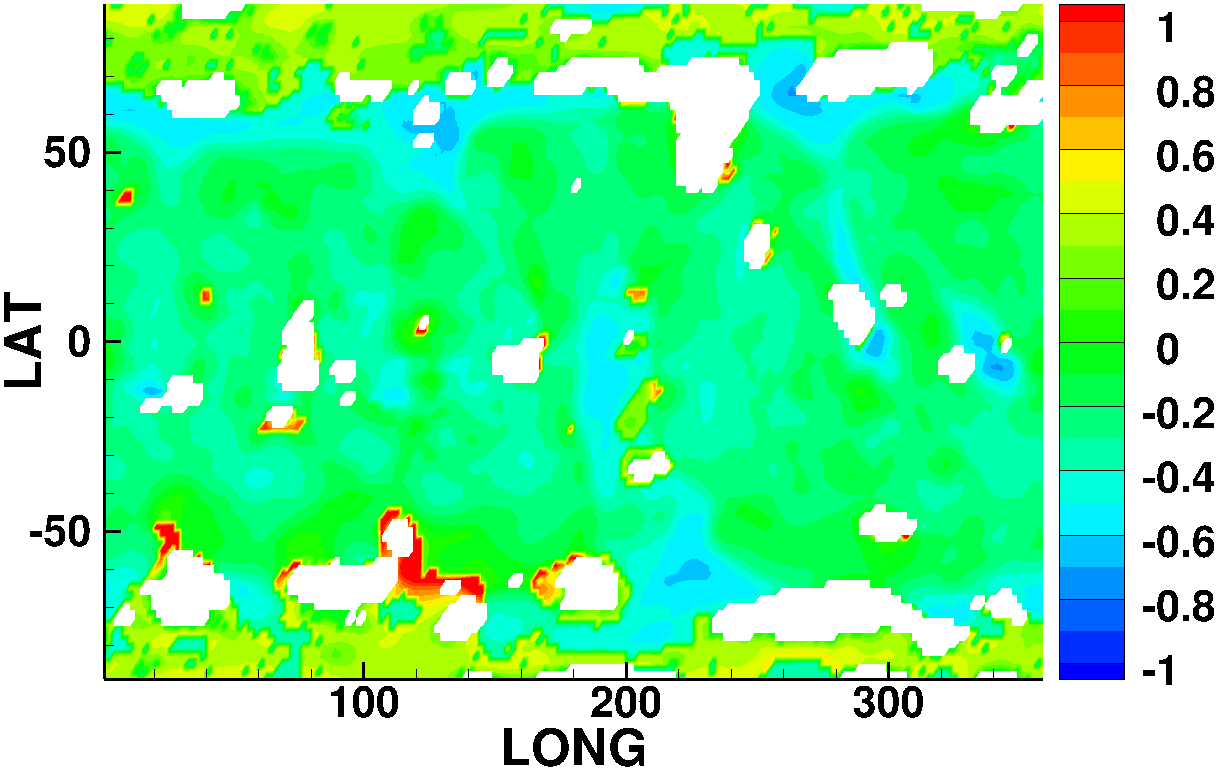}{0.33\textwidth}{Te Rel Diff}
          \fig{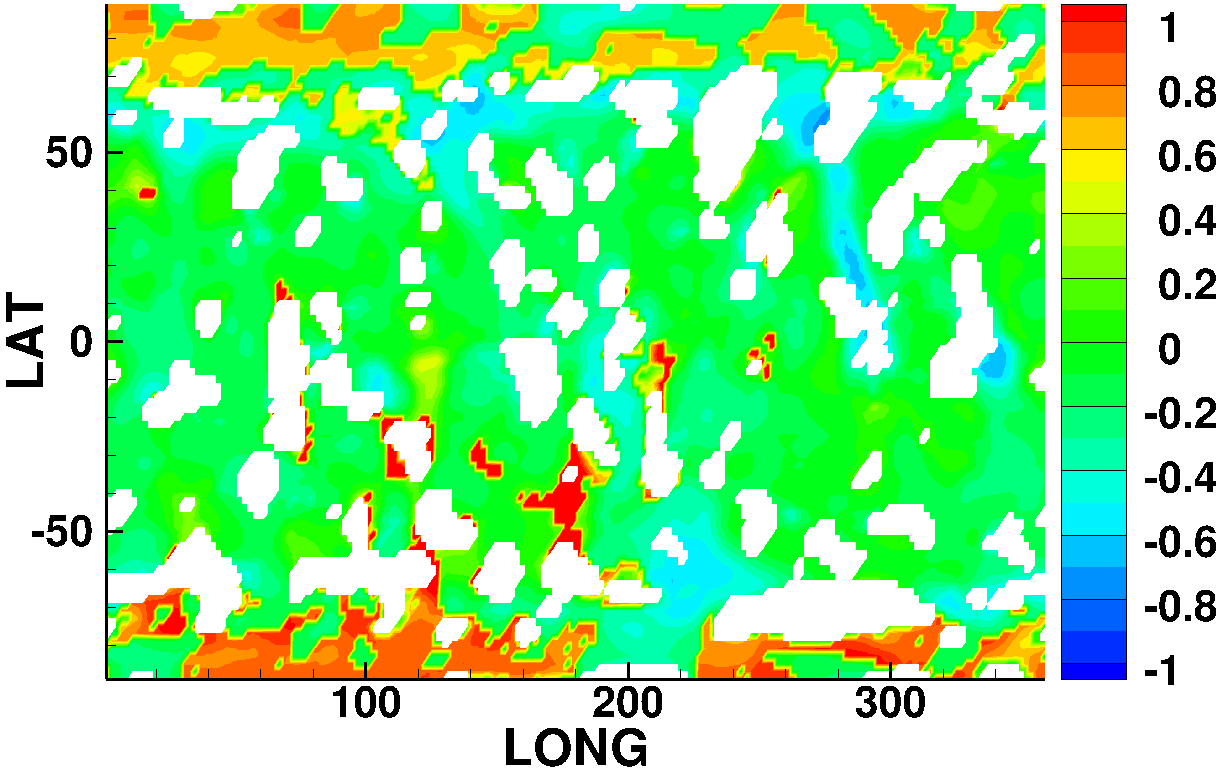}{0.33\textwidth}{Te Rel Diff}
          \fig{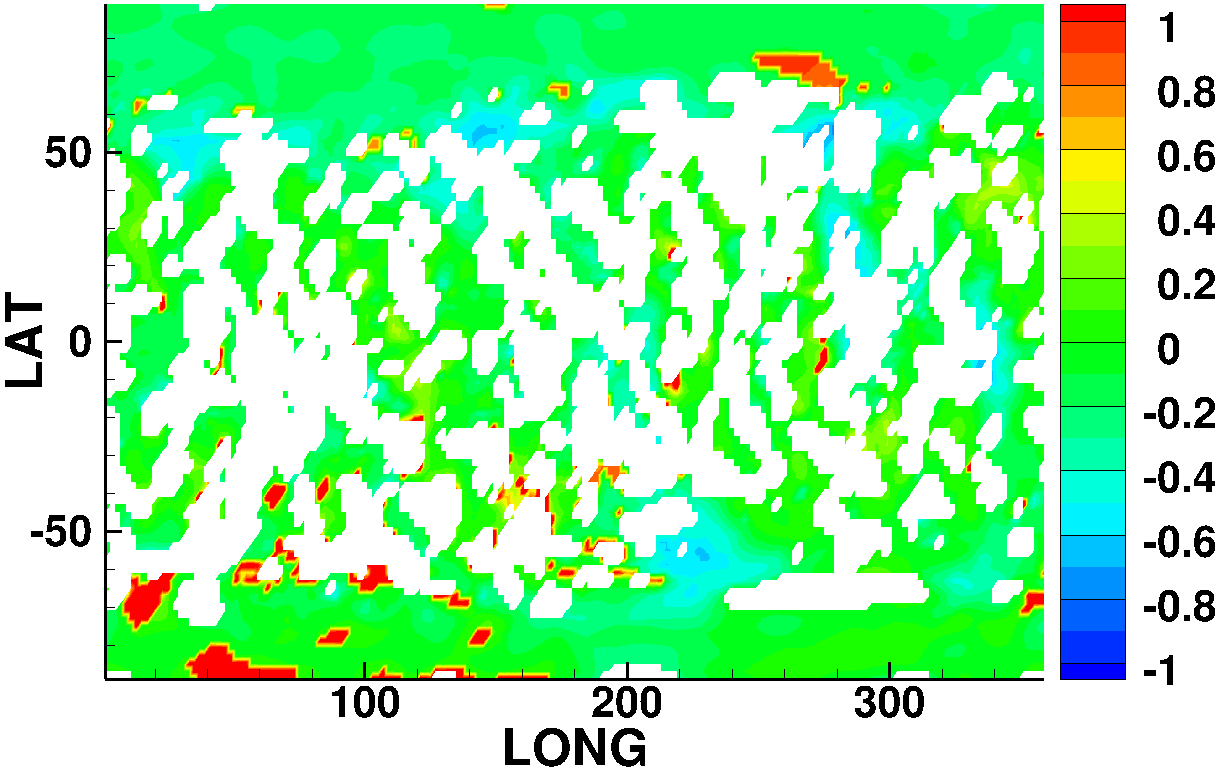}{0.33\textwidth}{Te Rel Diff}
          }
\gridline{\fig{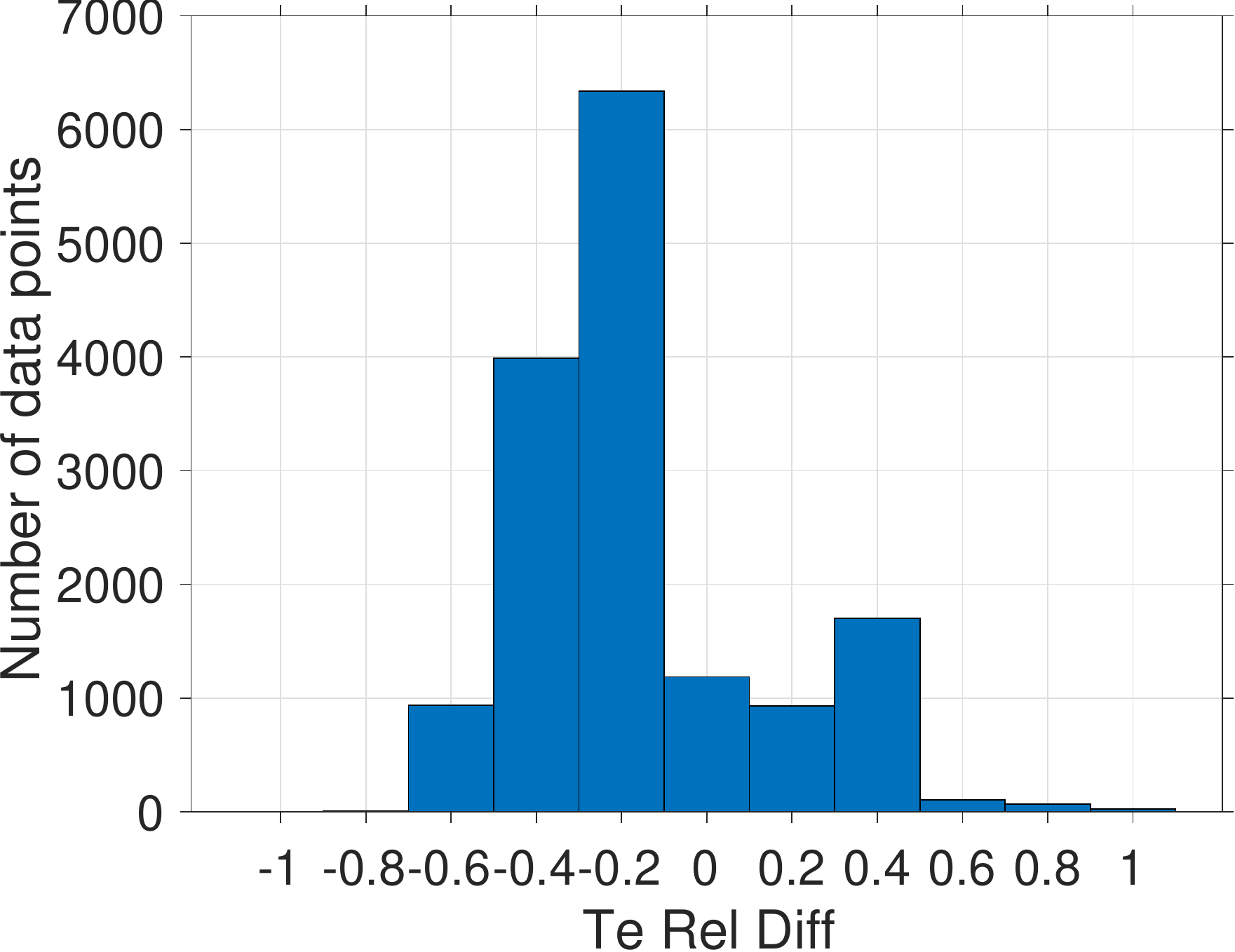}{0.3\textwidth}{(a) r=1.055 \Rs}
          \fig{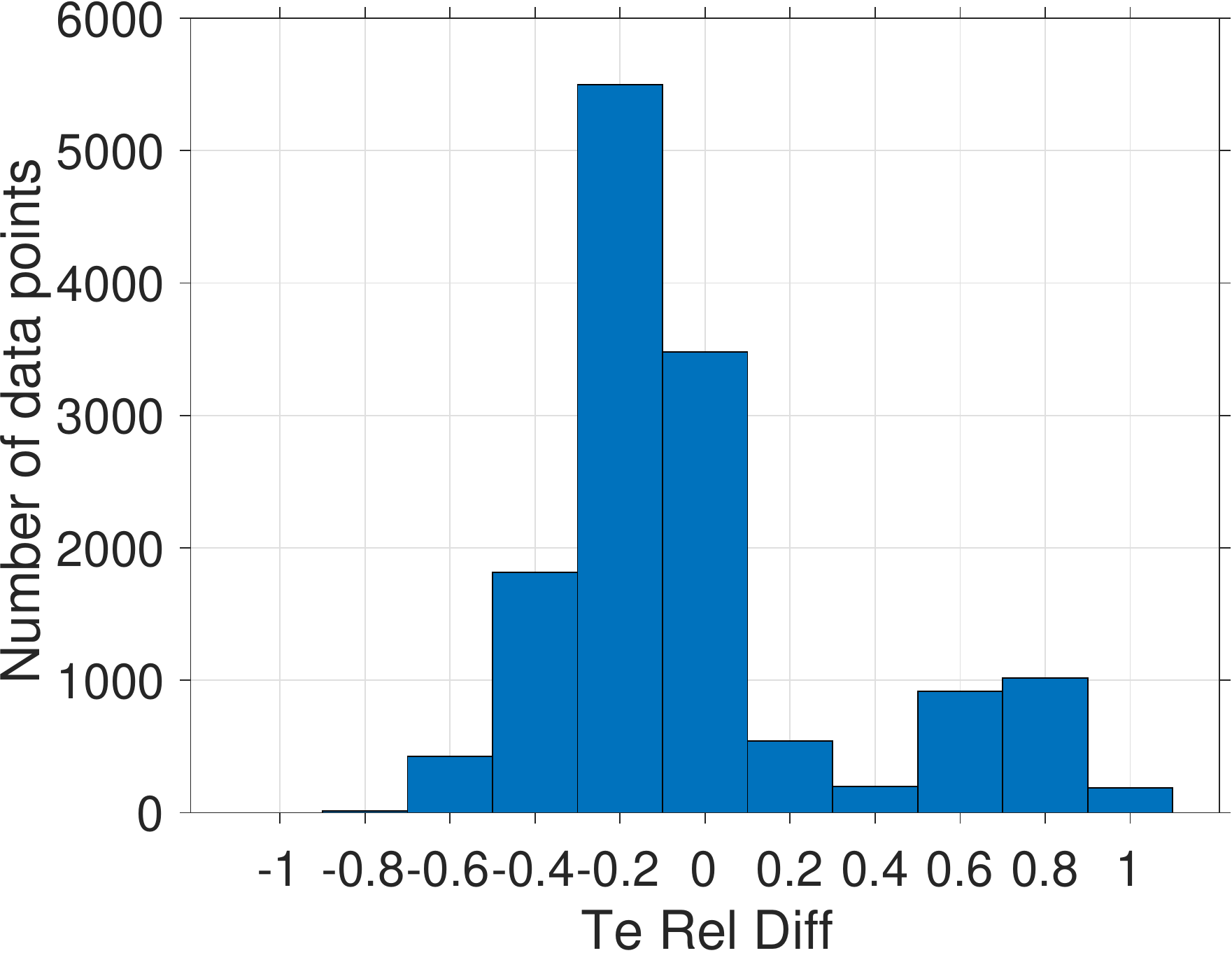}{0.3\textwidth}{(b) r=1.105 \Rs}
          \fig{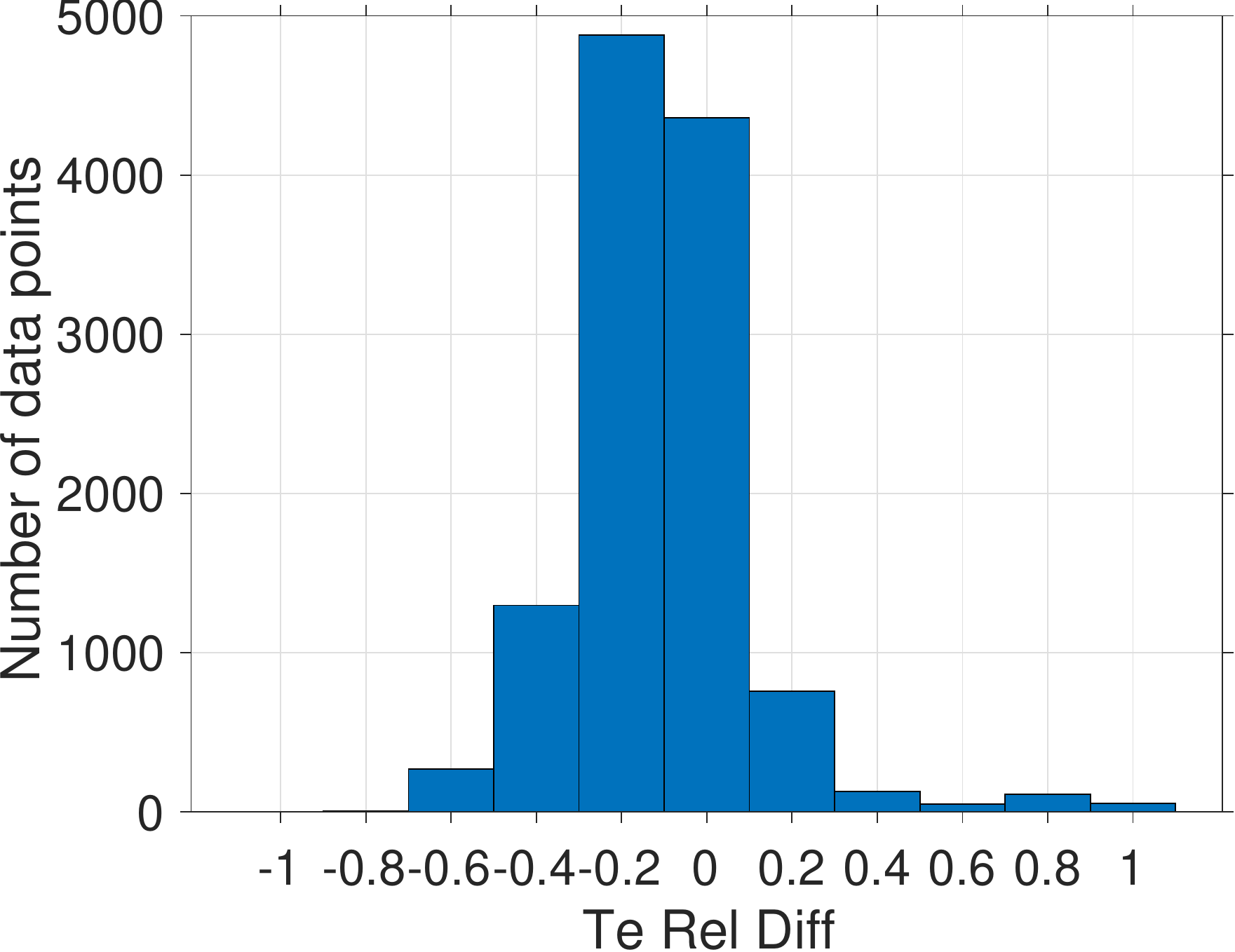}{0.3\textwidth}{(c) r=1.205 \Rs}
          }
\caption{Comparison of tomographic reconstructions of electron temperature from EUV observations and AWSoM model simulation results for CR2209 at (a) 1.055\,\Rs, (b) 1.105\,\Rs~and (c) 1.205\,\Rs. First and second rows show the 3D reconstructed temperature from SDO/AIA observations using DEMT (Te DEMT) and the model predicted temperature (Te AWSoM), respectively in units of $10^{6}$ K (MK). The third row depicts the relative difference between the observations and model results. The quantity shown is \rm{Te Rel Diff} $=\big(\frac{Te_{AWSoM}}{Te_{DEMT}}-1 \big)$. Bottom row shows the histogram distribution for \rm{Te Rel Diff}.}
\label{fig:demt4_09}
\end{figure*}
%%%%CR2208 X=0 DEMT
\begin{figure*}[ht!]
%\pdfscale{1}
\gridline{\fig{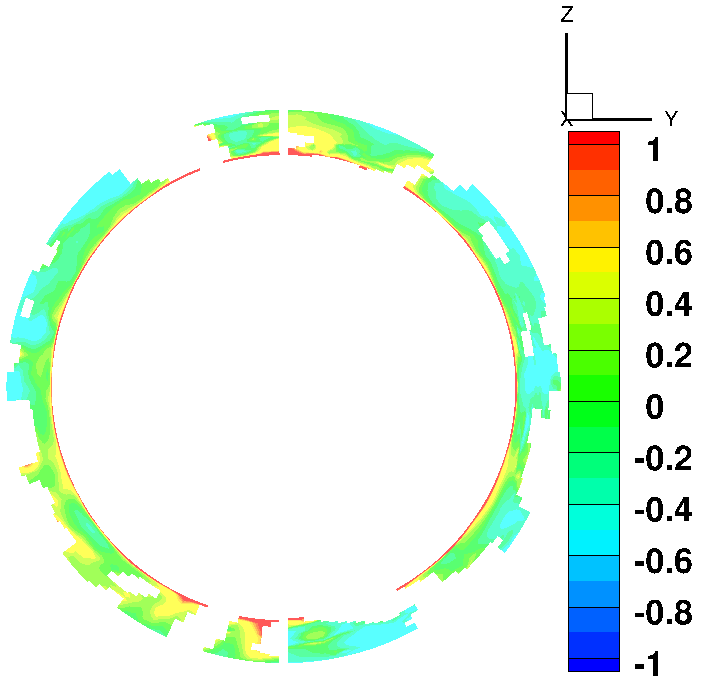}{0.32\textwidth}{(a) $\frac{Ne_{AWSoM}}{Ne_{DEMT}}-1$}
         \fig{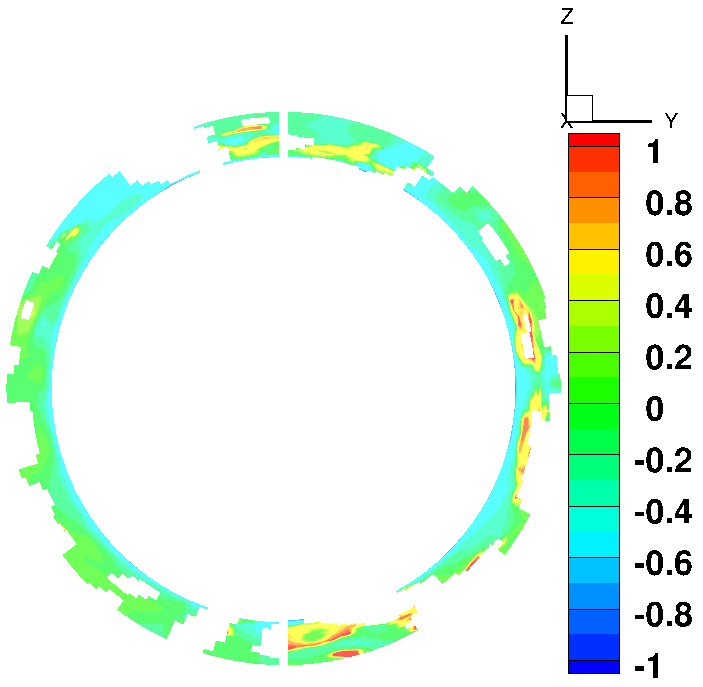}{0.32\textwidth}{(b) $\frac{Te_{AWSoM}}{Te_{DEMT}}-1$}
         }
\caption{X=0 slice for CR2208 showing the relative difference in (a) electron density and (b) electron temperature for $1.025<r<1.225$\,\Rs.}
\label{fig:demt5_08}              
\end{figure*}
%%%%CR2209 X=0 DEMT
\begin{figure*}[ht!]
%\pdfscale{1}
\gridline{\fig{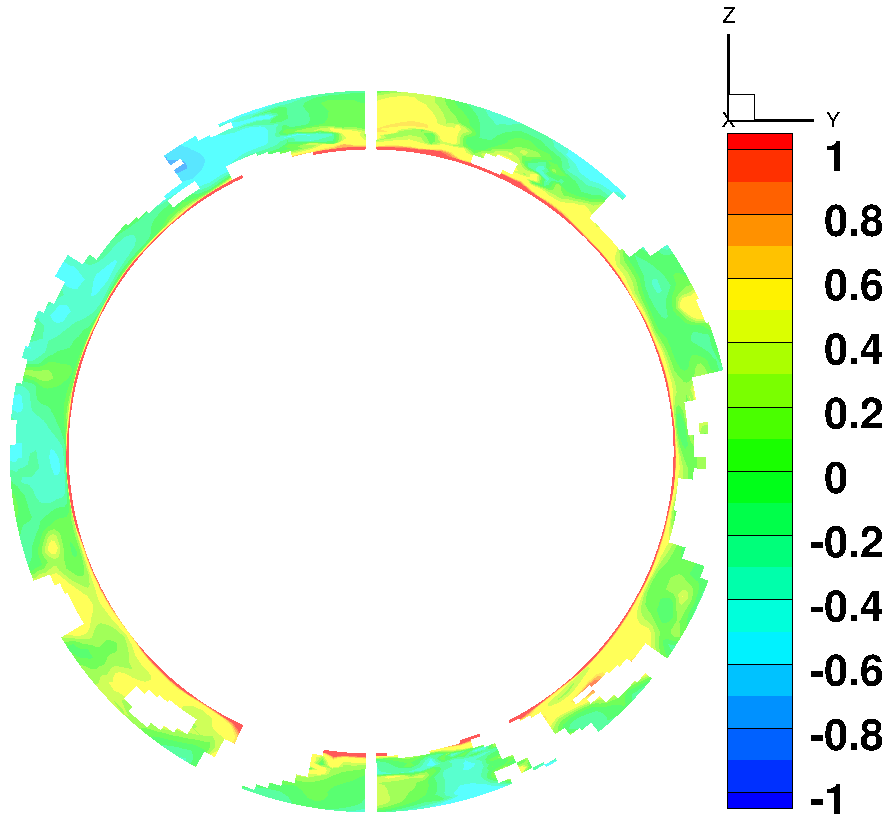}{0.32\textwidth}{(a) $\frac{Ne_{AWSoM}}{Ne_{DEMT}}-1$}
         \fig{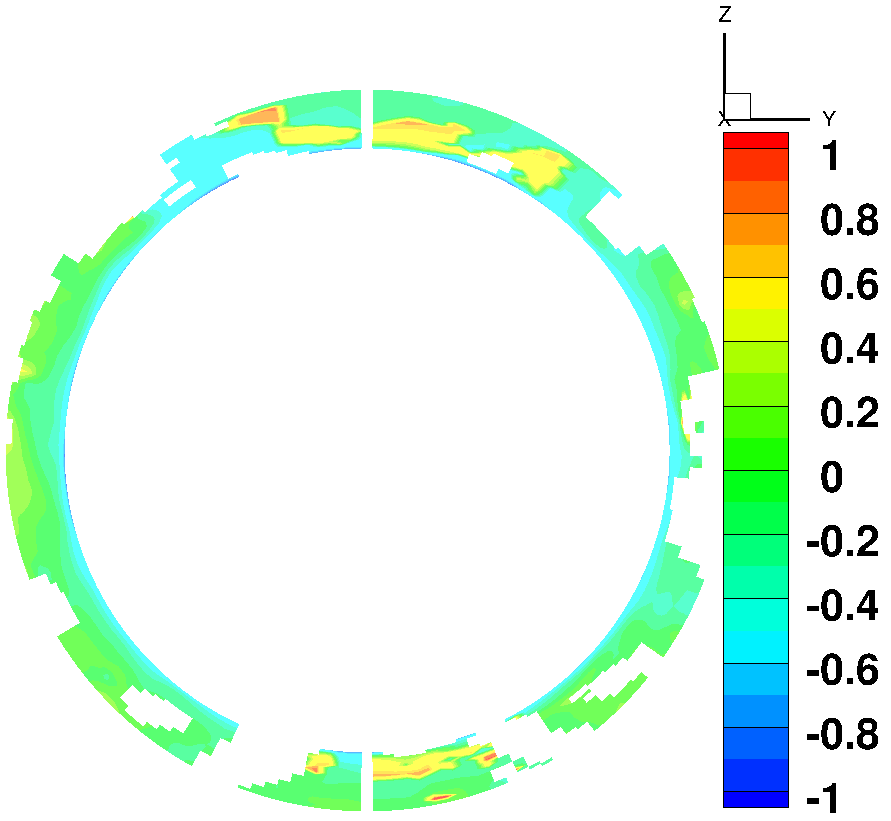}{0.32\textwidth}{(b) $\frac{Te_{AWSoM}}{Te_{DEMT}}-1$}
         }
\caption{X=0 slice for CR2209 showing the relative difference in (a) electron density and (b) electron temperature for $1.025<r<1.225$\,\Rs.}
\label{fig:demt6_09}              
\end{figure*}
\begin{figure*}[ht!]
%\pdfscale{1}
\gridline{\fig{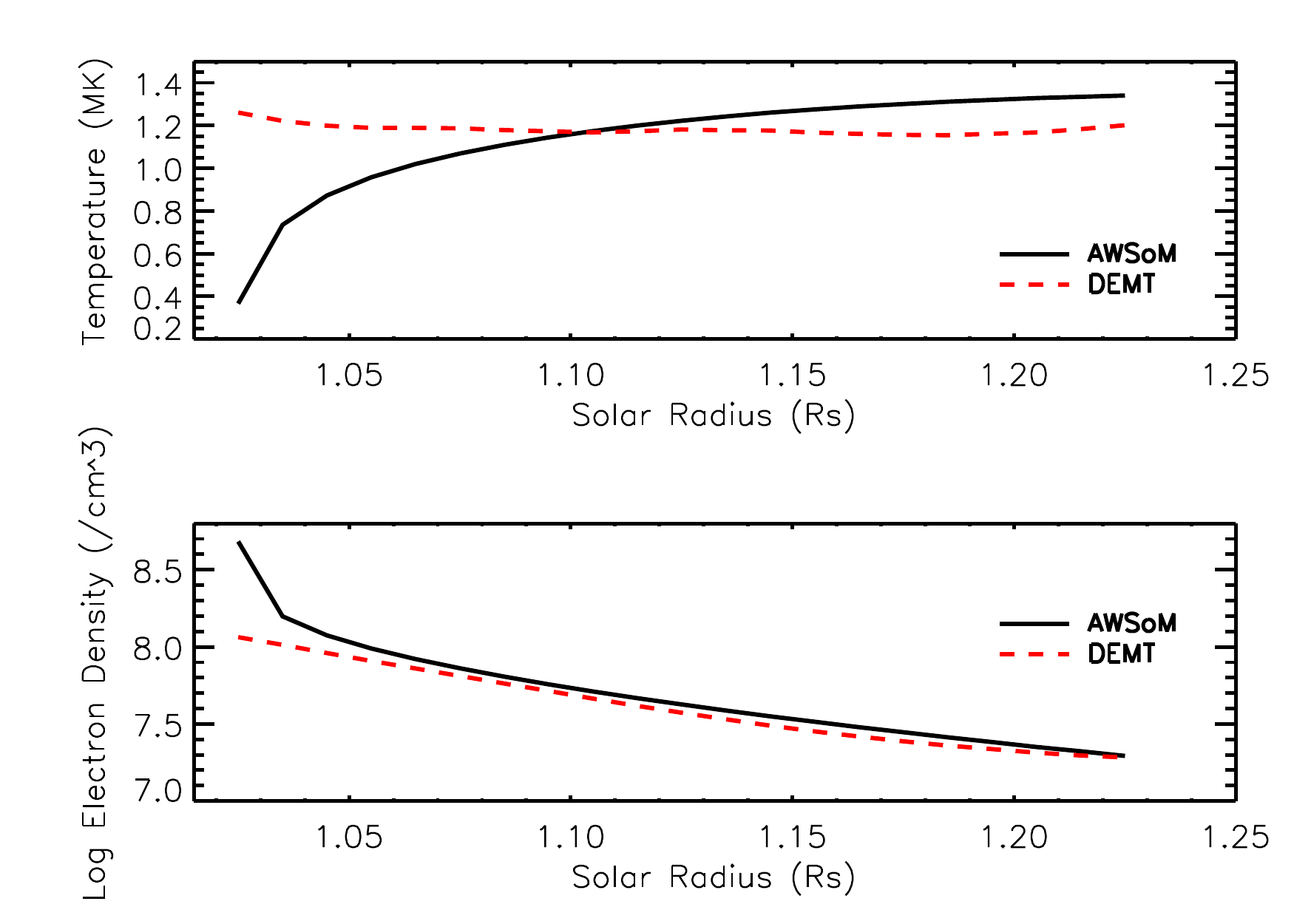}{0.38\textheight}{(a) CR2208}
         \fig{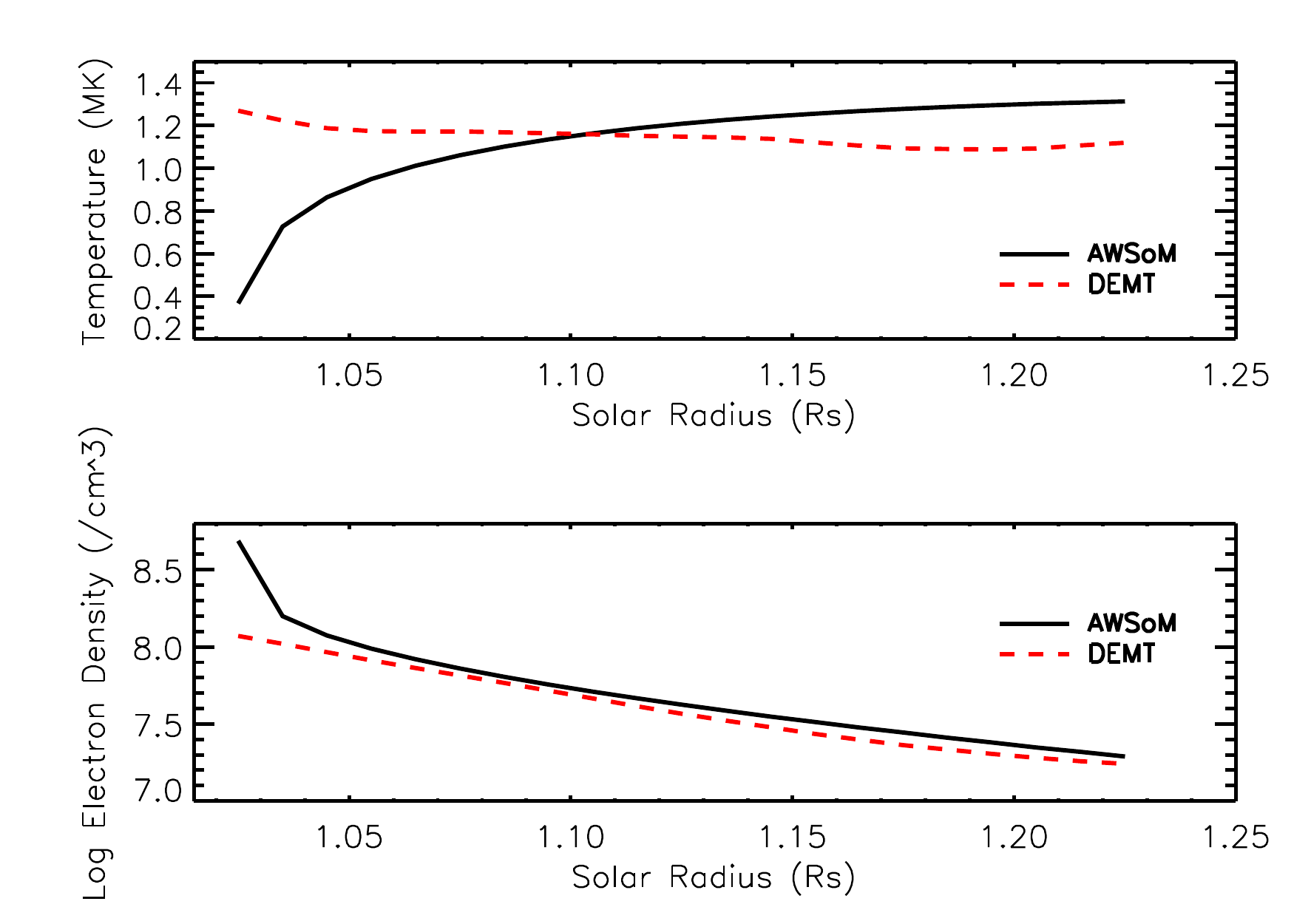}{0.38\textheight}{(b) CR2209}
         }
\caption{Variation of the longitude-latitude averaged electron temperature (in MK) and log electron density (in cm$^{-3}$) from AWSoM simulations (black) and DEMT reconstruction (red) for CR2208 (left) and CR2209 (right) with the radial distance ranging between 1.025 - 1.225\, \Rs.}
\label{fig:demt_awsom}              
\end{figure*}

\subsection{Differential Emission Measure Tomography (DEMT)}\label{sec:DEMT}
Differential Emission Measure Tomography (DEMT) is a solar rotational tomography technique which employs a time series of EUV images to reconstruct the 3D Differential Emission Measure (DEM) in the solar corona \citep{Fra2005,Fra2009,Vas2016}. DEMT combines the EUV tomography in several pass bands with local DEM analysis to produce 3D distributions of the coronal electron density and temperature in the radial range of 1.025 - 1.225\,\Rs. \citet{Vas2010} and \citet{Llo2017} used DEMT for a comparative analysis of the coronal structure during solar minima. 

%The coronal heating effects in the three-temperature AWSoM model due to non-linear turbulent cascade based on interaction of forward propagating and reflected Alfv\'{e}n waves can be studied by comparing the electron temperature and densities derived from EUV images using DEMT analysis.
In this study, the DEMT analysis for CR2208 and CR2209 uses the superior high-cadence SDO/AIA data to have better signal to noise ratio. In this work, the technique uses for the first time a newly implemented 3D regularization scheme instead of the latitude-longitude regularization scheme used in the previous DEMT efforts.
This implies that the tomography results are now more trustworthy at the lowest heights and boundary-induced artifacts are minimized. For each instrument, the DEMT analysis entails a cross-validation study to determine the optimal regularisation level. This level is different for each wavelength band and is sensitive to the activity level of the Sun.
We obtained tomographic reconstructions for each of the rotations using $1/2$ rotation of off-limb data, fully blocking the disk (hollow tomography). Here, we show comparisons with the hollow tomography reconstructions for both CR2208 and CR2209.

Figures \ref{fig:demt1_08} and \ref{fig:demt2_08} show the comparisons between the DEMT reconstructed electron density and temperature and the model output, respectively at three radial distances, r=1.055, 1.105 and 1.205\,\Rs. 
The top two panels in Figure \ref{fig:demt1_08} show the longitude-latitude maps for the tomographic electron density (Ne DEMT) and model output (Ne AWSoM) in units of $10^{8}$\,cm$^{-3}$. The white regions in the DEMT maps are zones not reliably reconstructed by the tomography, as discussed below. The bottom two panels show the relative difference in electron density, \rm{Ne Rel Diff}$=\big(Ne_{AWSoM}/Ne_{DEMT}\big)-1$ and the corresponding histogram distribution. Figure \ref{fig:demt2_08} shows the same results for the electron temperature in units of MK. Top two panels show Te DEMT and Te AWSoM, and the bottom two panels show \rm{Te Rel Diff}$=\big(Te_{AWSoM}/Te_{DEMT}\big)-1$. Figures \ref{fig:demt3_09} and \ref{fig:demt4_09} show the same quantities for CR2209. 

The white regions in the DEMT maps in Figures \ref{fig:demt1_08}-\ref{fig:demt6_09} are those for which the tomography can not provide a reliable reconstruction. These regions include cells where the reconstructed emissivity, forced to be positive, is null in at least one of the bands. These are called \emph{zero-density-artifacts}, which are caused by coronal dynamics not accounted by the DEMT technique (see, \citealp{Fra2009,Llo2017}).

In cells where DEMT provides positive emissivities, the local-DEM (LDEM) of each voxel is determined. The resulting DEM is then evaluated in each voxel for consistency with the tomographic reconstruction of the emissivity in all three bands. To that end, we define a quantity, R which is the fractional difference between the tomographic emissivity and the synthetic one predicted by the DEM of that voxel, averaged for three EUV bands. In other words, R is a measure of the degree of success of the LDEM in reproducing the tomographically reconstructed emissivity in all three bands. R lies between 0 and 1, where 0 means a good agreement. Regions that have R$> $0.25 are excluded, which are the white regions in the data-model comparisons.

We also show the X=0 slice for relative difference in density and temperature in Figure \ref{fig:demt5_08} for CR2208 and Figure \ref{fig:demt6_09} for CR2209. It can be seen that from the innermost boundary of the tomographic computational domain (r=1.025\,\Rs) up to about 1.055\, \Rs, the model electron density is overestimated compared to the DEMT results. This overestimate is the result of artificial broadening of the transition region to be consistent with our limited numerical resolution. This is also evident from Figure \ref{fig:demt_awsom} which shows the average (over all longitudes and latitudes) of temperature and density at different radial distances between 1.025\,\Rs~and 1.225\,\Rs. The DEMT reconstructed data are shown in red and AWSoM results are shown in black for CR2208 (left) and CR2209 (right). We see that the model temperature converges to reconstructed values at lower heights, but the density cannot catch up. The comparisons get significantly better as we go higher radially. 

%This overestimate is the result of artificial broadening of the transition region. %to be consistent with our limited numerical resolution. 
The steep gradients in temperature and density in the thin transition region require excessive numerical resources to resolve on a global scale. These gradients are a result of the balance of coronal heating, heat conduction and the radiative losses. Therefore, as described in \citet{Lio2009,Sok2013} the transition region is artificially broadened so as to be properly resolved with our finest grid resolution of $\approx$ 0.001\,\Rs. This broadening of the transition region pushes the corona outwards. In addition, if the chromospheric density is too low, the transition region may evaporate. As described in Section \ref{sec:Sim}, the density at the inner boundary (upper chromosphere) is taken to be an overestimate,
which ensures that the base is not affected by chromospheric evaporation and the density of the upper chromosphere falls rapidly to correct (lower) values. At this level, the radiative losses are sufficiently low so that the temperature can increase monotonically with height and form the transition region.
Thus, at low radial distances of about 1.025\,\Rs~to 1.055\,\Rs~the AWSoM predicted density is still an overestimate compared to the DEMT reconstructed values using EUV data. \\
In addition, Alfv\'{e}n wave heating also affects energy balance in the transition region. This heating can be improved upon, as the reflection physics through the transition region is not fully accounted for. Currently, we set an artificial upper bound for the wave reflection in the transition region based on the cascade rate (details in \citealp{Van2014}). Hence, the coronal heating might be underestimated at the transition region which can further lower the temperature compared to the DEMT reconstructed values.

\subsection{LASCO-C2 solar tomography}\label{sec:LASCO}
Time-dependent solar rotational tomography (SRT) is applied to white-light coronal images obtained with the LASCO-C2 coronagraph to produce the three-dimensional electron density distributions \citep{Fra2010,Vil2016}. We compare these tomographic reconstructions to the model simulated densities at heights between 2.55 and 6\,\Rs. The LASCO-C2 images use most up-to-date superior instrumental corrections and calibration \citep{Gar2013,Lam2017} as provided by the Laboratoire d'Astrophysique de Marseille (LAM).

\begin{figure*}[ht!]
%\pdfscale{1}
\gridline{\fig{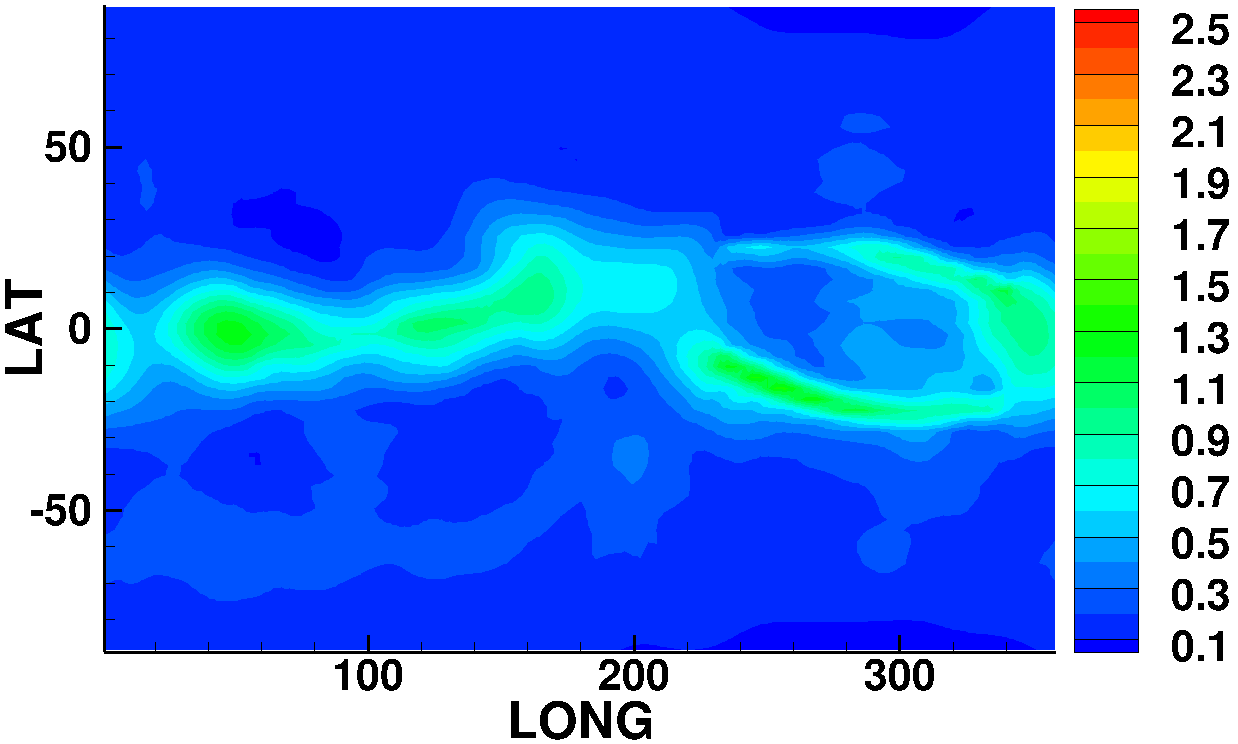}{0.3\textwidth}{Ne LASCO (10$^{5}$ cm$^{-3}$)}
         \fig{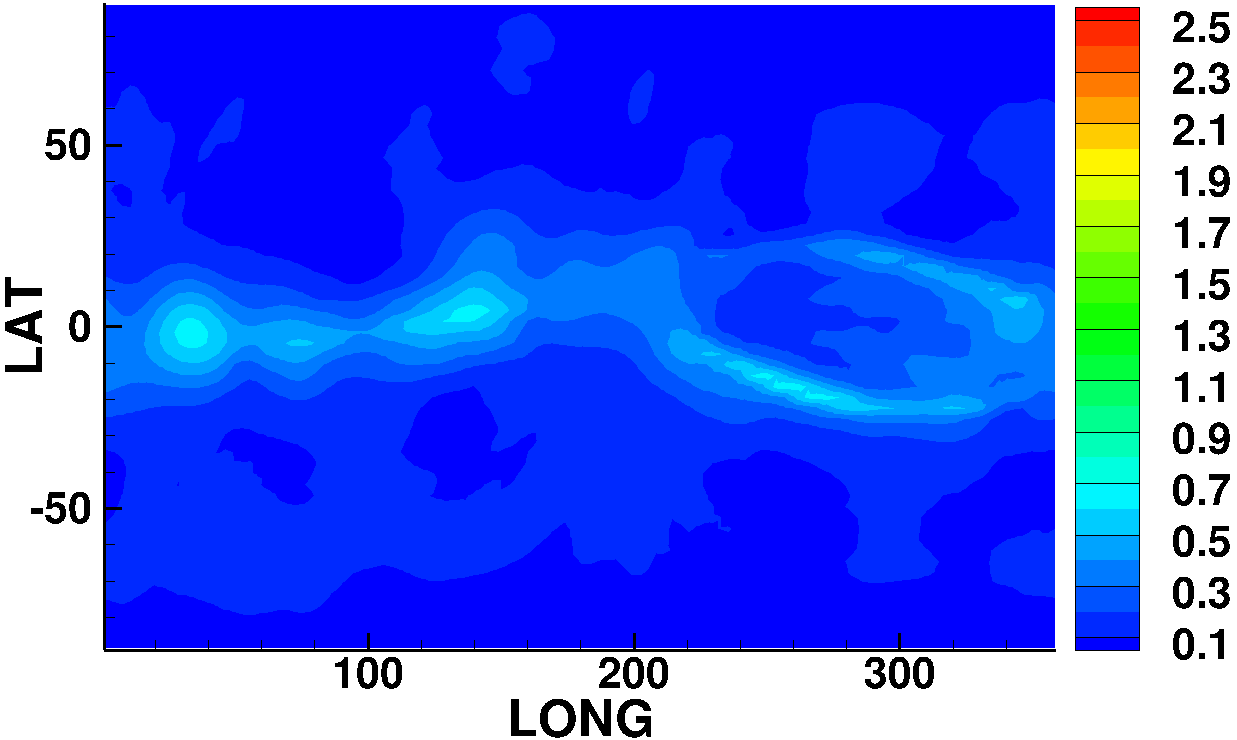}{0.3\textwidth}{Ne LASCO (10$^{5}$ cm$^{-3}$)}
               }
\gridline{\fig{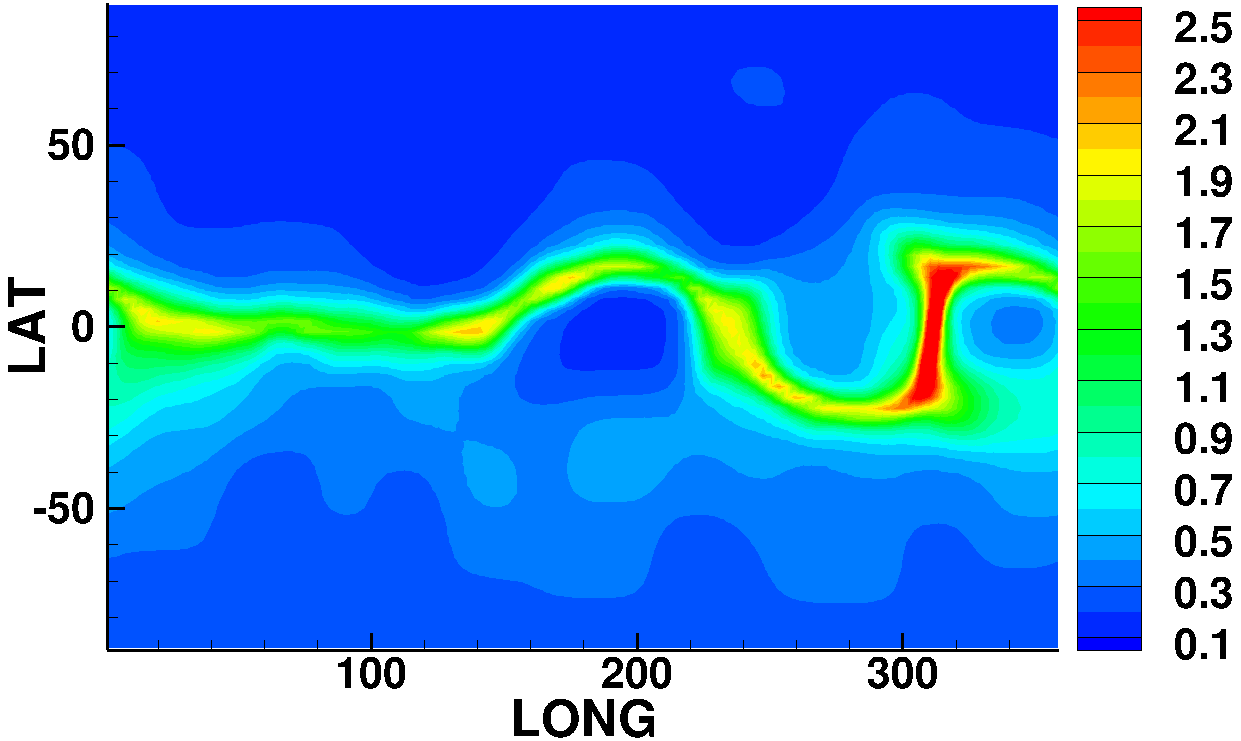}{0.3\textwidth}{Ne AWSoM (10$^{5}$ cm$^{-3}$)}
          \fig{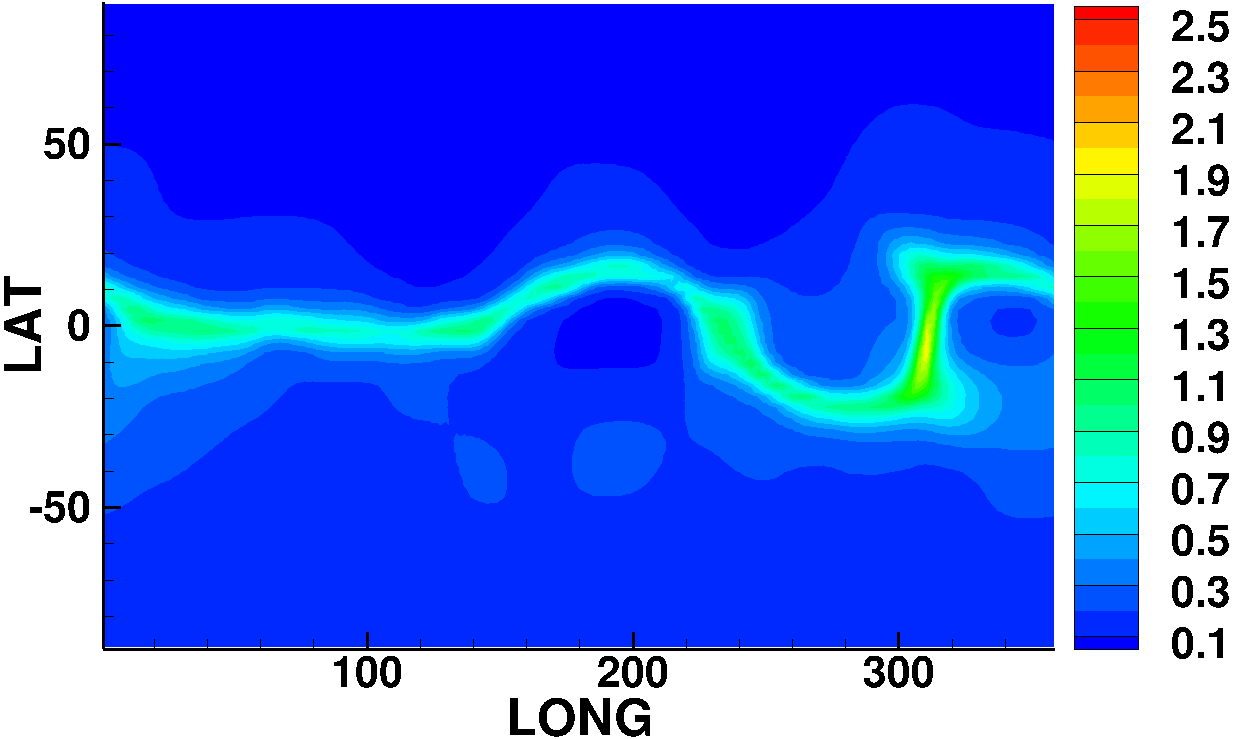}{0.3\textwidth}{Ne AWSoM (10$^{5}$ cm$^{-3}$)}
          }
\gridline{\fig{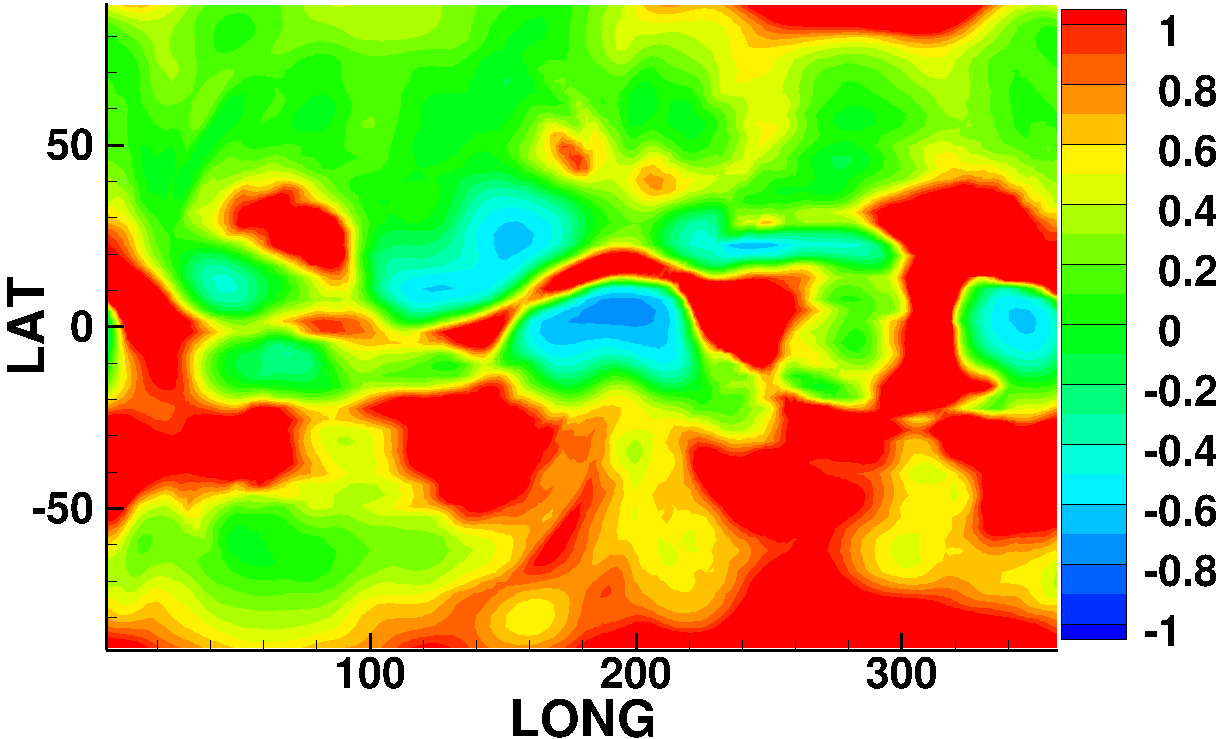}{0.3\textwidth}{Ne Rel Diff}
          \fig{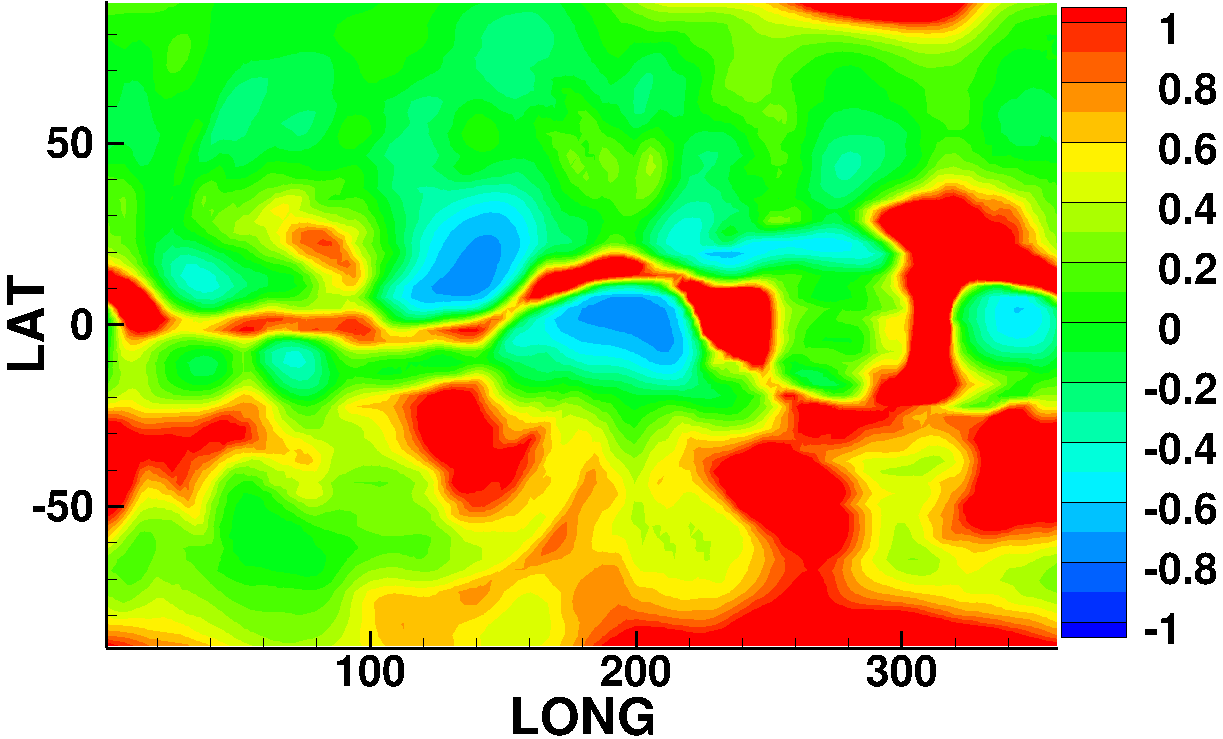}{0.3\textwidth}{Ne Rel Diff}
          }          
\gridline{\fig{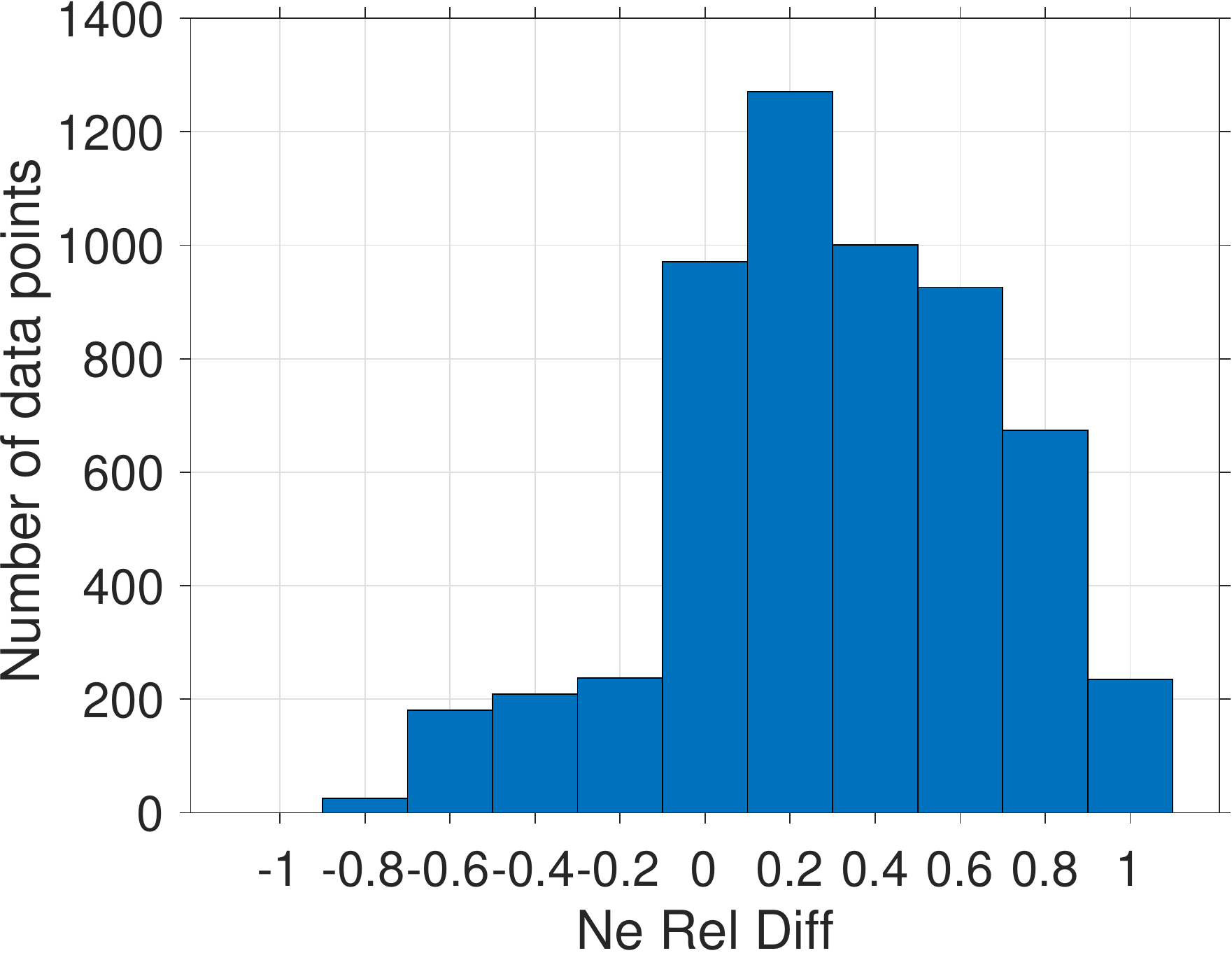}{0.3\textwidth}{(a) r=4 \Rs}
          \fig{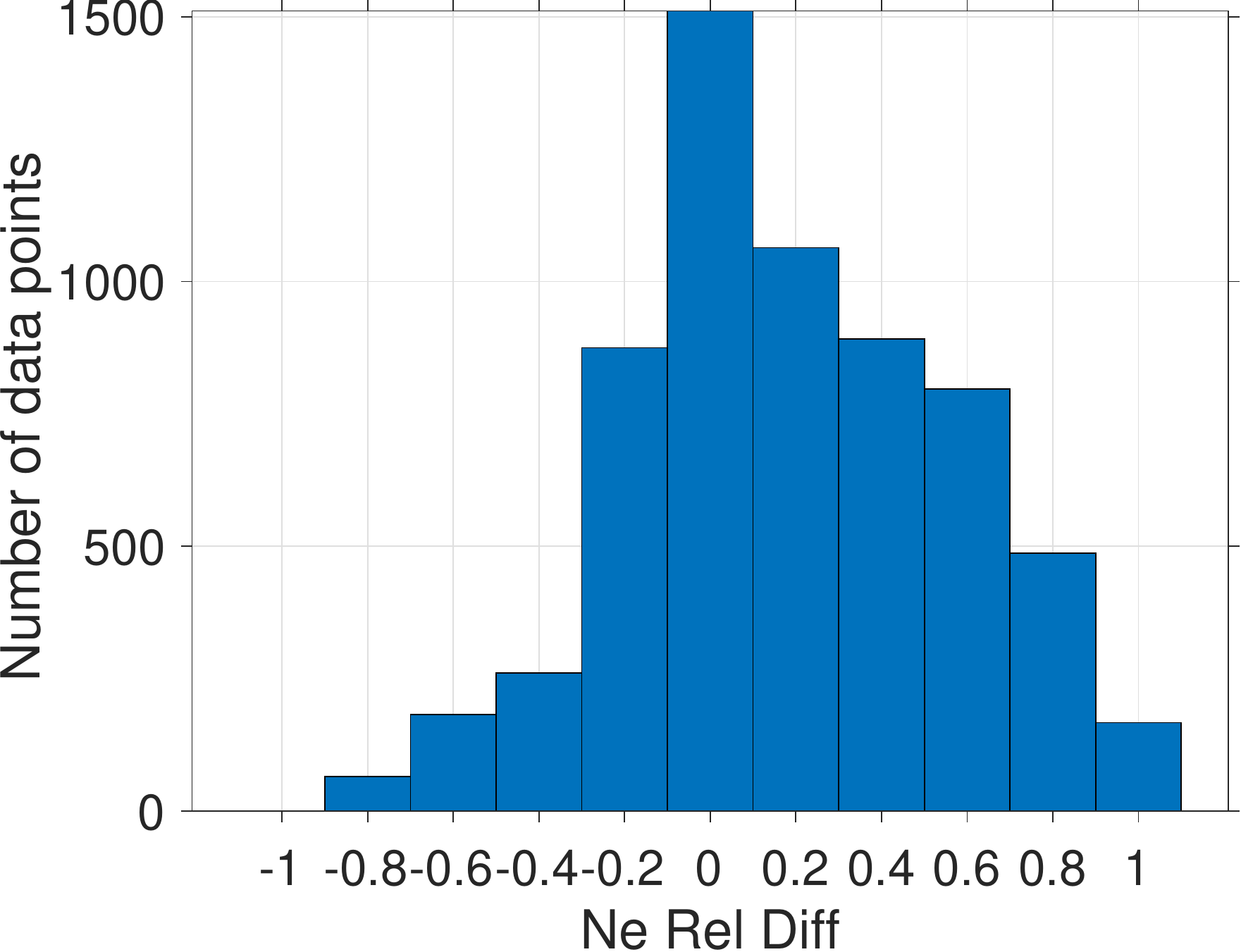}{0.3\textwidth}{(b) r=5 \Rs}
          }
\caption{Comparison of LASCO-C2 reconstructed electron density and AWSoM model simulations for CR2208 at (a) 4\,\Rs~and (b) 5\,\Rs. First and second rows show the LASCO 3D reconstructed density and the density predicted by the model, respectively in units of $10^{5}$ cm$^{-3}$. Bottom two rows depict the quantity, \rm{Ne Rel Diff}= \big($\frac{Ne_{AWSoM}}{Ne_{LASCO}}-1 \big)$, which is the relative difference between the model density and observations in the form of a latitude-longitude plot and a histogram distribution.}
\label{fig:lasco1_08}
\end{figure*}

\begin{figure*}[ht!]
%\pdfscale{1}
\gridline{\fig{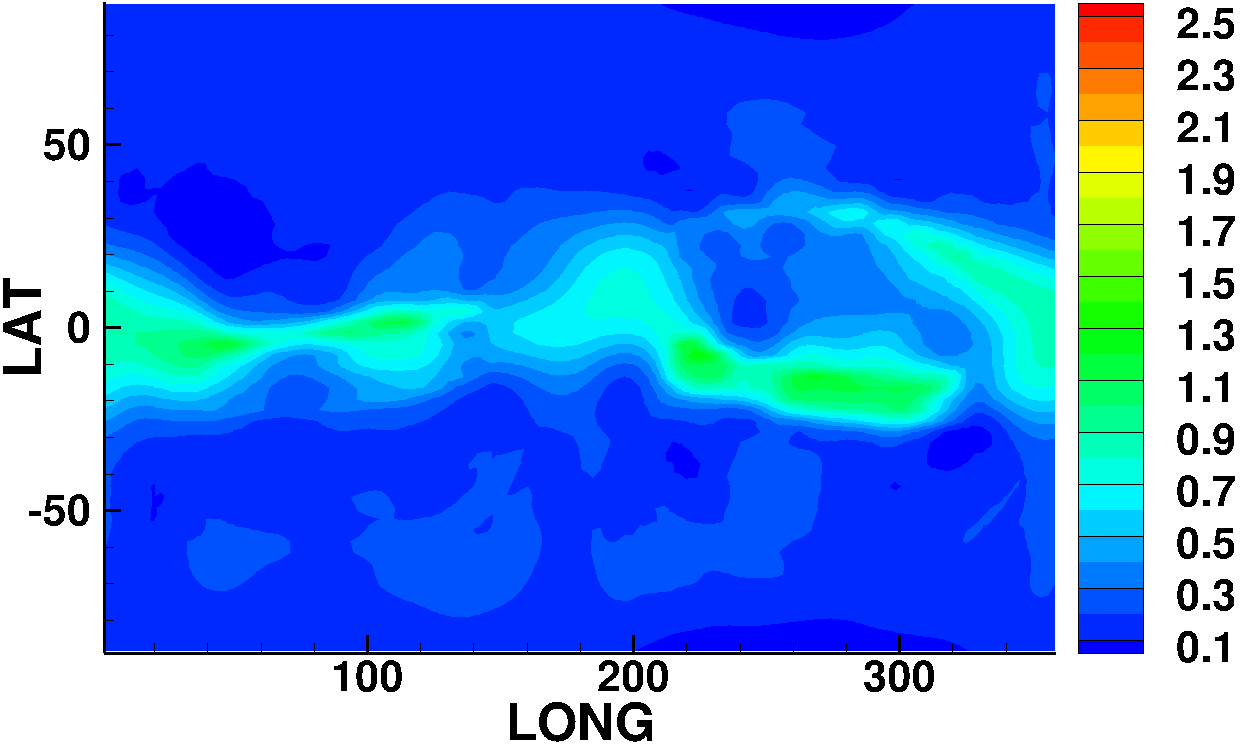}{0.3\textwidth}{Ne LASCO (10$^{5}$ cm$^{-3}$)}
         \fig{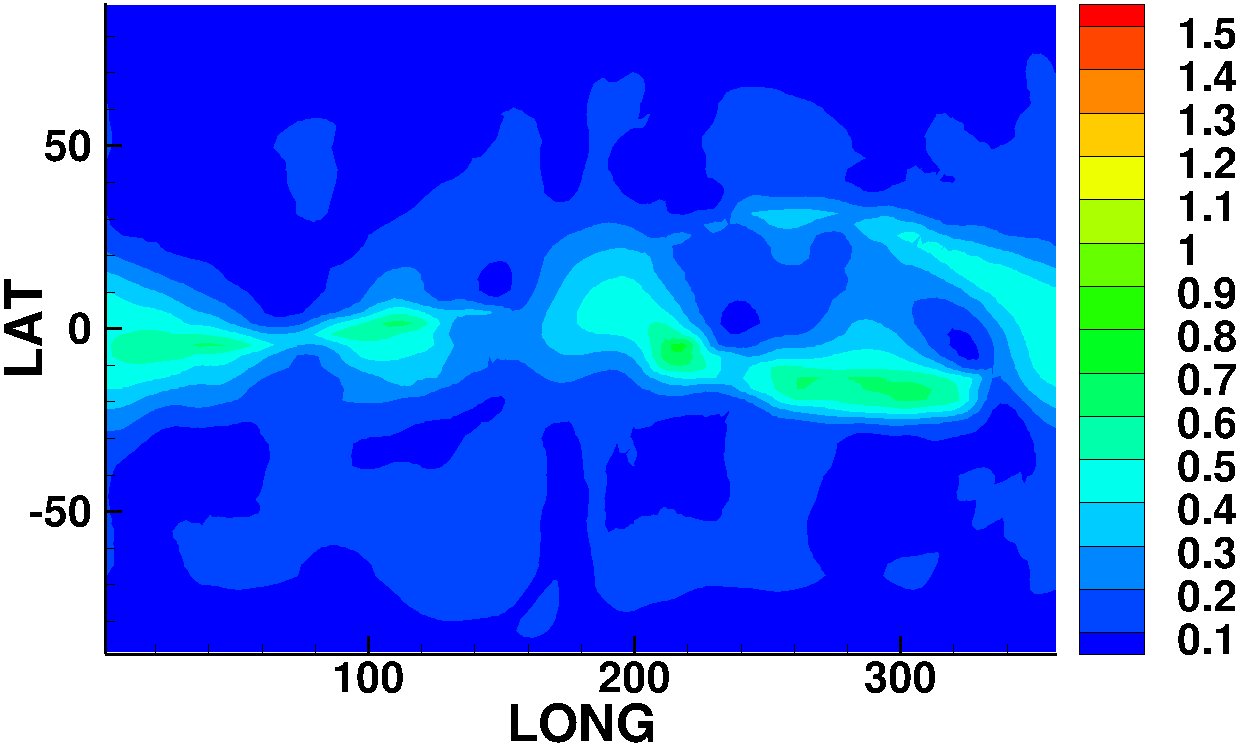}{0.3\textwidth}{Ne LASCO (10$^{5}$ cm$^{-3}$)}
               }
\gridline{\fig{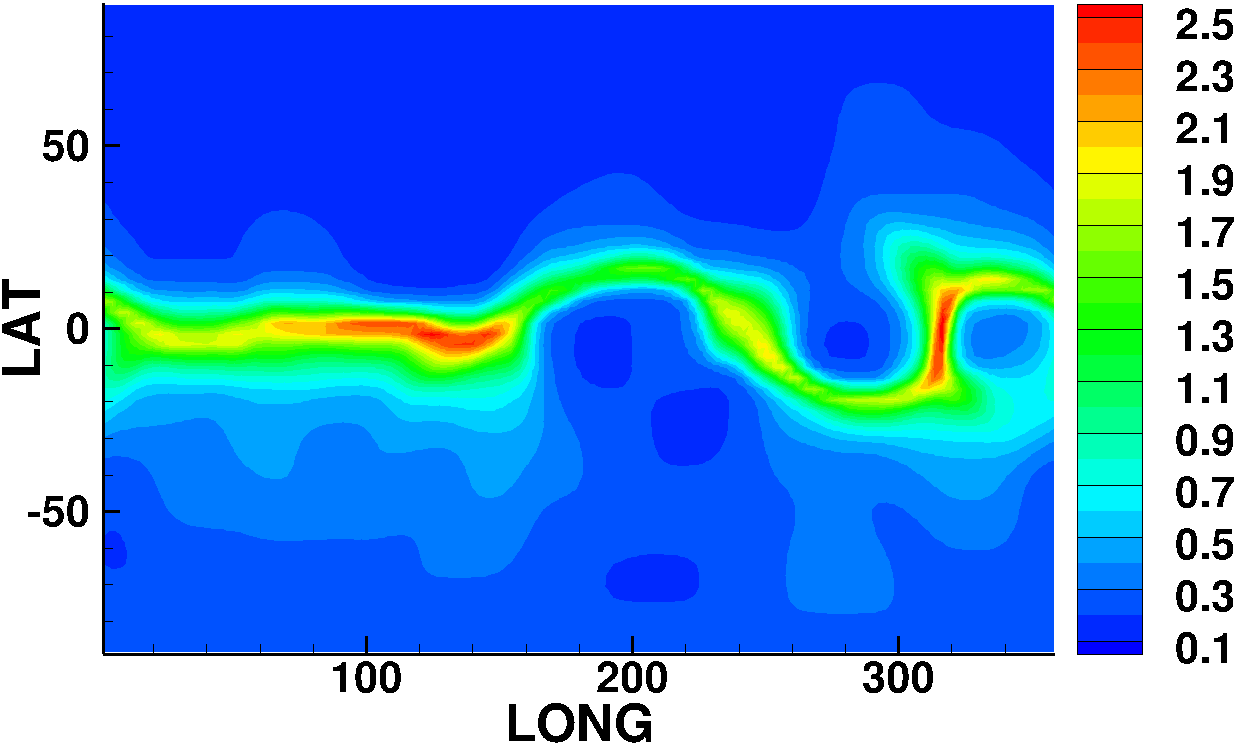}{0.3\textwidth}{Ne AWSoM (10$^{5}$ cm$^{-3}$)}
          \fig{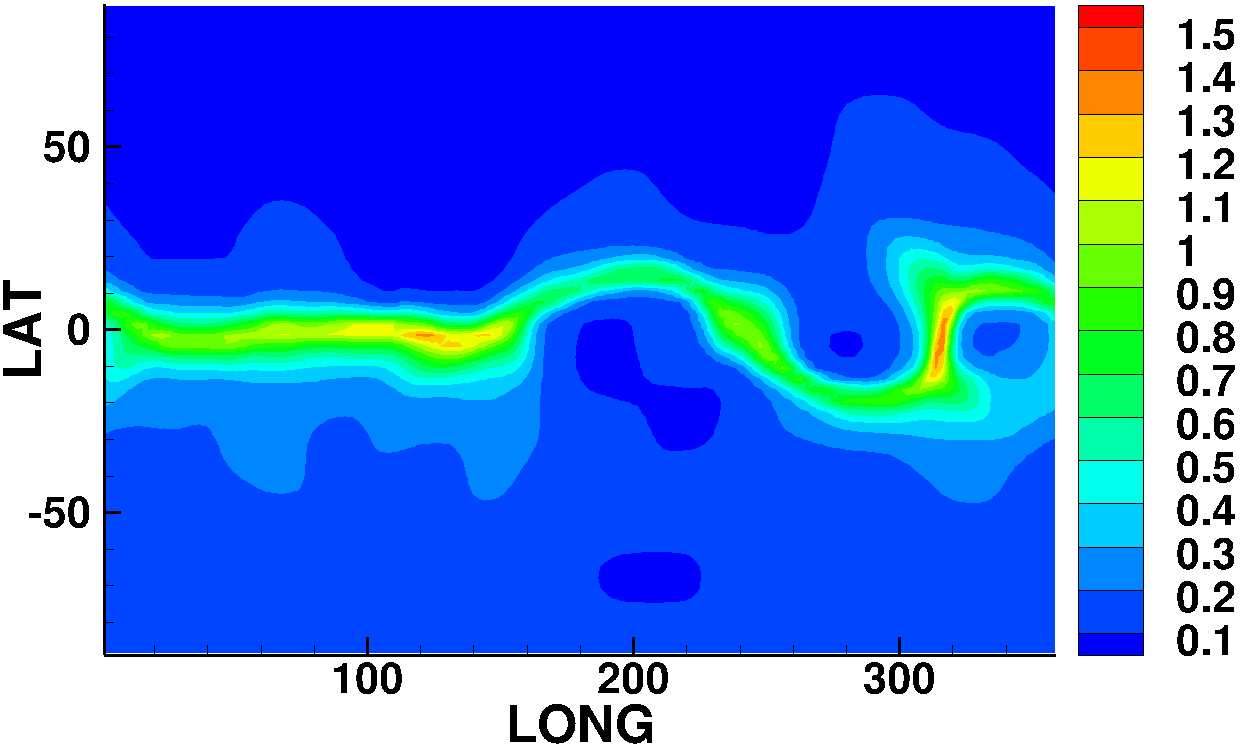}{0.3\textwidth}{Ne AWSoM (10$^{5}$ cm$^{-3}$)}
          }
\gridline{\fig{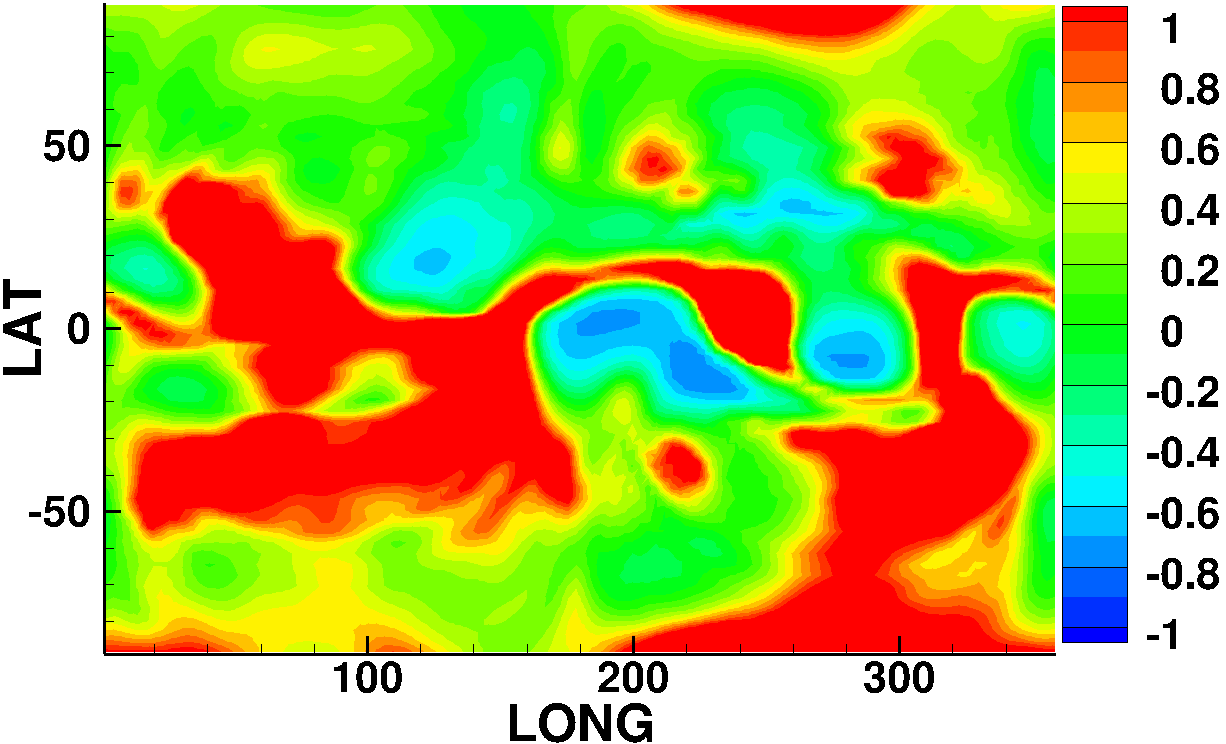}{0.3\textwidth}{Ne Rel Diff}
          \fig{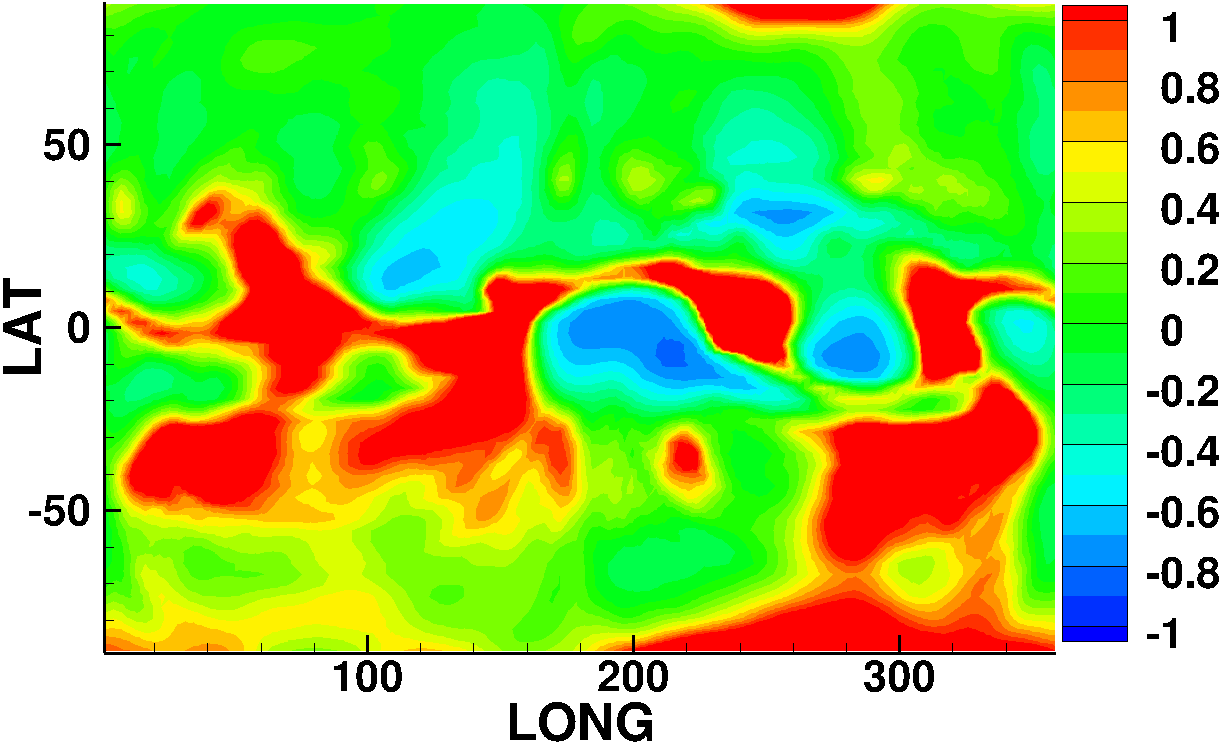}{0.3\textwidth}{Ne Rel Diff}
          }          
\gridline{\fig{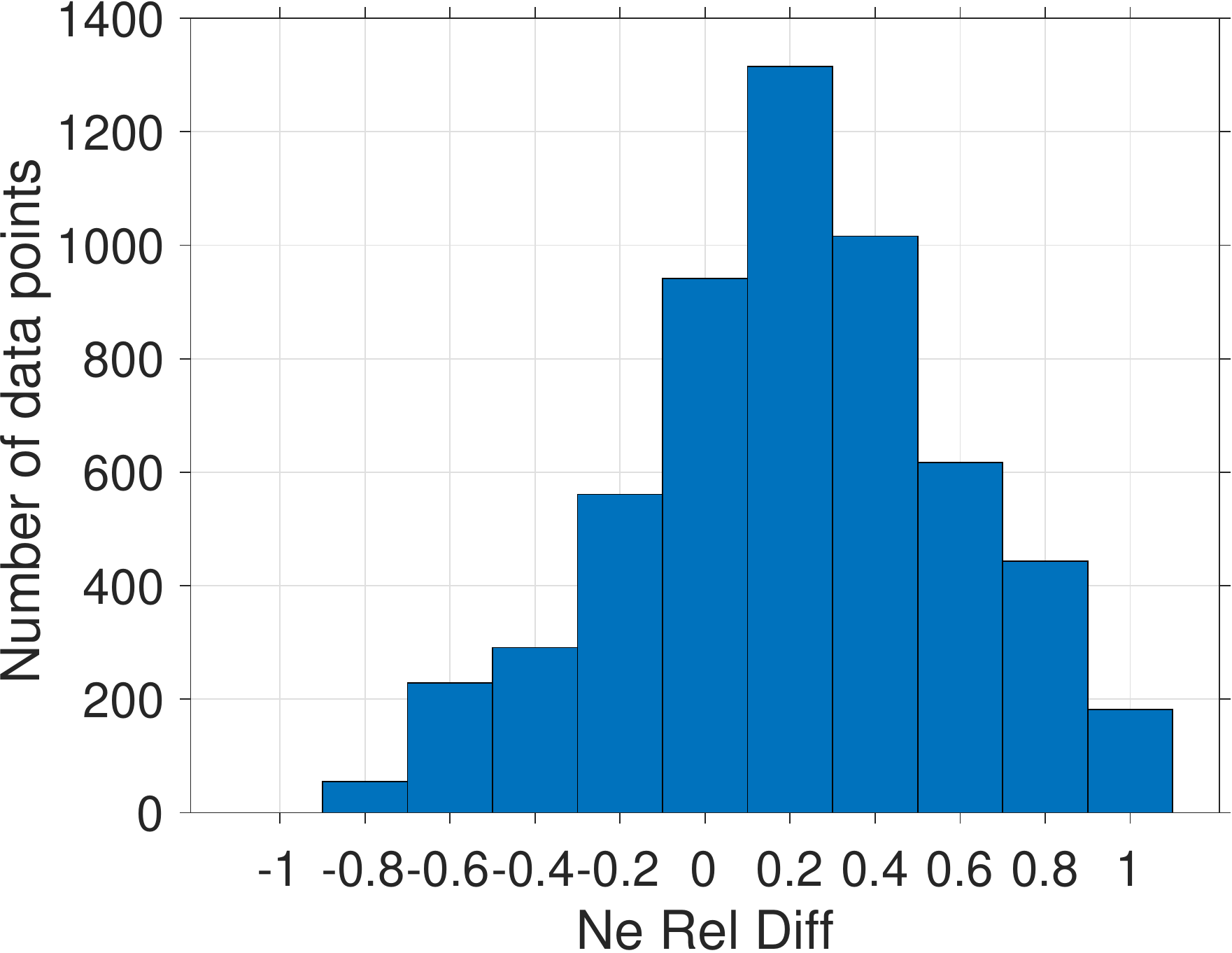}{0.3\textwidth}{(a) r=4 \Rs}
          \fig{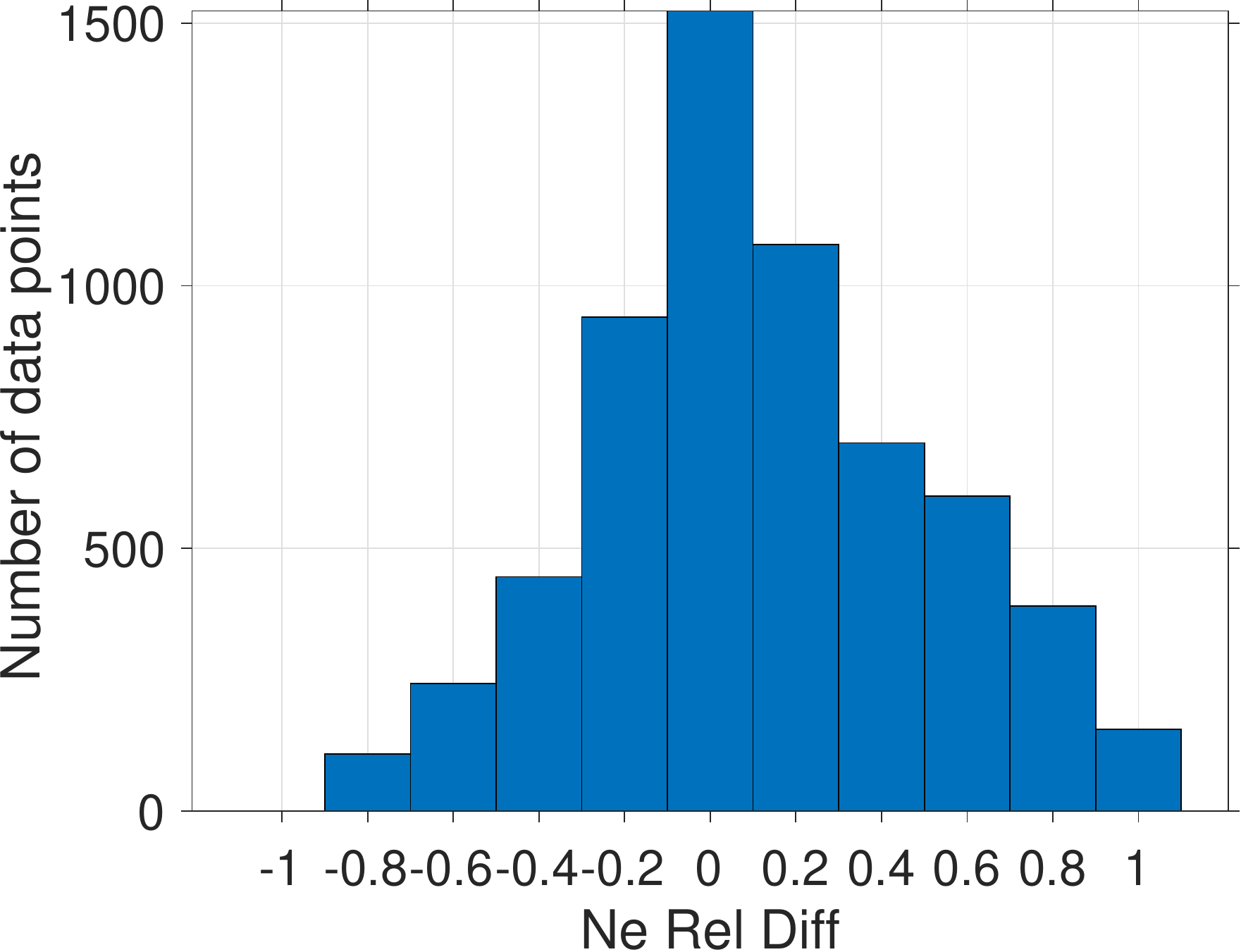}{0.3\textwidth}{(b) r=5 \Rs}
          }
\caption{Comparison of LASCO-C2 reconstructed electron density and AWSoM model simulations for CR2209 at (a) 4\,\Rs~and (b) 5\,\Rs. First and second rows show the LASCO 3D reconstructed density and the density predicted by the model, respectively in units of $10^{5}$ cm$^{-3}$. Bottom two rows depict the quantity, \rm{Ne Rel Diff}= \big($\frac{Ne_{AWSoM}}{Ne_{LASCO}}-1 \big)$, which is the relative difference between the model density and observations in the form of a latitude-longitude plot and a histogram distribution.}
\label{fig:lasco2_09}
\end{figure*}

Figures \ref{fig:lasco1_08} and \ref{fig:lasco2_09} show the relative difference between the reconstructed coronal density and model results for CR2208 and CR2209 respectively. In each figure, the first two rows show the density obtained from tomography (Ne LASCO) and the density from AWSoM model results (Ne AWSoM) respectively, in units of 10$^{5}$ cm$^{-3}$. Bottom two rows show the comparisons between tomography data and model solutions at (a) 4\,\Rs~and (b) 5\,\Rs~and the corresponding histograms. The quantity shown here for comparison is the density difference relative to the observed tomographic density, \rm{Ne Rel Diff}$=\big(Ne_{AWSoM}/Ne_{LASCO}\big) -1$. We find that the predicted densities in the range of heliocentric heights within the LASCO FOV lie within $\pm$ 20\,-\,30\,\% of the observed densities reconstructed from LASCO C2. The larger discrepancy along the streamer cusp can be attributed to the underresolved features in the LASCO reconstructions. AWSoM results show a highly resolved thin current sheet with high density regions, compared to the features in LASCO that seem to be smeared out along the current sheet. Therefore, a cell-by-cell comparison shows differences that are way off in this region.

We find that that the AWSoM model produces an asymmetric density distribution between the two hemispheres, which is direct consequence of the different sizes of the northern and southern coronal holes as seen in the EUV images. The polar asymmetry originates in the magnetic field maps for the two Carrington rotations, where the unipolar magnetic fields of the northern polar regions extend to lower latitudes compared to the southern pole. As a result, a more narrow coronal hole form in the south for which magnetic field has larger expansion, which in turn leads to a comparatively slower and denser solar wind. This can explain the over dense regions in the southern hemisphere of the AWSoM model results in the LASCO FOV compared to the LASCO reconstructions. This asymmetry in density (and speed) is also seen further out in the inner heliosphere (Section \ref{sec:IPS}).

%Due to the heliospheric current sheet folding differently in observations and the model, we see major difference in the low latitudes.}

%The histograms show that the AWSoM model overpredicts the electron density by a {\bf what is the factor value} small factor for both CR2208 and CR2209. Therefore, our model has a good capability to reproduce the observed electron densities within 20-30\% of the expected values as we move farther out from the low corona.

%%%%%%%%%%%%%%%%%%% LASCO FIGURES %%%%%%%%%%%%
%%%%%% LASCO 2208
%%%%%%%%%% LASCO CR2209

\subsection{InterPlanetary Scintillation (IPS)}\label{sec:IPS}
We use the InterPlanetary Scintillation (IPS) time-dependent, kinematic 3D reconstruction technique to obtain the solar wind parameters in the inner heliosphere. Time-dependent results can be extracted at any radial distance within the reconstructed volume. Here, we show the IPS data and AWSoM model comparisons at r=20\,\Rs, 100\,\Rs~and 1 AU. The University of California, San Diego (UCSD) have developed an iterative Computer Assisted Tomography (CAT) program \citep{Hic2004,Jac1998,Jac2003,Jac2010,Jac2011,Jac2013,Yu2015}, that incorporates remote sensing data from Earth to a kinematic solar wind model to provide 3D reconstructed velocity distributions over the inner heliosphere.
%%%%%%%%%%%%%%%%%%%%%%% IPS FIGURES %%%%%%%%%%%
%%%%CR2208 figure IPS
\begin{figure*}[ht!]
%\pdfscale{1}
\gridline{\fig{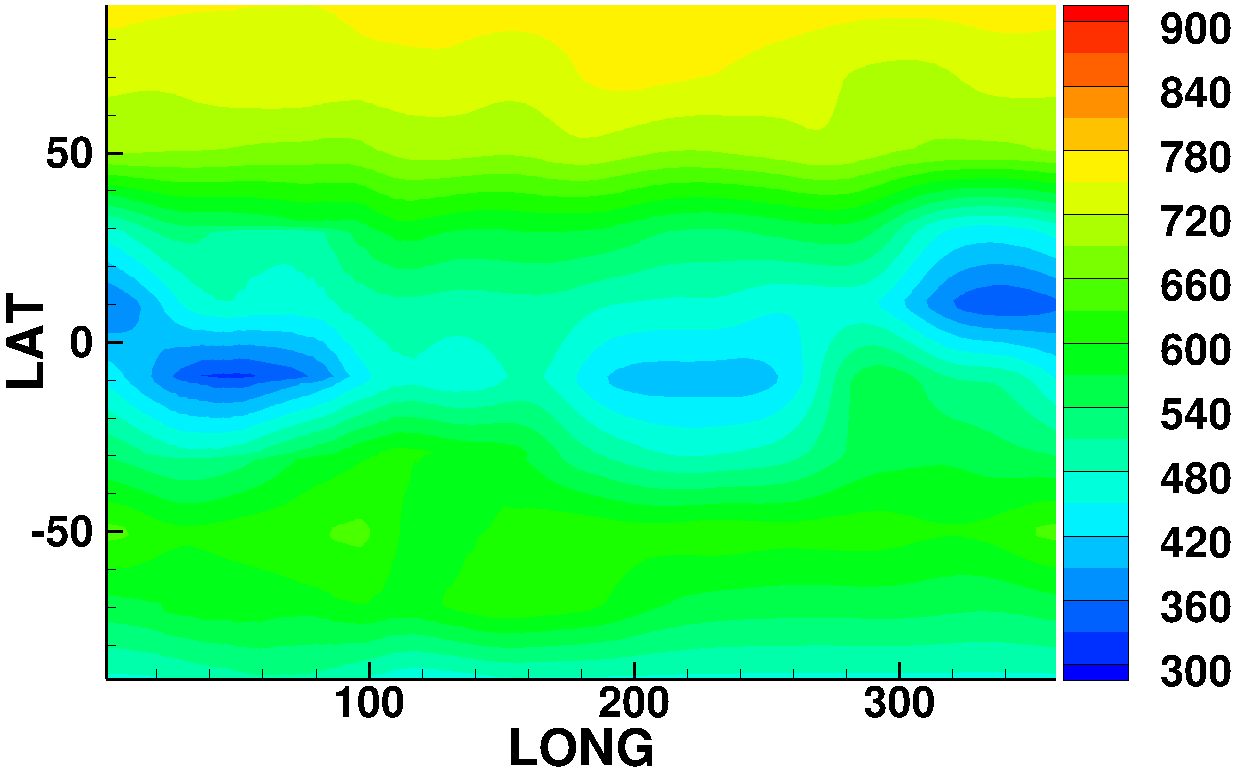}{0.33\textwidth}{V IPS (kms$^{-1}$)}
          \fig{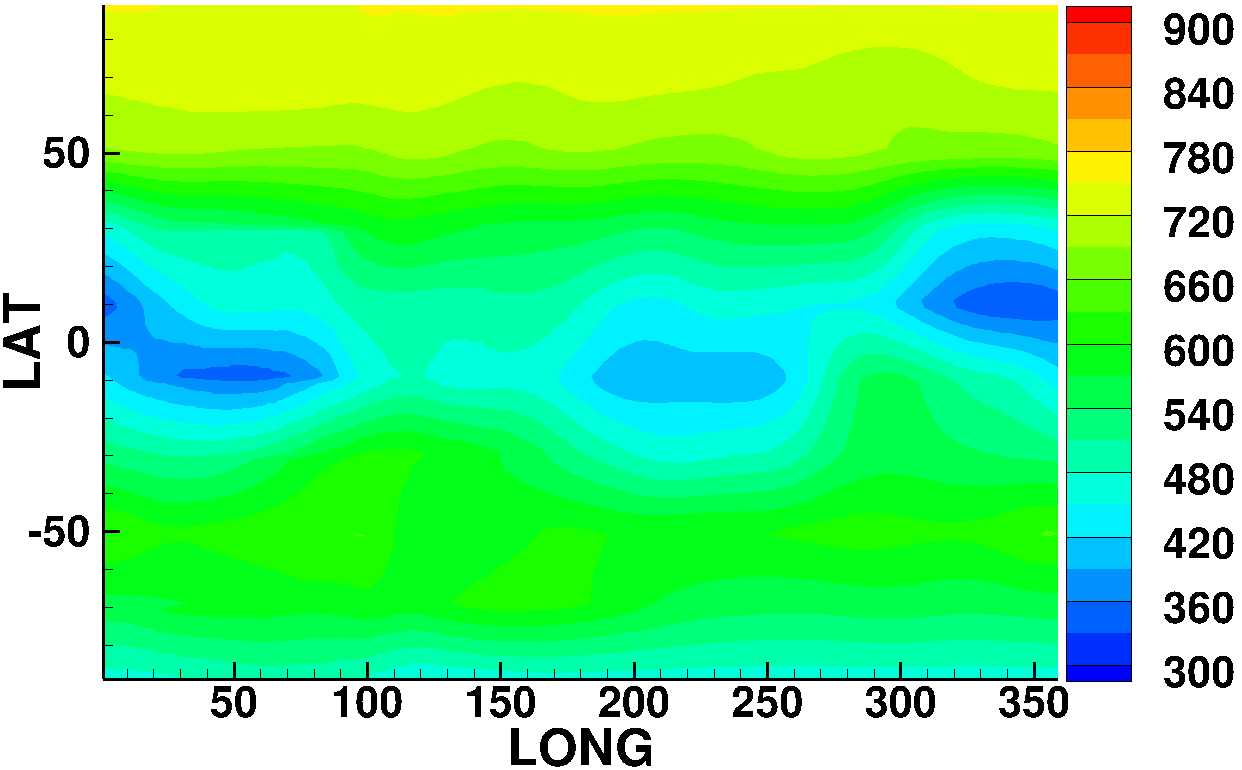}{0.33\textwidth}{V IPS (kms$^{-1}$)}
         \fig{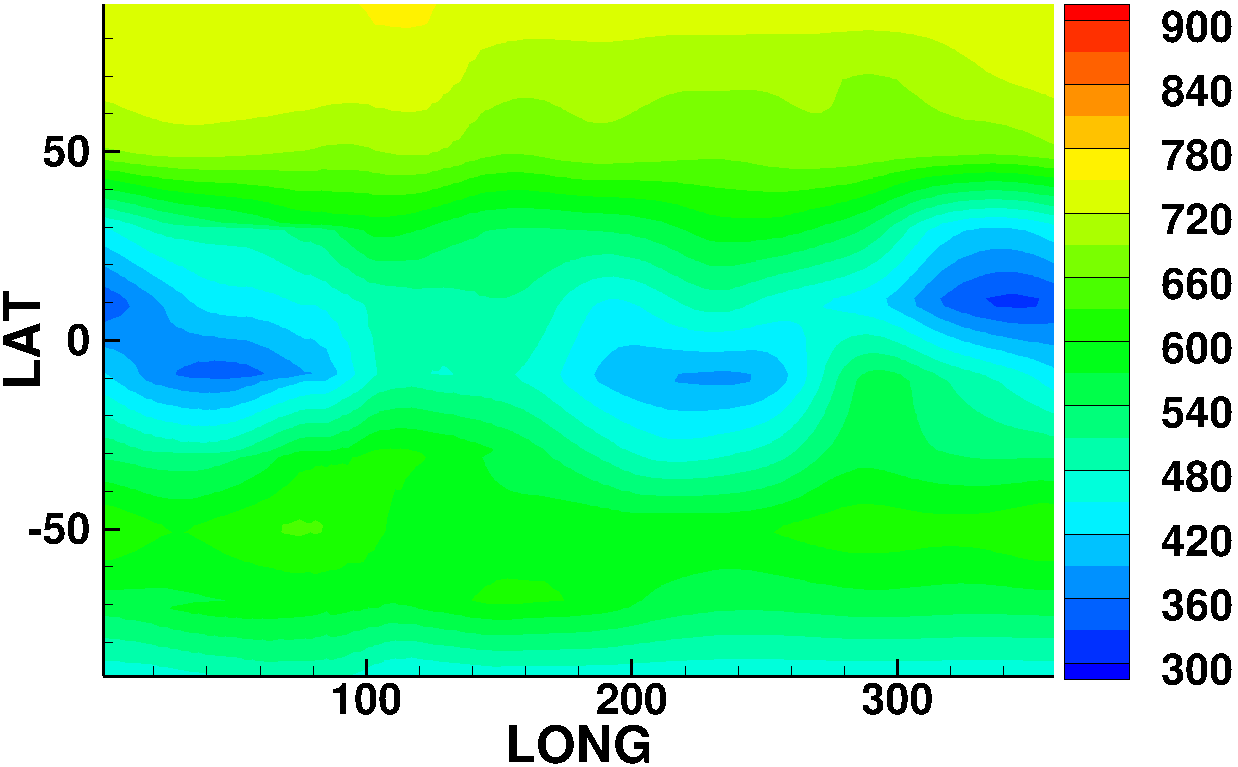}{0.33\textwidth}{V IPS (kms$^{-1}$)}
          }
\gridline{\fig{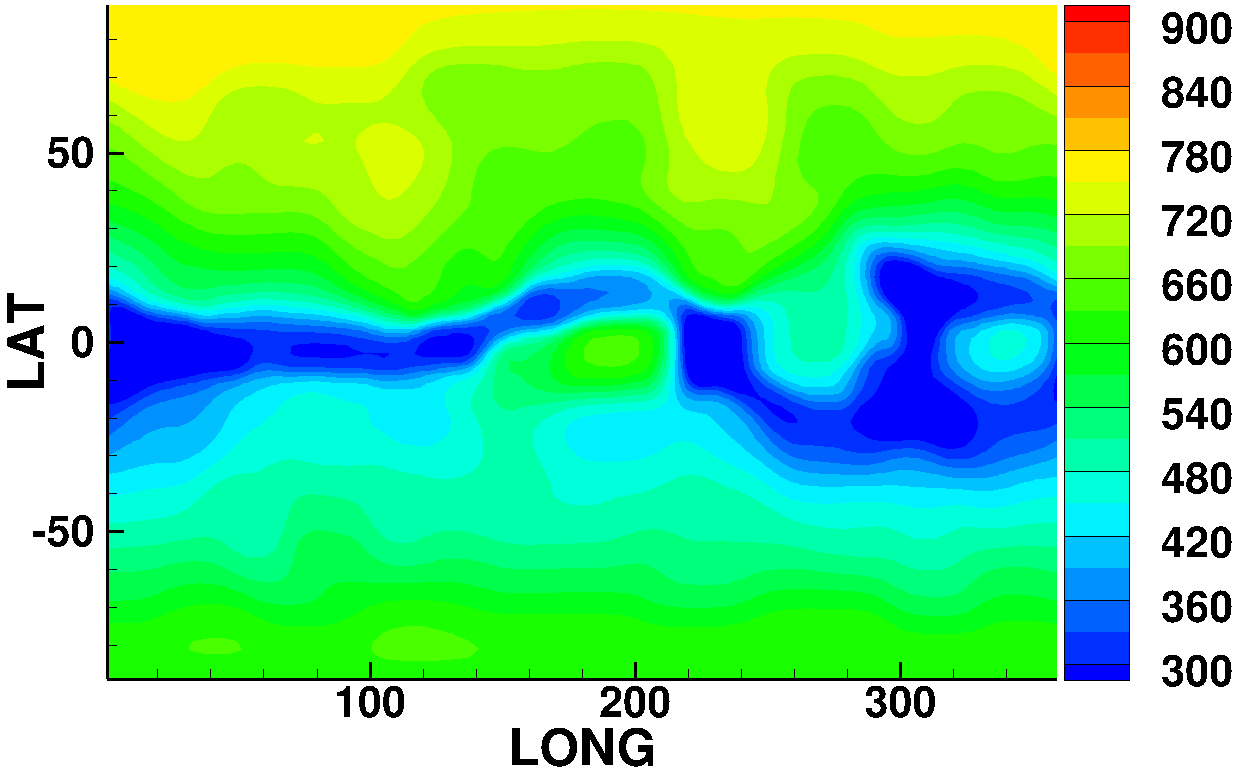}{0.33\textwidth}{V AWSoM (kms$^{-1}$)}
          \fig{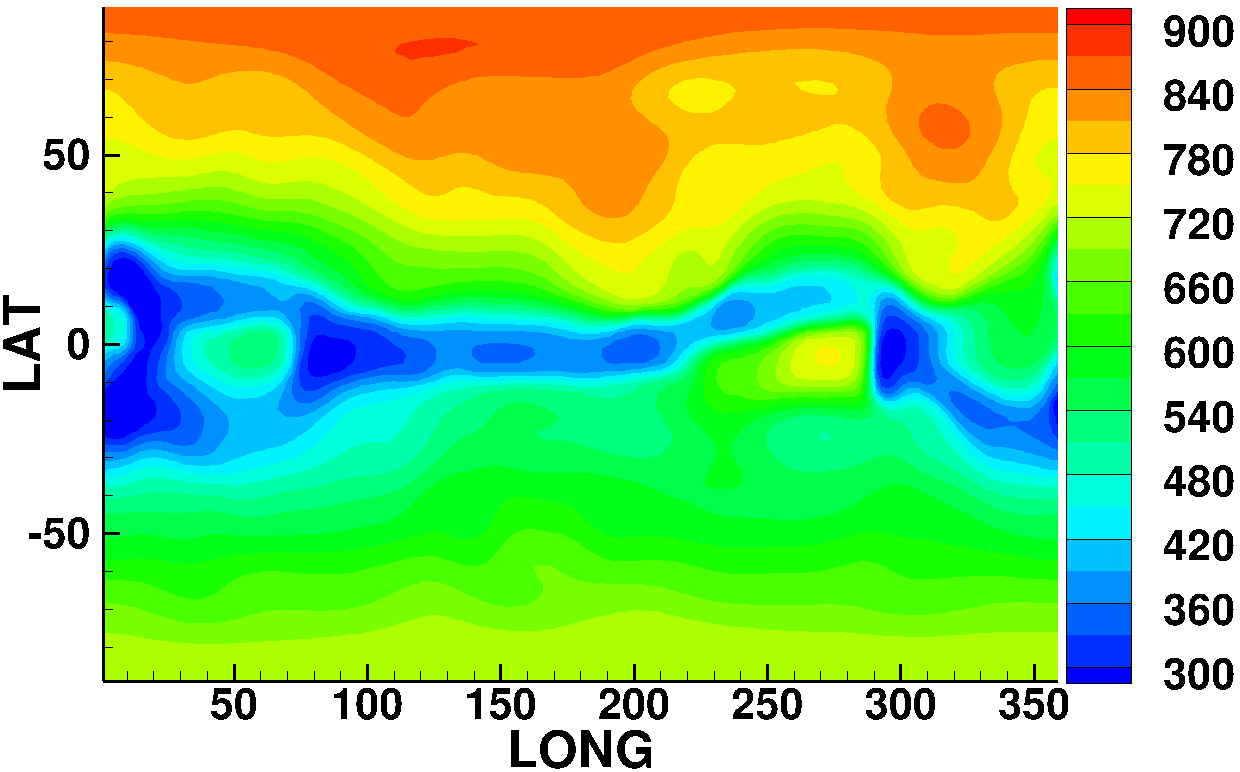}{0.33\textwidth}{V AWSoM (kms$^{-1}$)}
          \fig{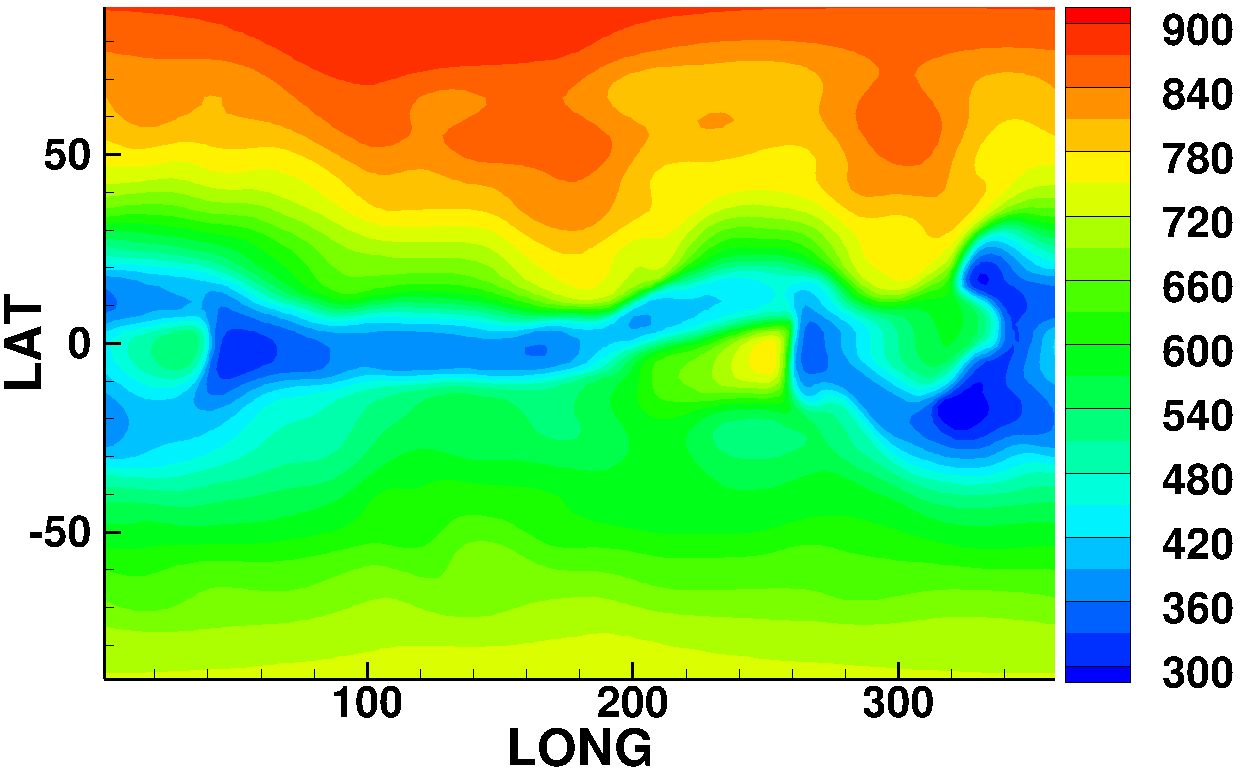}{0.33\textwidth}{V AWSoM (kms$^{-1}$)}
          }
\gridline{\fig{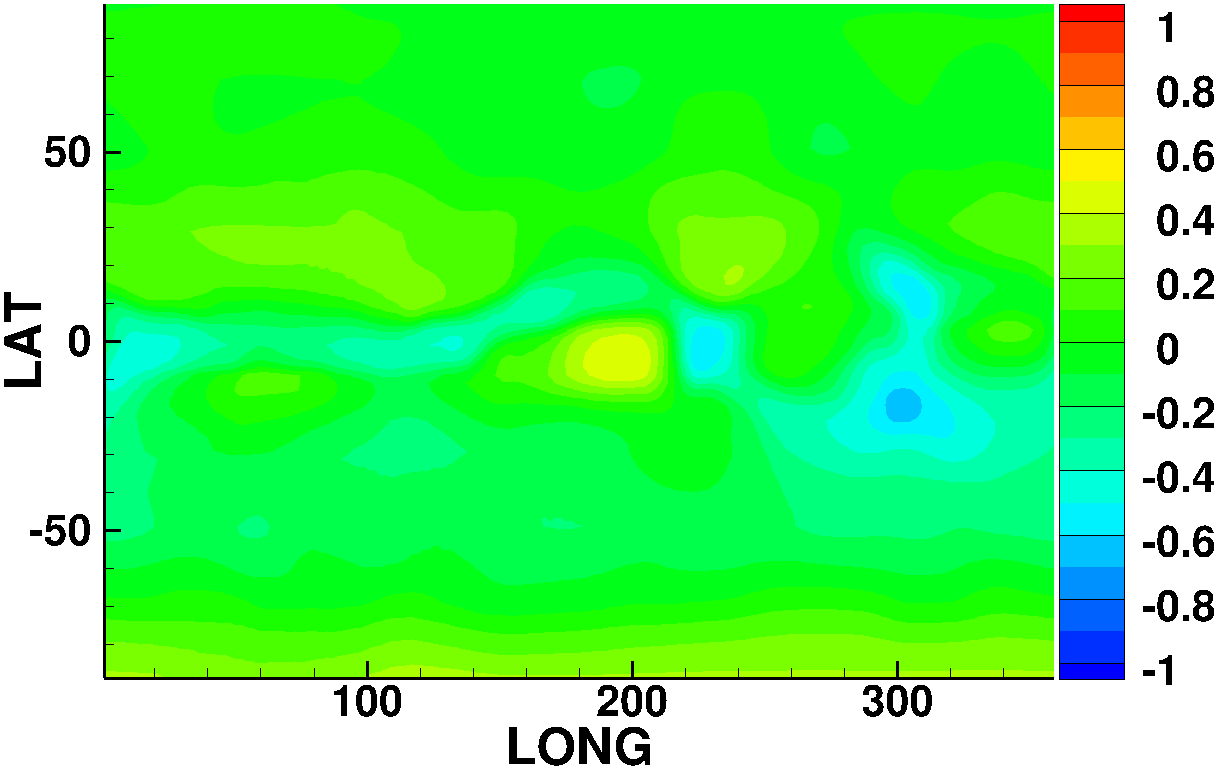}{0.33\textwidth}{V Rel Diff}
          \fig{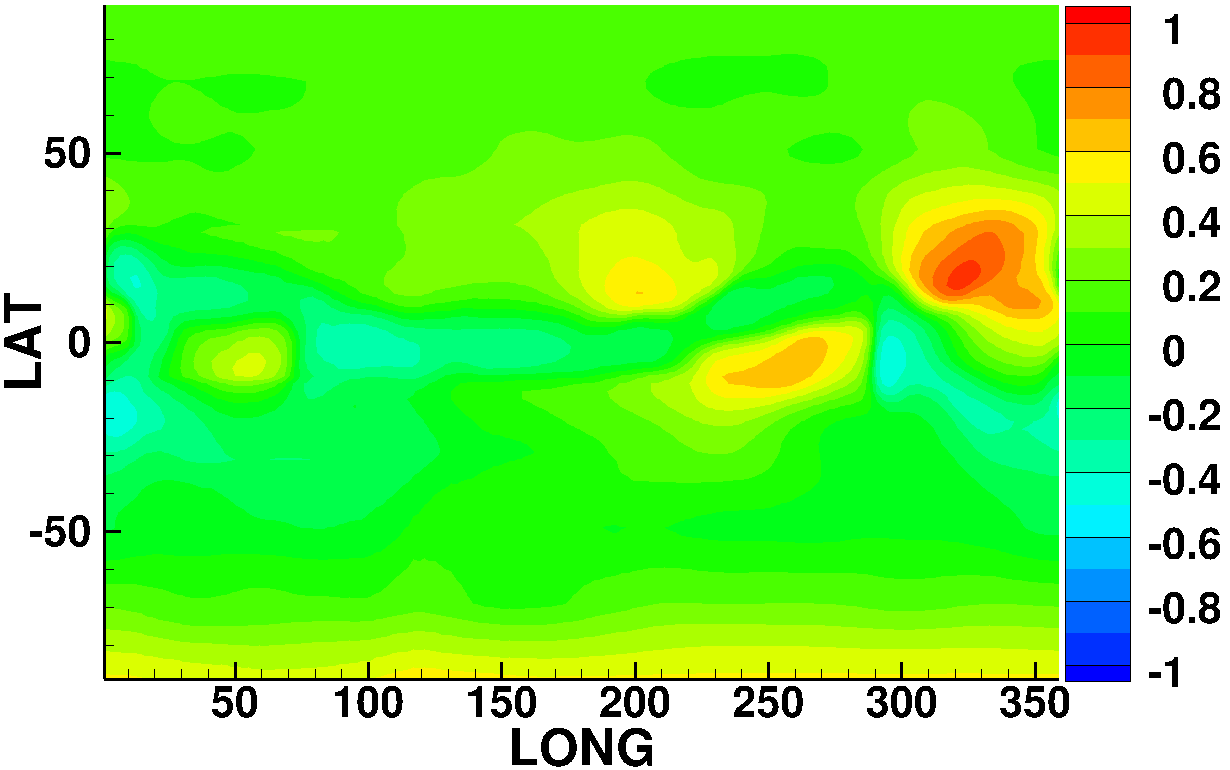}{0.33\textwidth}{V Rel Diff}
          \fig{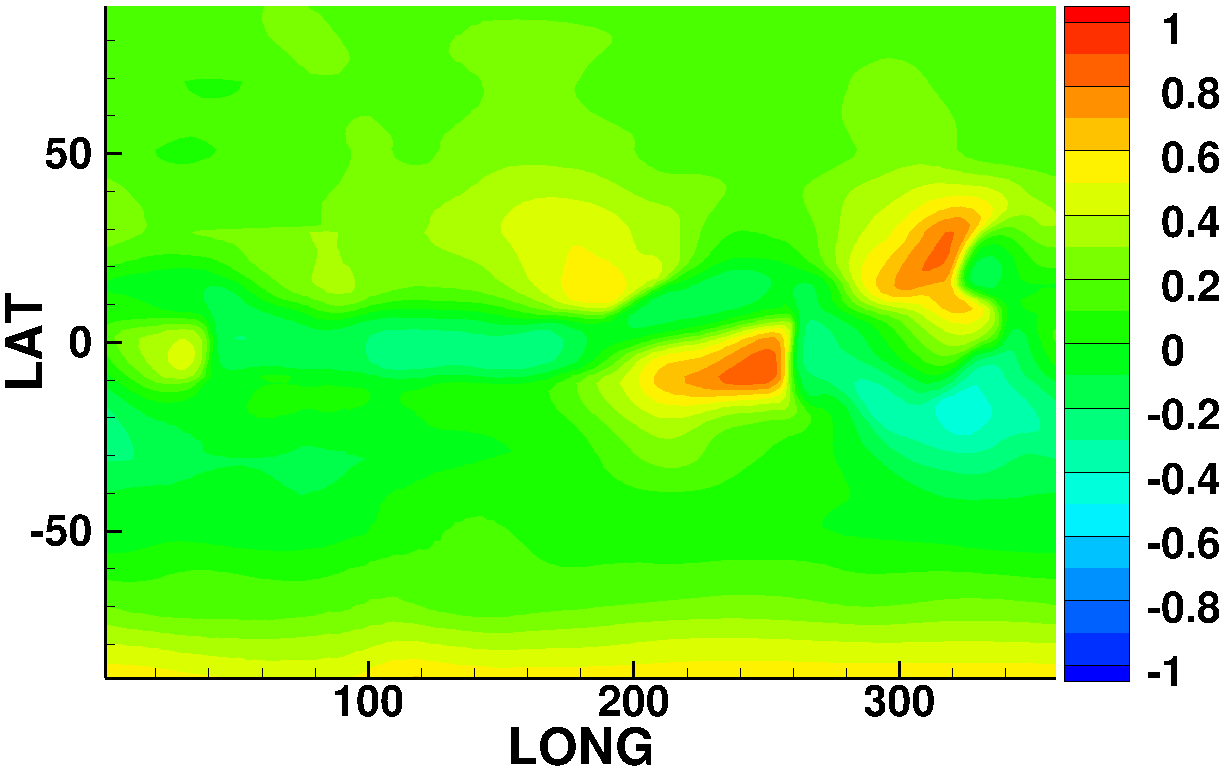}{0.33\textwidth}{V Rel Diff}
          }
\gridline{\fig{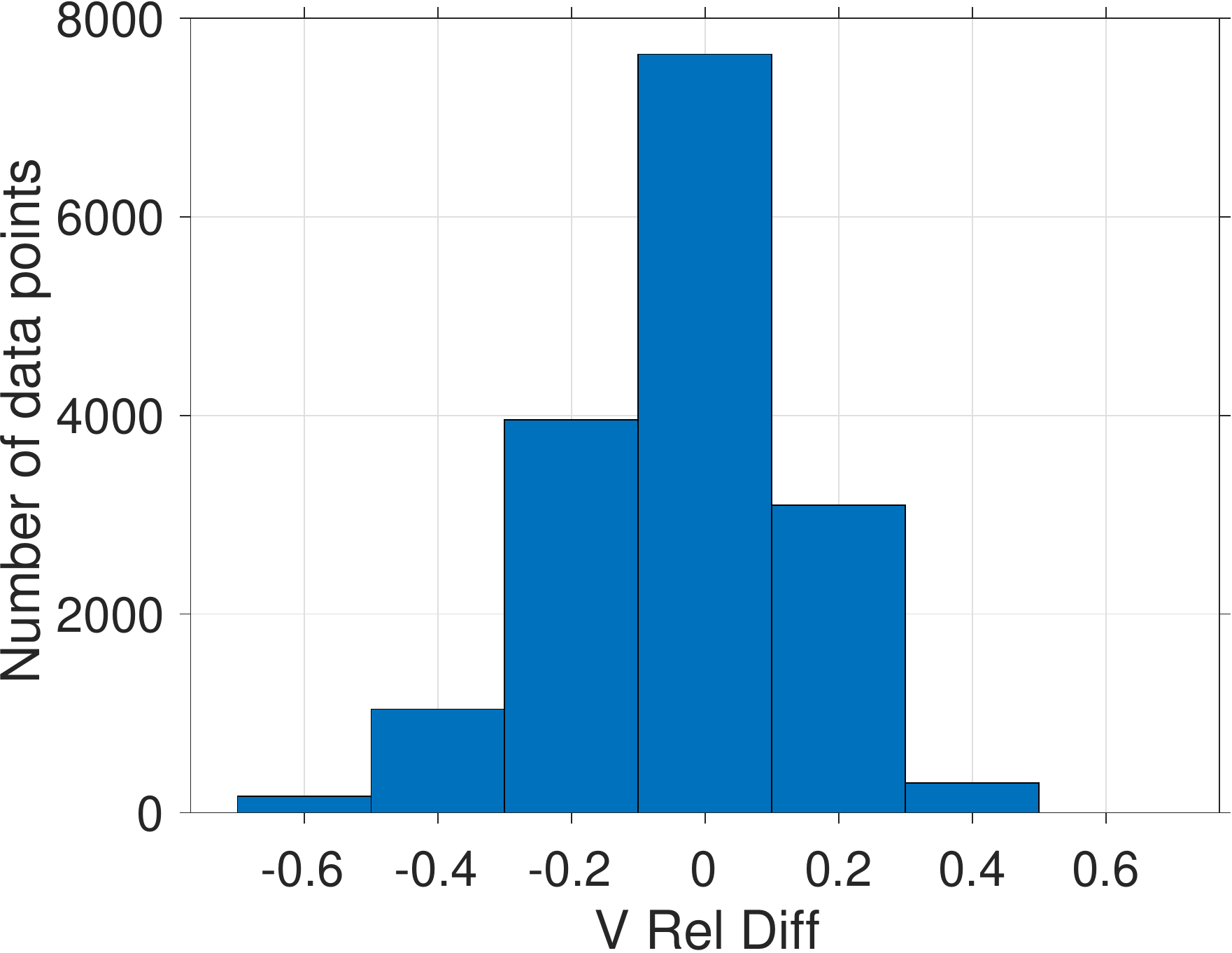}{0.33\textwidth}{(a) 20 \Rs}
          \fig{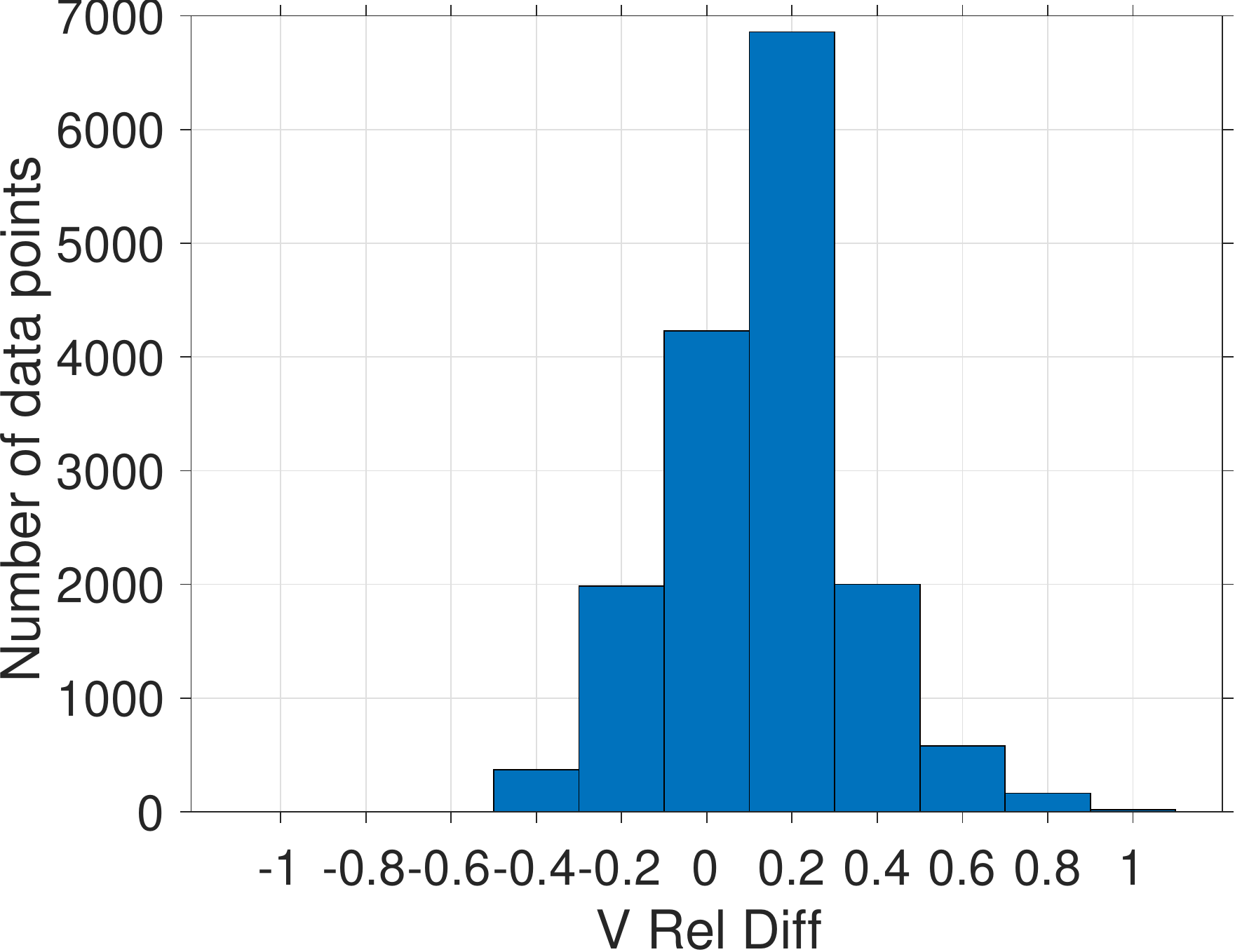}{0.33\textwidth}{(b) 100 \Rs}
          \fig{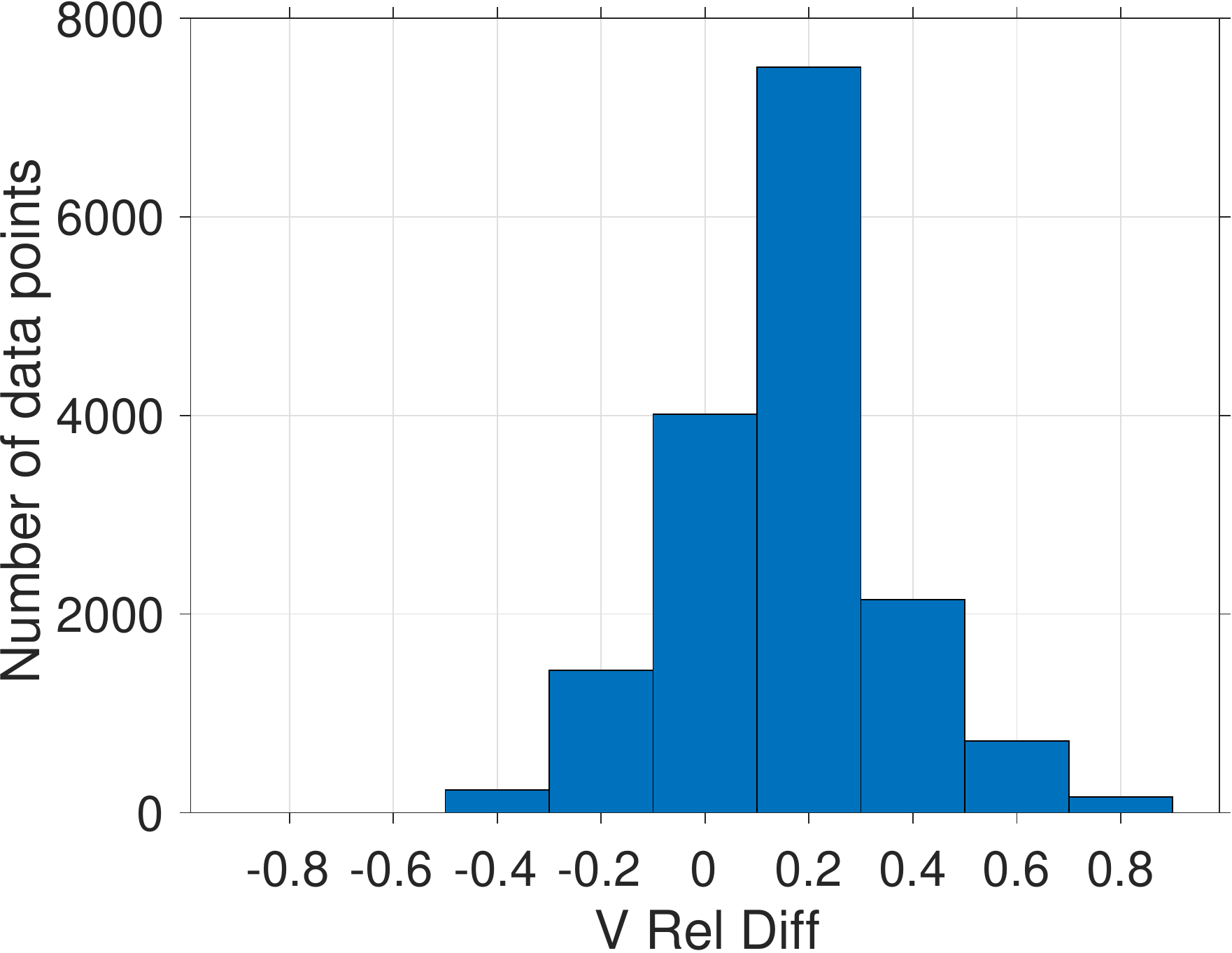}{0.33\textwidth}{(c) 1 AU}
          }
\caption{Comparison of reconstructed IPS velocity with AWSoM model simulations for CR2208 at three radial distances. The three columns correspond to results at (a) 20\,\Rs, (b) 100\,\Rs~and (c) 1 AU respectively. The following quantities are shown in each succeeding row - IPS reconstructed solar wind velocity in kms$^{-1}$ (V IPS), AWSoM predicted velocity in kms$^{-1}$ (V AWSoM), relative velocity difference between IPS observations and model output, \rm{V Rel Diff}=\big($\frac{V_{AWSoM}}{V_{IPS}}-1\big)$ and the histogram which shows how the relative difference is distributed.
\label{fig:ips_08}}
\end{figure*}

%%%%%%CR2209 IPS
\begin{figure*}[h!]
\gridline{\fig{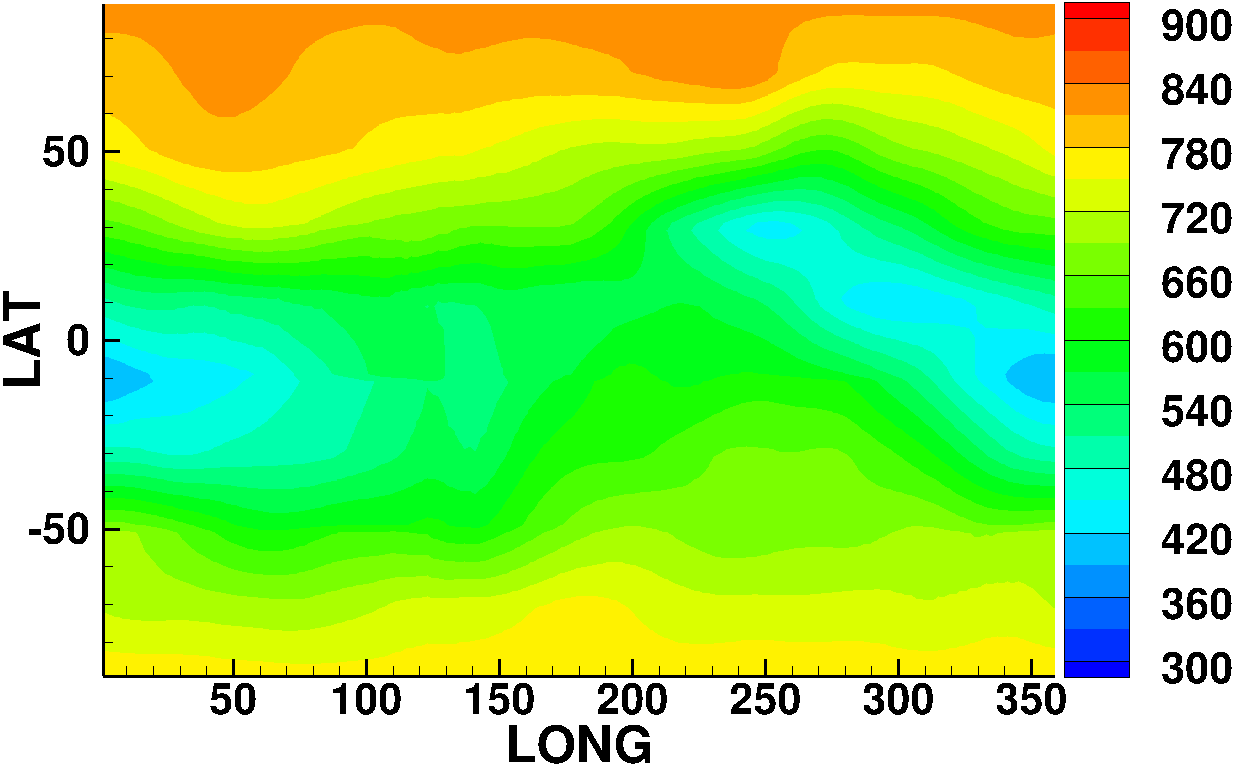}{0.33\textwidth}{V IPS (kms$^{-1}$)}
          \fig{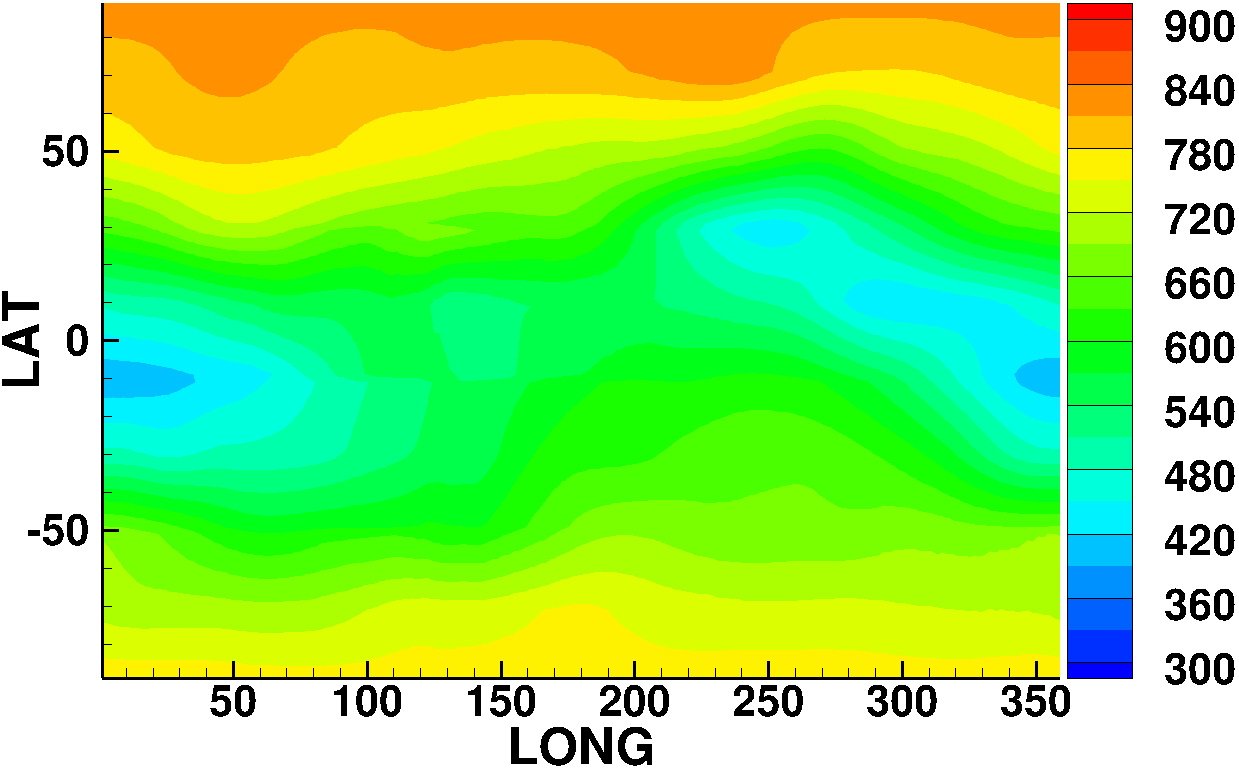}{0.33\textwidth}{V IPS (kms$^{-1}$)}
         \fig{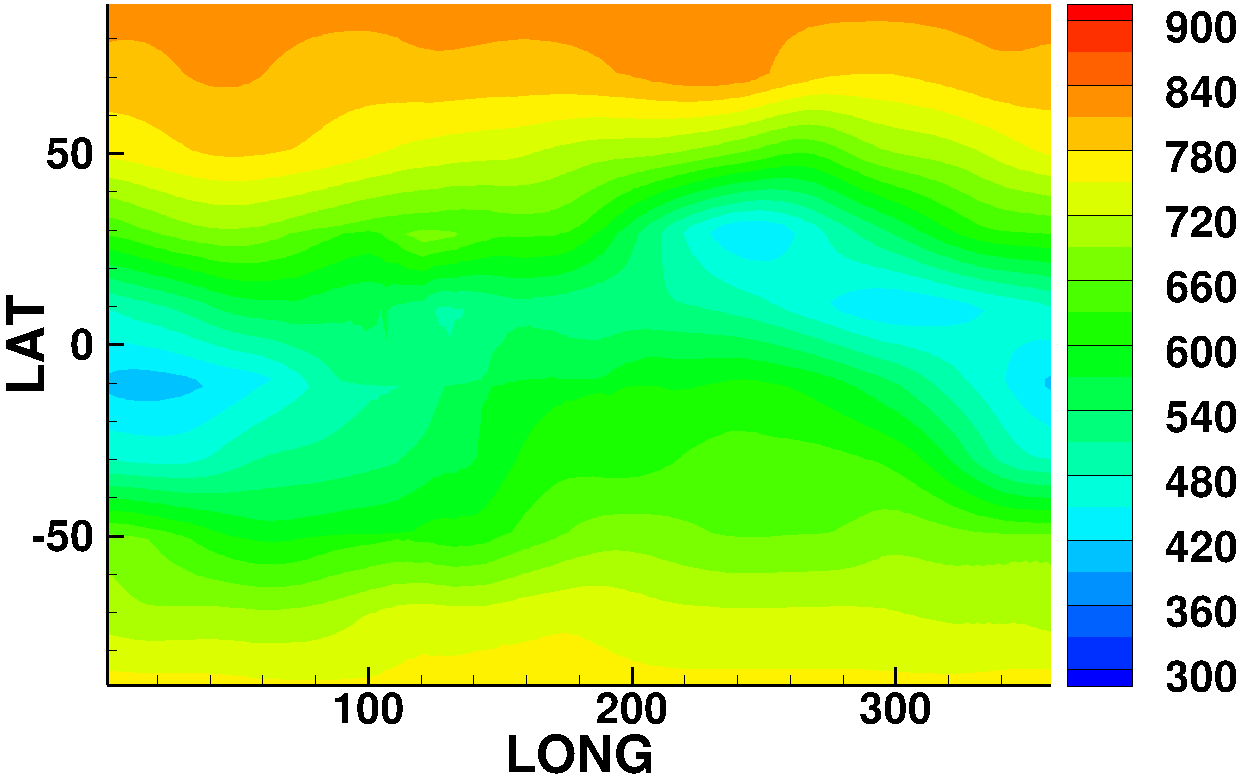}{0.33\textwidth}{V IPS (kms$^{-1}$)}
          }
\gridline{\fig{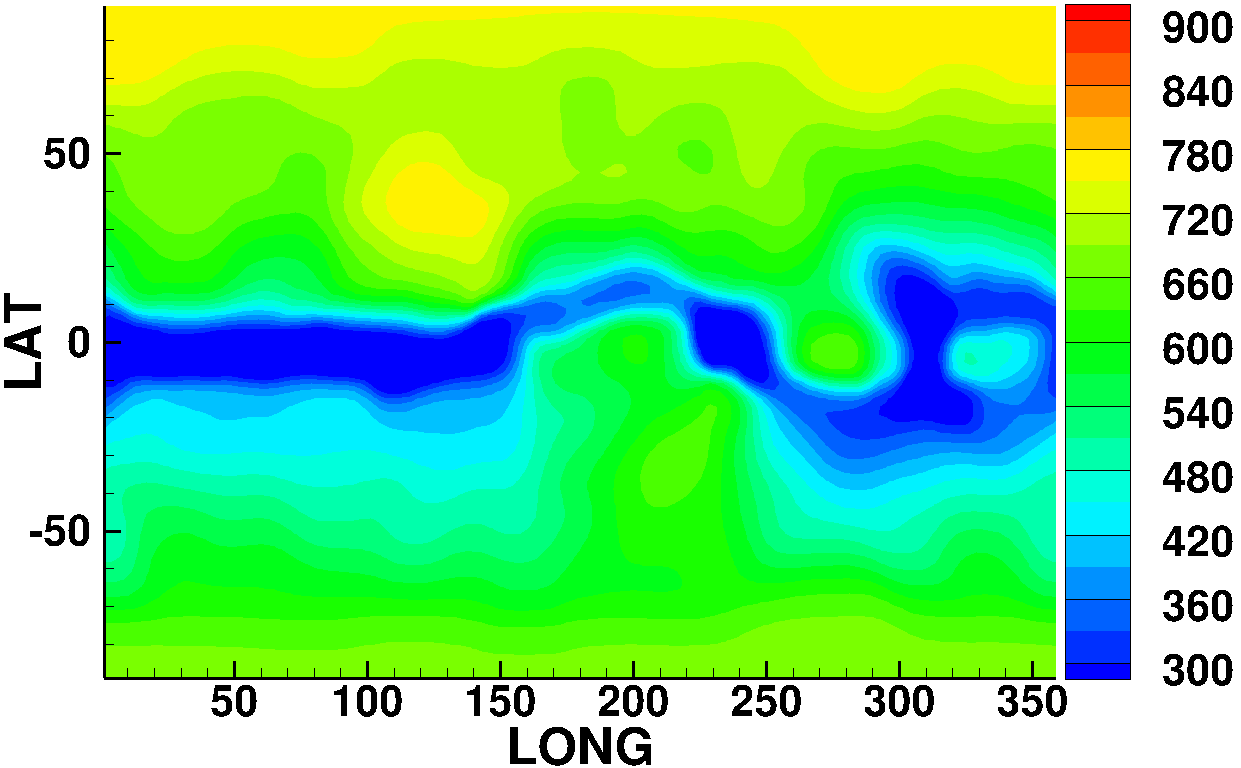}{0.33\textwidth}{V AWSoM (kms$^{-1}$)}
          \fig{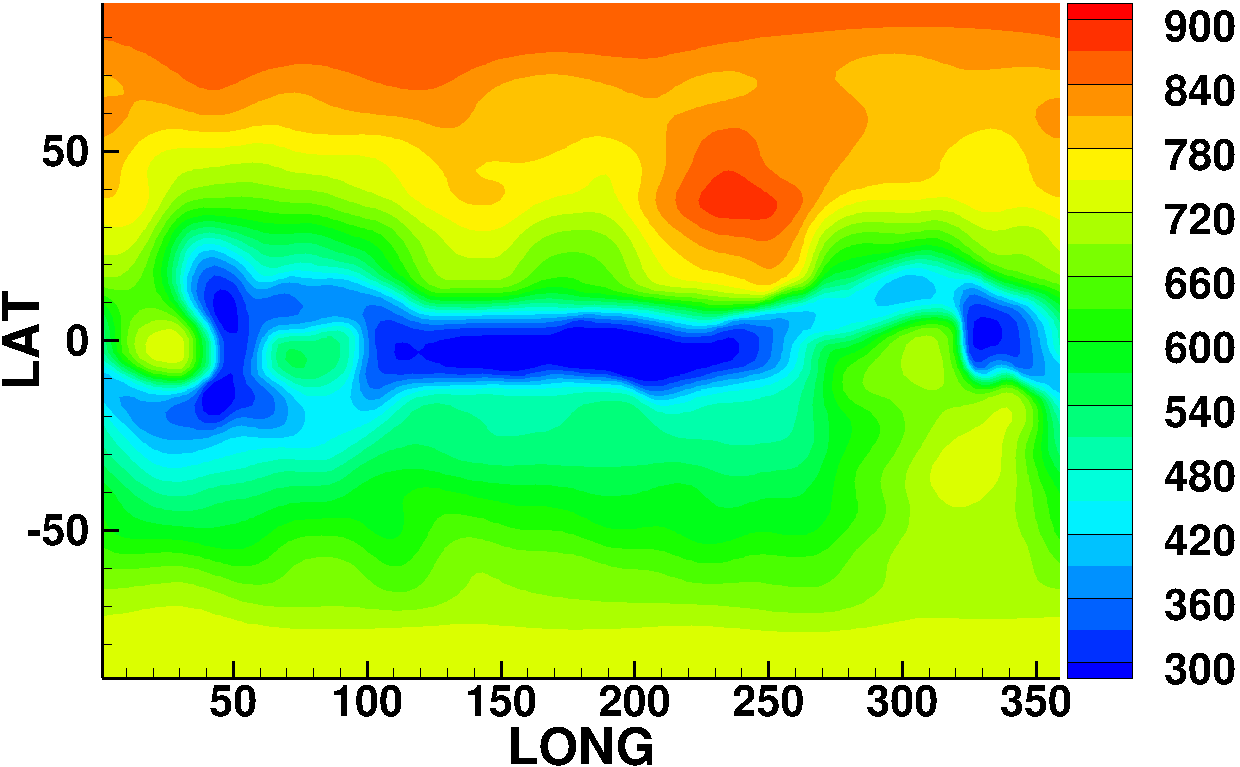}{0.33\textwidth}{V AWSoM (kms$^{-1}$)}
          \fig{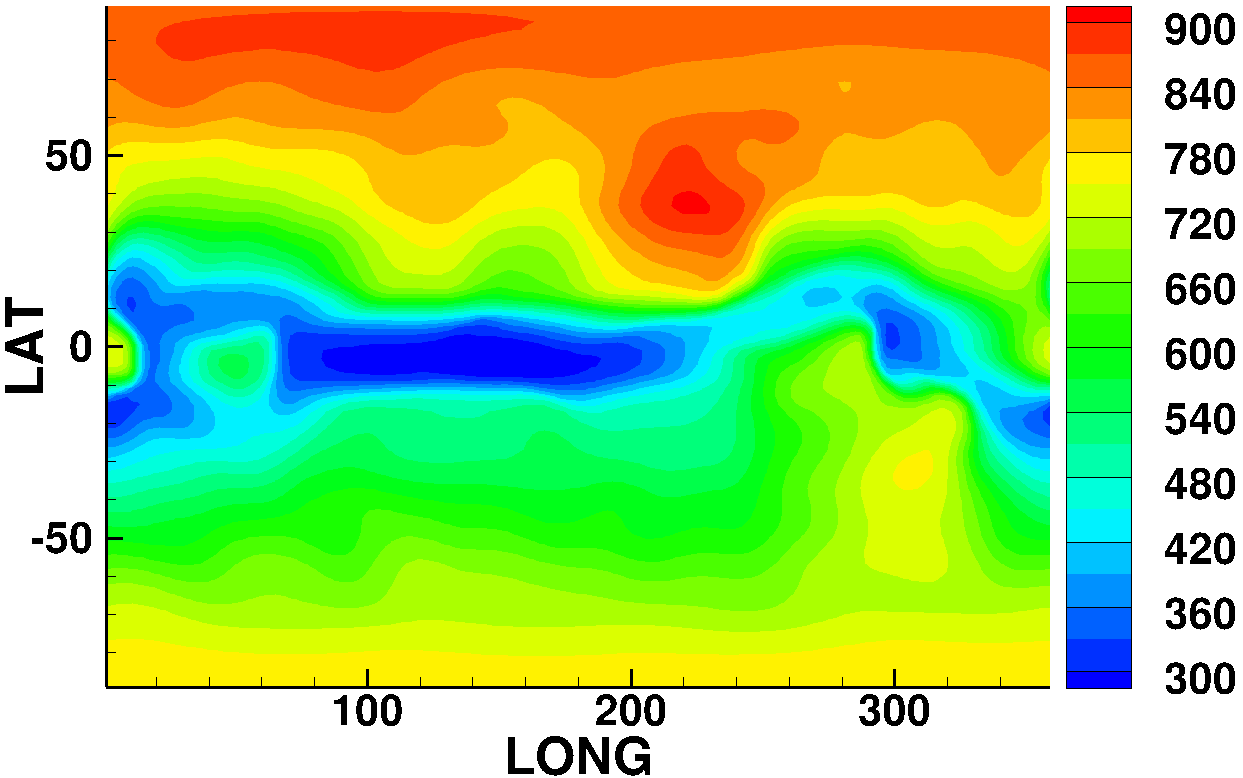}{0.33\textwidth}{V AWSoM (kms$^{-1}$)}
          }
\gridline{\fig{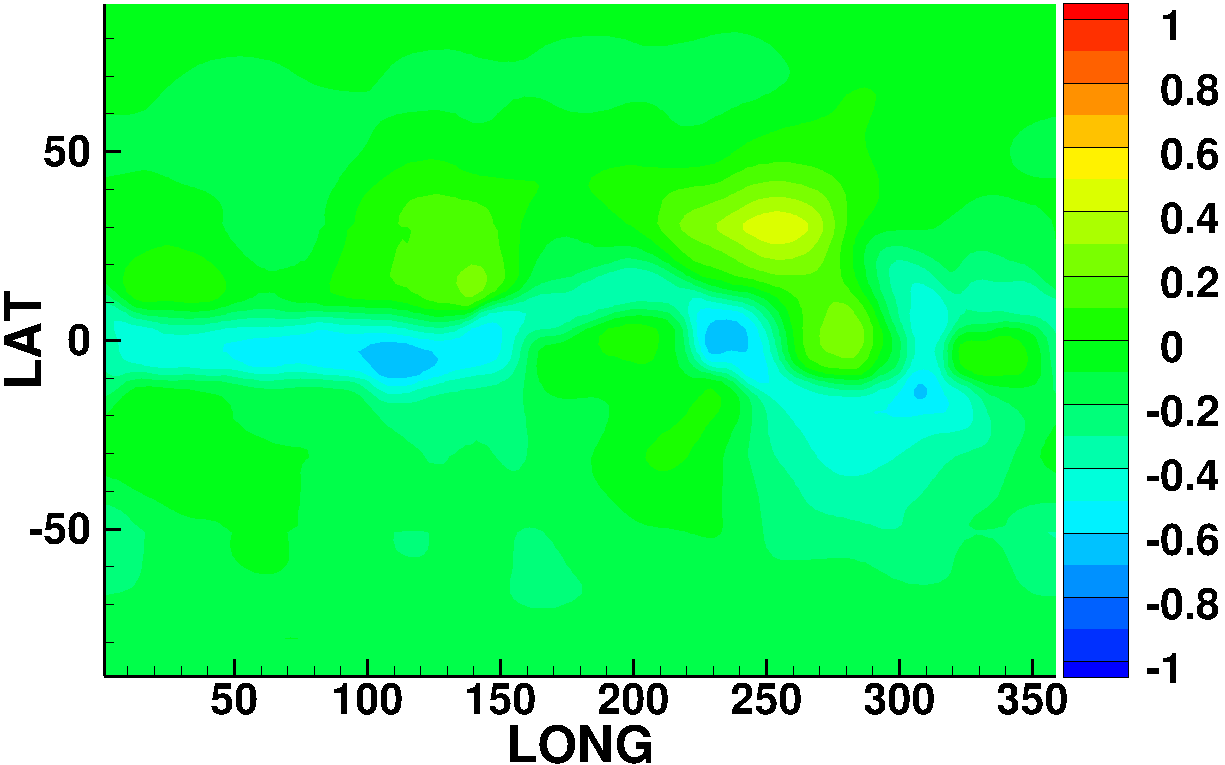}{0.33\textwidth}{V Rel Diff}
          \fig{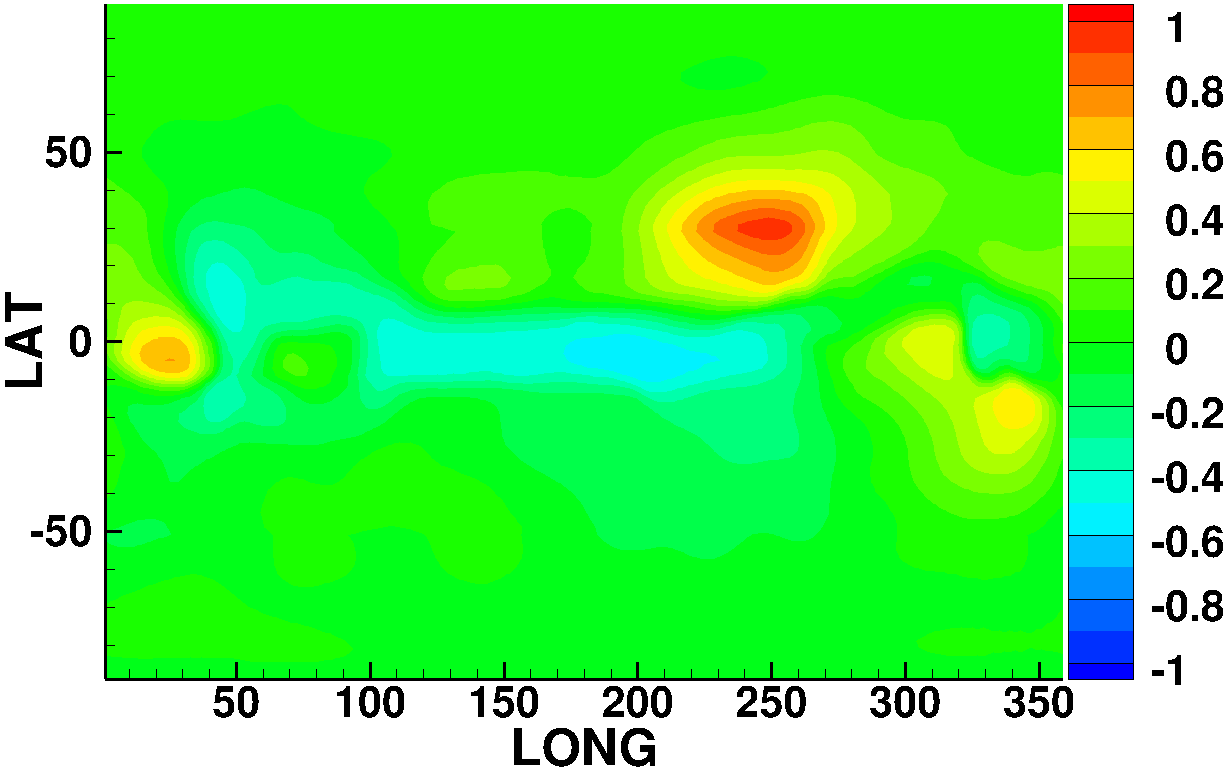}{0.33\textwidth}{V Rel Diff}
          \fig{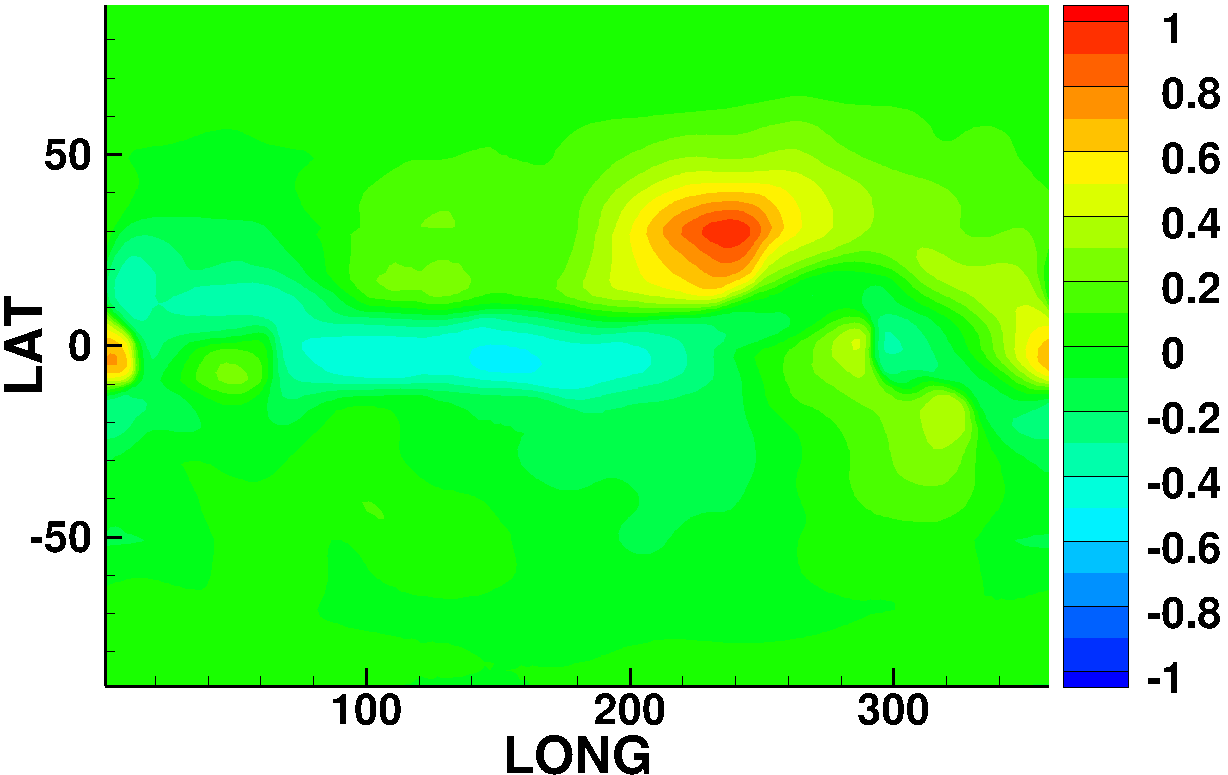}{0.33\textwidth}{V Rel Diff}
          }
\gridline{\fig{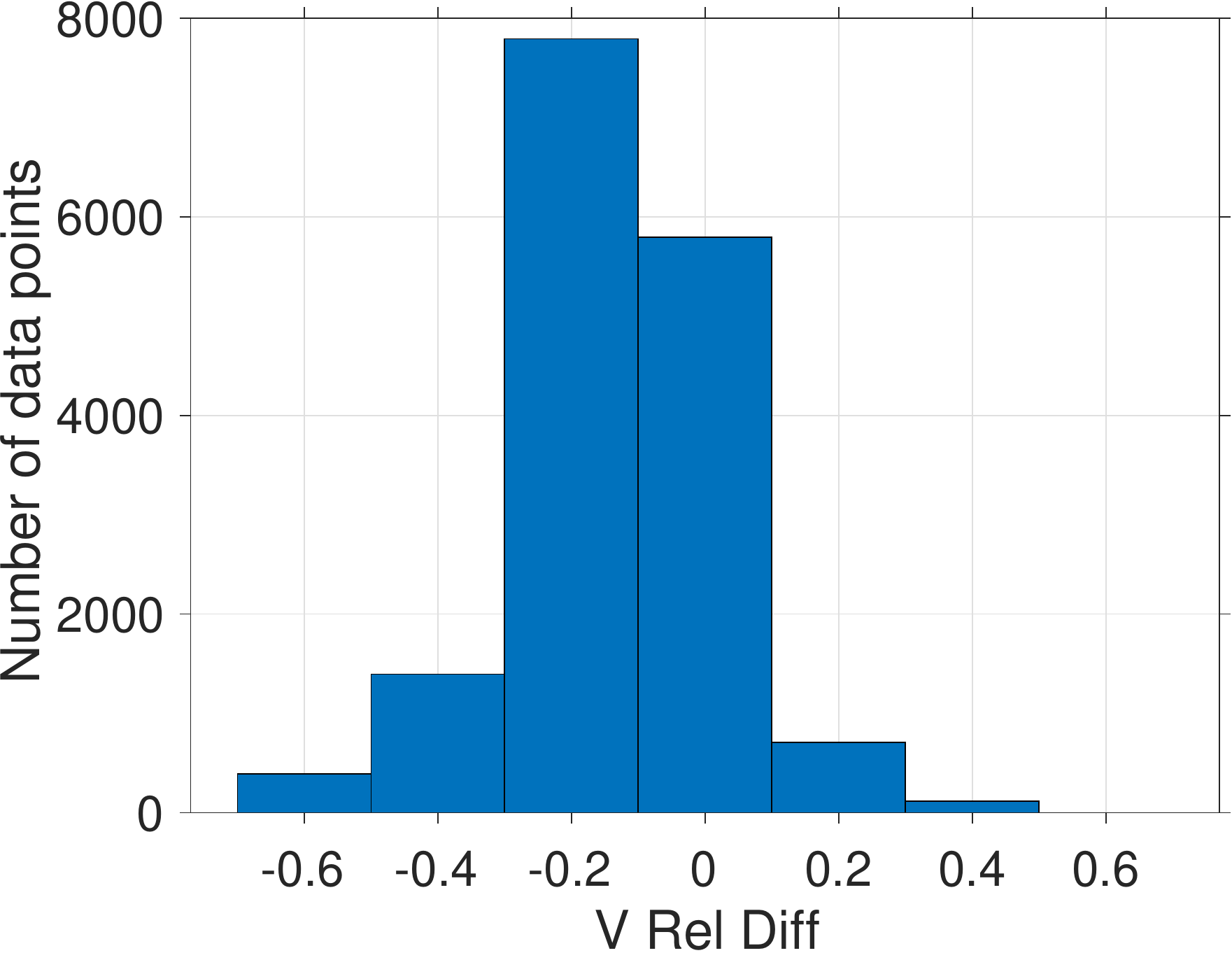}{0.33\textwidth}{(a) 20 \Rs}
          \fig{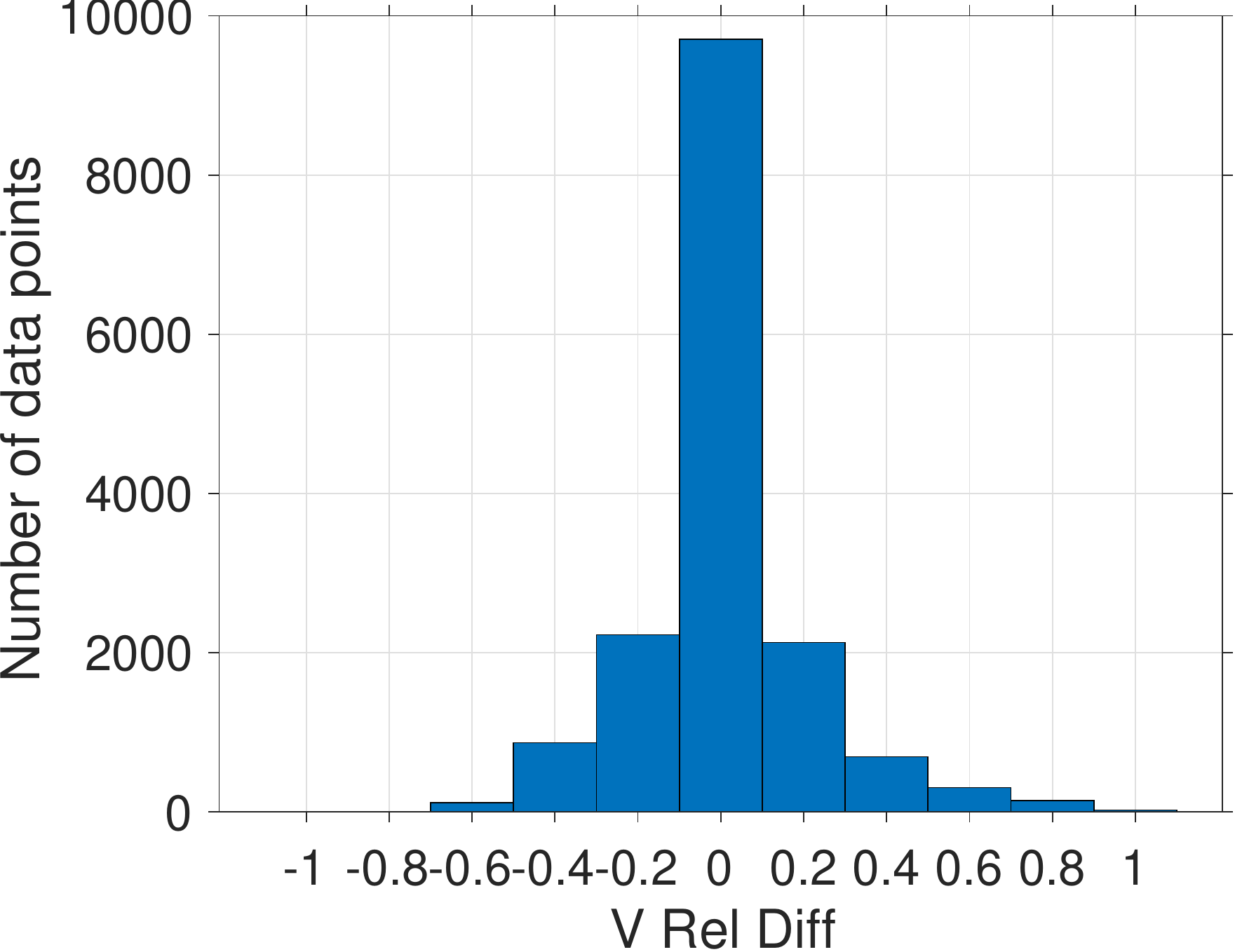}{0.33\textwidth}{(b) 100 \Rs}
          \fig{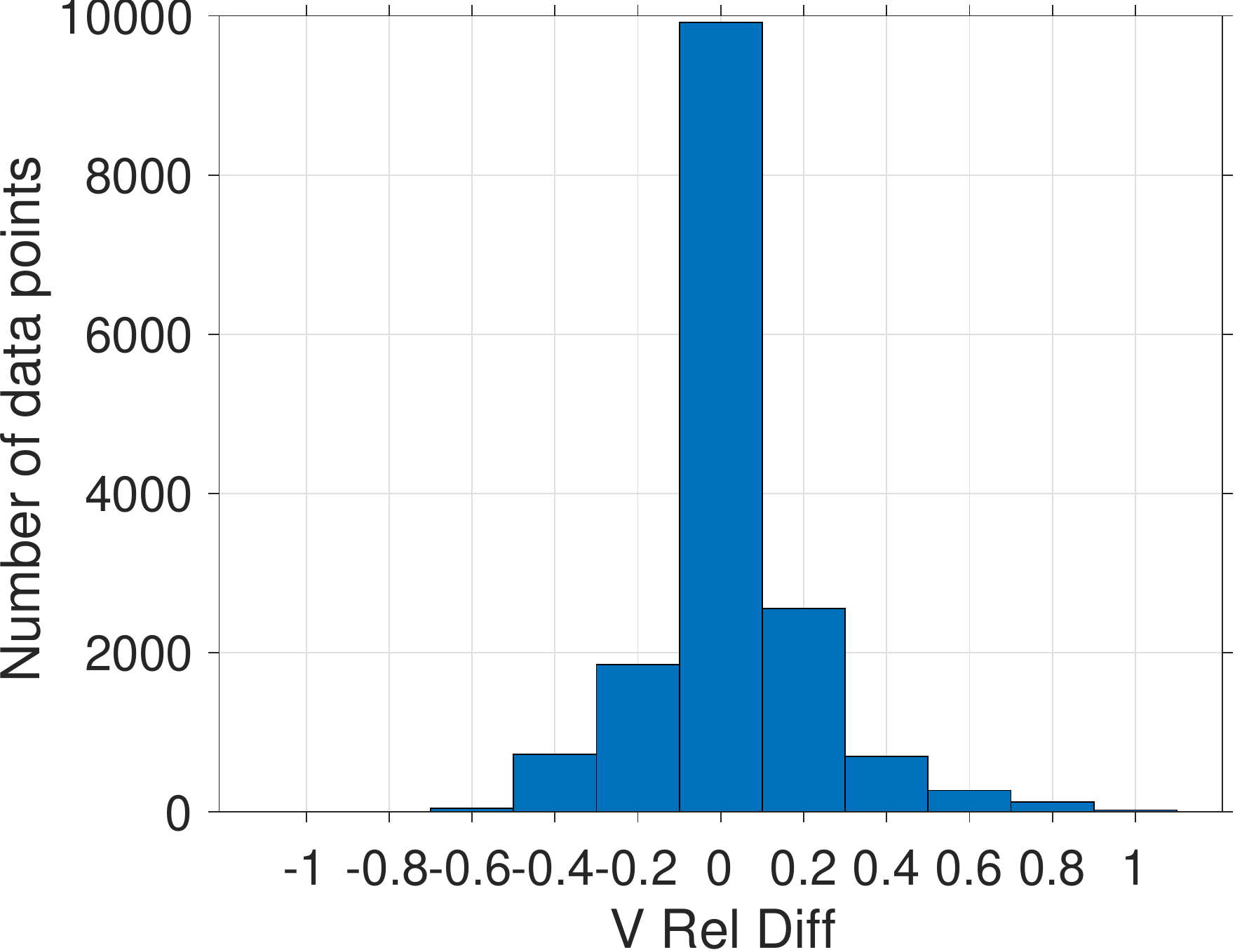}{0.33\textwidth}{(c) 1 AU}
          }
\caption{Comparison of reconstructed IPS velocity with AWSoM model simulations for CR2209 at three radial distances. The three columns correspond to results at (a) 20\,\Rs, (b) 100\,\Rs~and (c) 1 AU respectively. The following quantities are shown in each succeeding row - IPS reconstructed solar wind velocity in kms$^{-1}$ (V IPS), AWSoM predicted velocity in kms$^{-1}$ (V AWSoM), relative velocity difference between IPS observations and model output, \rm{V Rel Diff}=\big($\frac{V_{AWSoM}}{V_{IPS}}-1\big)$ and the histogram which shows how the relative difference is distributed.
\label{fig:ips_09}}
\end{figure*}

Figures \ref{fig:ips_08} and \ref{fig:ips_09} show the velocity comparisons of AWSoM model results for CR2208 and CR2209 respectively, with the IPS reconstructions at three radial distances, 20\,\Rs, 100\,\Rs~and 1 AU. At each distance, the first row shows the IPS reconstructed velocity (V IPS in km s$^{-1}$) and the second row shows the AWSoM model simulated velocity (V AWSoM in km s$^{-1}$). The third and fourth rows show the longitude-latitude maps and the histogram, respectively, of the relative difference in the velocity, given by the quantity 
\rm{V Rel Diff}$=\big(V_{AWSoM}/V_{IPS}\big)-1$. Each column depicts the results corresponding to (a) 20\,\Rs, (b) 100\,\Rs~and (c) 1 AU. The radial evolution of velocities can also be seen from the figures. The major difference between AWSoM and IPS velocities arises in the low latitude regions, which is where the heliospheric current sheet is located. The histograms indicate that the relative difference is very close to zero, that is, the model predictions agree quite well with the IPS reconstructions, specially at 100\,\Rs  and 1 AU. At 20\,\Rs, the agreement is within $20\,\%\,-\,30\,\%$. In particular, the excellent agreement near the poles corrects the large discrepancy found in previous AWSoM models in the inner heliosphere \citep{Jia2016}. We also see that the model predicts slower solar wind speeds in the southern hemisphere compared to the northern hemisphere which can be attributed to the input magnetic field maps that show asymmetric north and south polar regions.

The IPS data shown here is averaged over the entire Carrington rotation for each radial distance. Data from remotely-sensed IPS is the best near the Earth, since this is where the lines of sight emanate from, and the resolution of the tomography is only about 20 $\times$ 20 degrees in longitude and latitude. Therefore, the analysis gets worse away from Earth. The analysis fits the in-situ observations at Earth, but the OMNI data uses a mix of DSCOVR and ACE data, and sometimes these data sets differ greatly from one for another even at these low resolutions by a factor of 2 or sometimes more \citep{Lug2018}.
%Velocities are better than densities. I have found that to fit values at Earth the fast events (CMEs) don’t decelerate fast enough to agree with their onset times or the IPS. The actual deceleration is not simply a matter of mass and mass flux conservation of our kinematic modeling. I think this is a topic for a lot of discussion.

%{\bf In general a  $r^{-2}$ fall-off relative to 1 AU is applied to densities. This is done to make structures look somewhat similar whether they are viewed near the Sun or far from it. 
%A fall-off other than this $r^{-2}$ is sometimes different in MHD analysis, since often the solar wind is thought to accelerate beyond 20 Rs, at least in the quiet solar wind. Acceleration of this type (if it exists) would imply that densities should be somewhat higher at 20RS  than presented in the  $r^{-2}$ fall-off corrected in order to fit densities at 1 AU; the IPS kinematic model does not take this type acceleration into account.}

\subsection{OMNI data}\label{sec:OMNI}
\begin{figure*}[ht!]
\plottwo{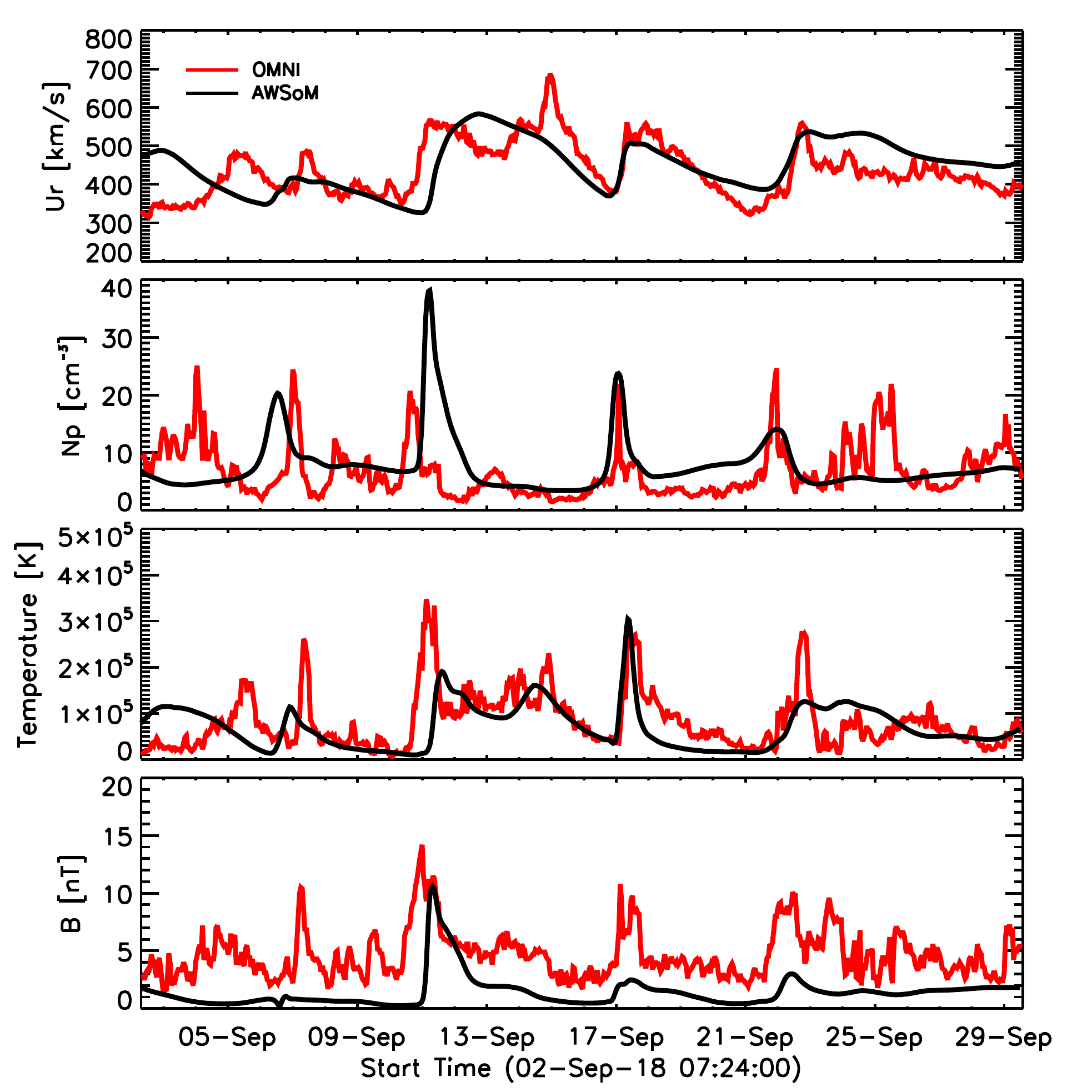}{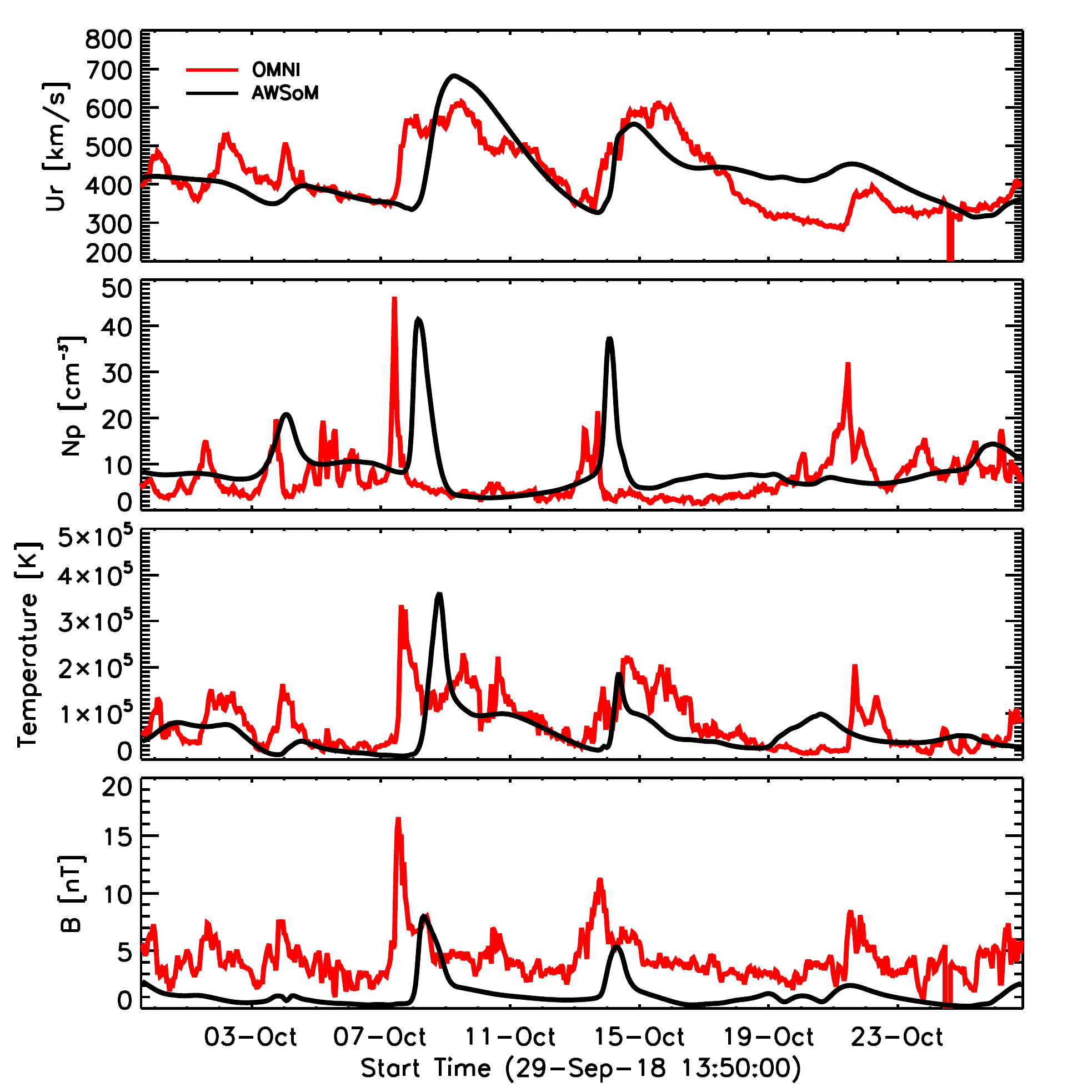}
% \put(-119.8,215.9){{\bf CR2209}}
 % \put(-365.8,215.9){{\bf CR2208}}
\caption{OMNI data (red) and AWSoM simulated solar wind parameters based on one realization of the ADAPT maps (black) at 1 AU for CR2208 (left) and CR2209 (right).}
\label{fig:omni}
\end{figure*}

We compare the model predicted solar wind properties at 1 AU with satellite observations using data from the OMNI database of the National Space Science Data Center [NSSDC].
Figure \ref{fig:omni} shows the comparisons of simulation results at 1 AU for CR2208 and CR2209 with the hourly averaged OMNI data. The observation data set consists of near-Earth solar wind magnetic field and plasma parameter {\it in-situ} data measured by several missions in L1 (Lagrange point) orbit. These spacecraft include the {\it Advanced Composition Explorer (ACE)}, WIND, and Geotail. The spatial distance between the location of the L1 point and the Earth is taken to be negligible on heliospheric scales.
Figure \ref{fig:omni} shows the comparison of radial flow speed (Ur), proton number density (Np), proton temperature and magnetic field strength (B) from OMNI data (red) with the AWSoM predicted results (black) at the end of the SC-IH simulations. 
%{\bf (Note: Both CR2208 and 09 in 2nd order scheme now.)}
We find that the model successfully reproduces the observed solar wind conditions at 1 AU. Most of the peaks in density, temperature and magnetic field are successfully reproduced. The AWSoM results systematically overestimate the proton density for both Carrington rotations and underestimate the magnetic field. However, the overall flow speeds match reasonably well with the observations. The model also reproduces the Co-rotating Interaction Regions (CIR's) represented as peaks in the density and temperature parameters quite well.

\begin{figure*}[ht!]
\plottwo{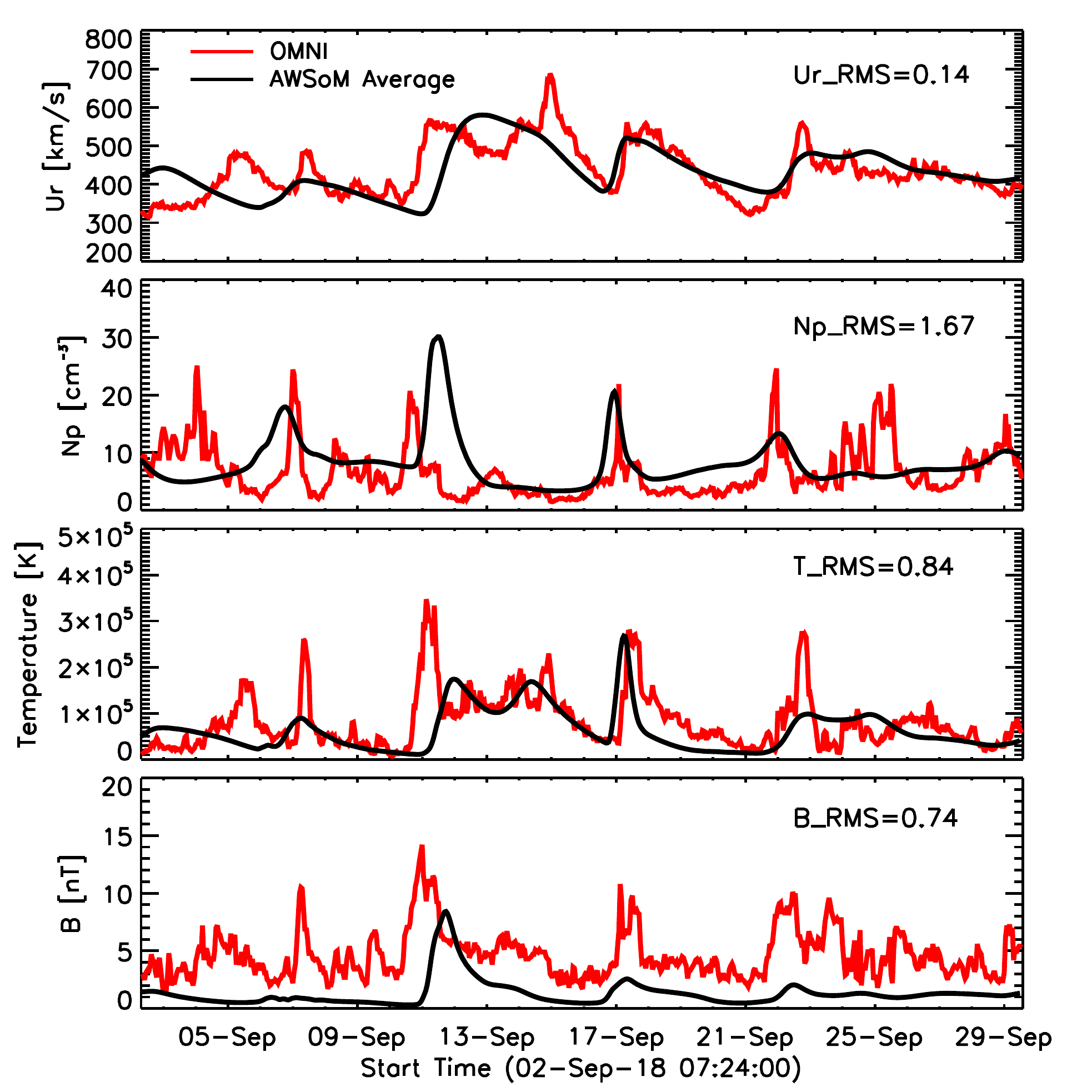}{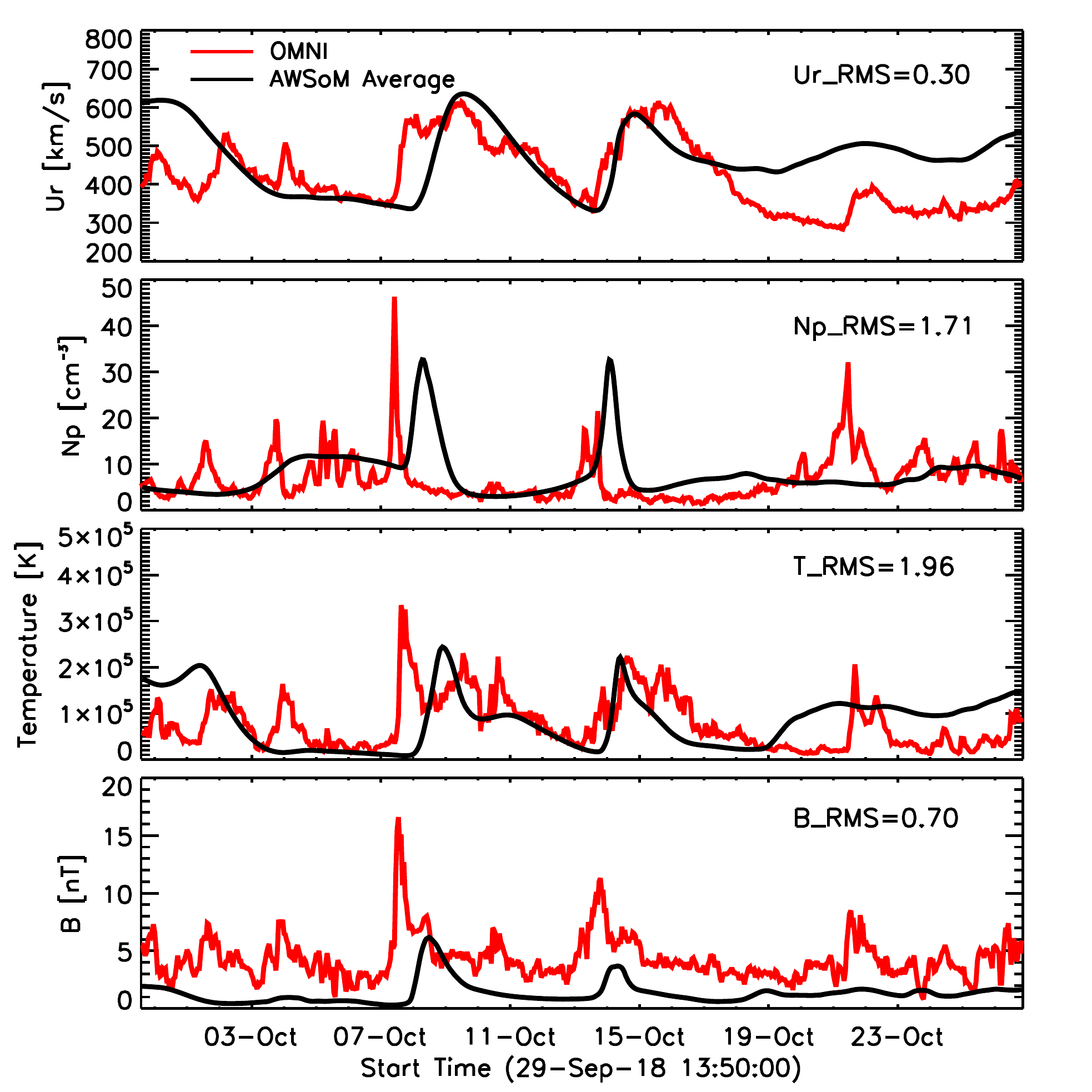}
\caption{Simulation results averaged over all realizations of ADAPT-GONG maps (black) compared with OMNI observations (red) at 1 AU. The corresponding \rm{RMSE} values for each parameter are informed in each panel for both CR2208 and CR2209.}
\label{fig:omniavg}
\end{figure*}

\begin{figure*}[ht!]
\plottwo{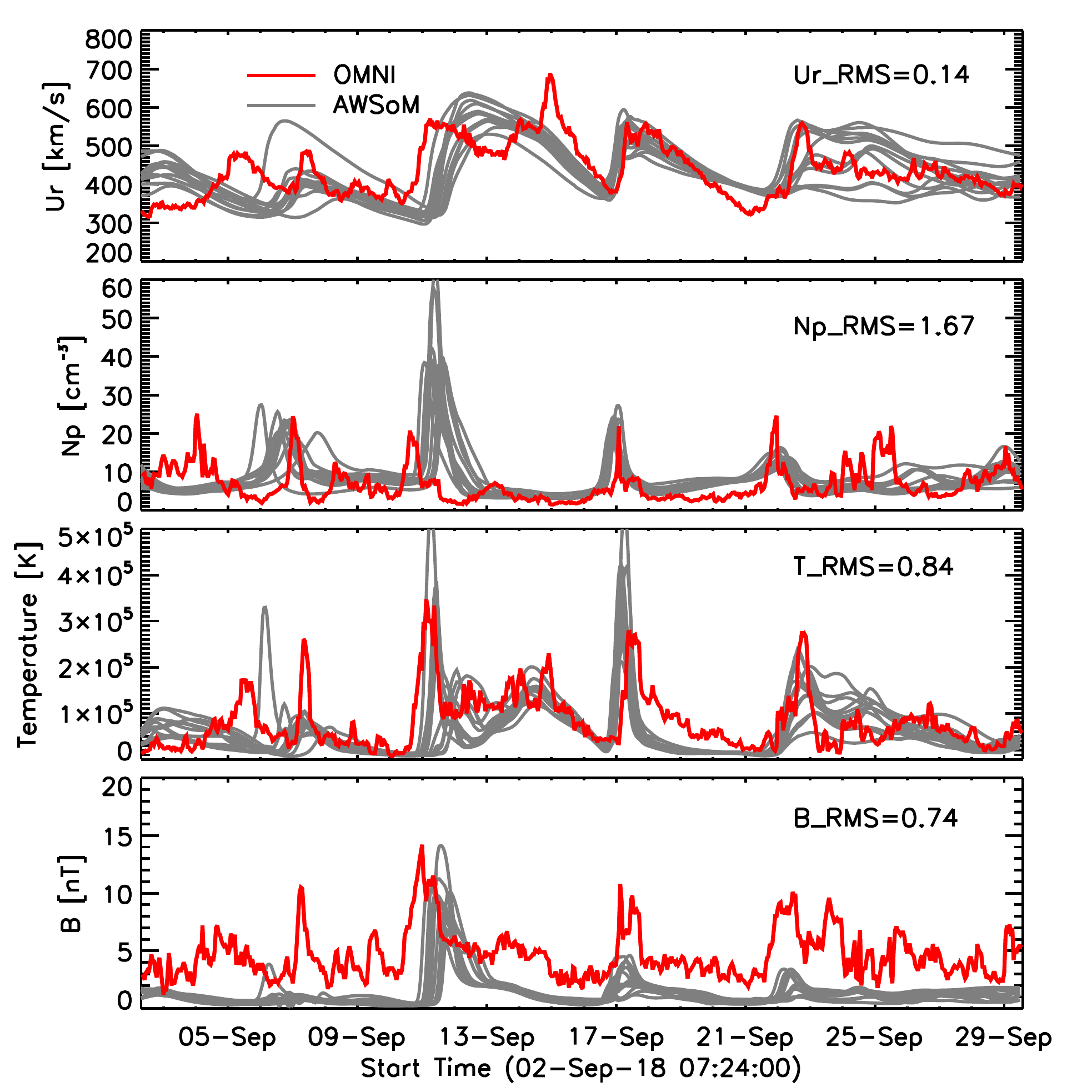}{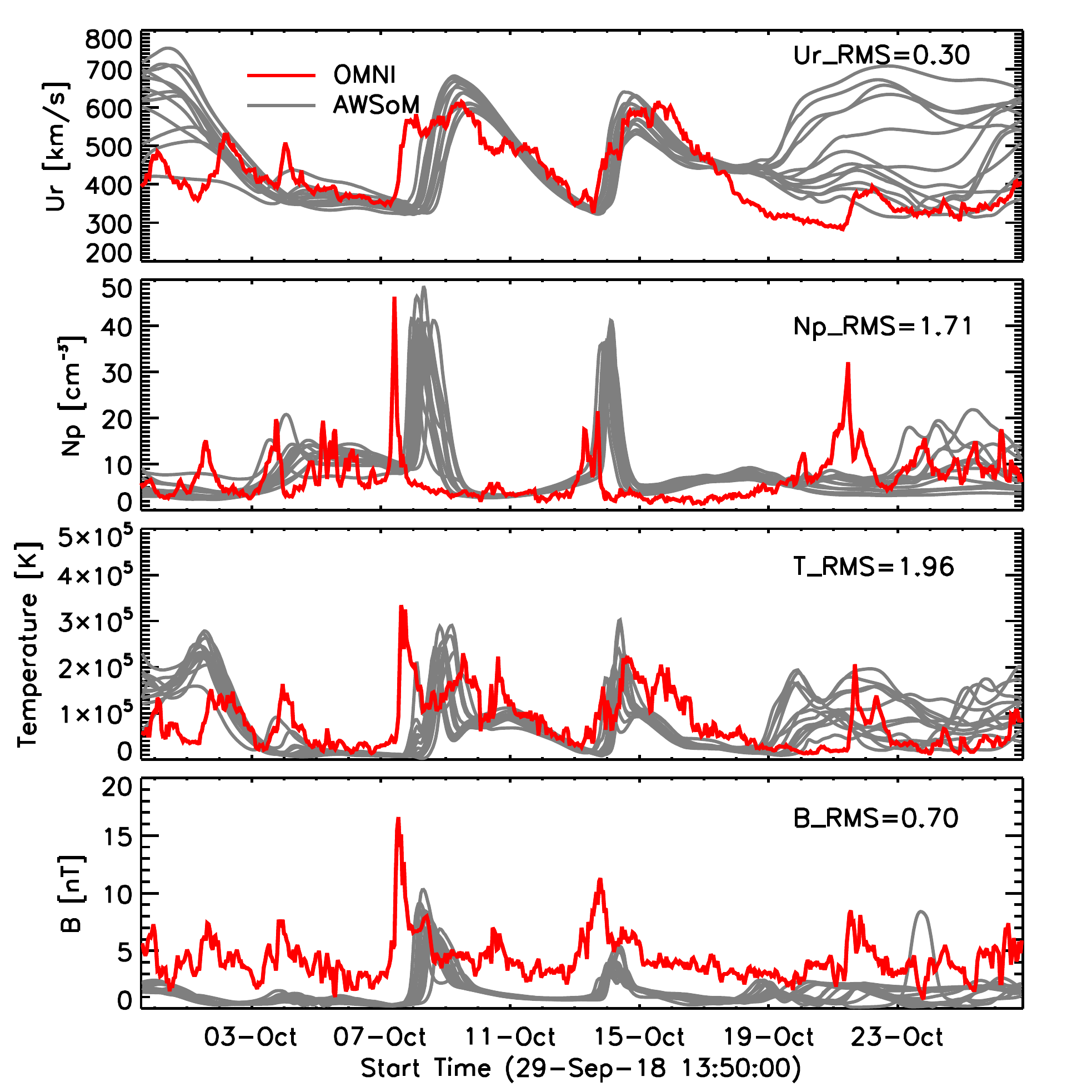}
\caption{Simulation results from all 12 realizations of ADAPT maps (grey) compared with OMNI observations (red) at 1 AU for both CR2208 and CR2209.}\label{fig:omniall}
\end{figure*}

ADAPT-GONG maps have multiple realizations of the global magnetic field maps. For each rotation, the simulation results shown in Figure \ref{fig:omni} are based on one realization of the ADAPT-GONG maps (shown in Figure \ref{fig:maps}). Figure \ref{fig:omniavg} displays the comparison of the OMNI data (red) with the average of simulations using all 12 realizations of the ADAPT-GONG maps (black). To quantify the uncertainty in the simulation results due to the different realizations of the ADAPT maps, each panel in Figure \ref{fig:omniavg} indicates the root mean square (RMS) error between the observed (OMNI data) and the average of the simulation results using all 12 realizations of the ADAPT-GONG maps for each of the rotations. For each observed plasma parameter $q$, we calculate the relative RMS error as
\begin{equation*}
\rm{RMS}= \sqrt{\frac{1}{n}\sum_{t=1}^{n}\bigg(\frac{q(t)-\bar q(t)}{q(t)}\bigg)^{2}}, 
\end{equation*}
where $\bar q$ denotes the average of the simulation results based on all 12 ADAPT-GONG realizations. The plot shows OMNI data (red) and the average of all ADAPT map results (black) for all plasma parameters. The small RMS values indicate that our model fits the observations quite well. Figure \ref{fig:omniall} shows the comparison of the OMNI data (red) to the results of the AWSoM model runs based on all 12 realizations of the ADAPT maps (grey), individually. An increase in the ensemble velocity spread is typically because of different current sheet crossing times between the realizations. That is, when the current sheet has a notable north-south alignment, however, there can also be periods when the current sheet is very close to the ecliptic. Most of the difference within the ensemble is driven by the poles, which can greatly influence the current sheet position.

In general, the root mean square error
\begin{equation*}
    E = \sqrt{\frac{1}{T} \int_0^T dt\, [q_1(t)-q_2(t)]^2},
\end{equation*}
between model results $q_1(t)$ and observations $q_2(t)$ over a time period $T$ can be misleading if the curves have sharp peaks and are shifted relative to each other in time. Here $q$ corresponds to one of the quantities of interest: density, velocity, temperature or magnetic field. For example, in Figure \ref{fig:omni}, while the data and model results look reasonably close (for a single realization), the errors can be large because the peaks in density and temperature are shifted. %Therefore, although intuitively the results look good, a small shift in x-axis (time for our case) can create a large error in the y-axis comparisons since it is coordinate dependent. 
We have defined a measure that evaluates the deviation between model results and observations in a more intuitive manner. We define a distance $D$ between two curves in a plane that is independent of the coordinate system, so that the temporal and amplitude errors are treated the same way:
\begin{equation*}
    D = \frac{D_{1,2} + D_{2,1}}{2}
\end{equation*}
Here $D_{1,2}$ is the average of the minimum distance between two curves integrated along curve 1:
\begin{eqnarray*}
D_{1,2} &=&\frac{1}{L_1} \int_0^{L_1} dl_1 \\
    &&\min_{l_2} \sqrt{[x_1(l_1)-x_2(l_2)]^2 +[y_1(l_1)-y_2(l_2)]^2}
    \nonumber
\end{eqnarray*}
where, $l_1$ and $l_2$ are the coordinates along the two curves described by the $(x_1, y_1)(l_1)$ and $(x_2, y_2)(l_2)$ functions.
The lengths of the curves are $L_1$ and $L_2$. $D_{2,1}$ is defined similarly as the average minimum distance integrated along curve 2, so that $D$ is a symmetric function of the two curves. Since time and the quantities of interest have different physical units, one needs to normalize them to the $x$ and $y$ coordinates. We choose $X=10$ days as the normalization for time and $Y=\max(q) - \min(q)$ for the normalization of quantity $q$, so that $x=t/X$ and $y=q/Y$. This means that a time shift of 10 days is considered as bad as the difference between the smallest and largest amplitudes. We will use the above defined distance $D$ to characterize
the error between the observations and a particular model run.

\begin{figure*}[ht!]
\plottwo{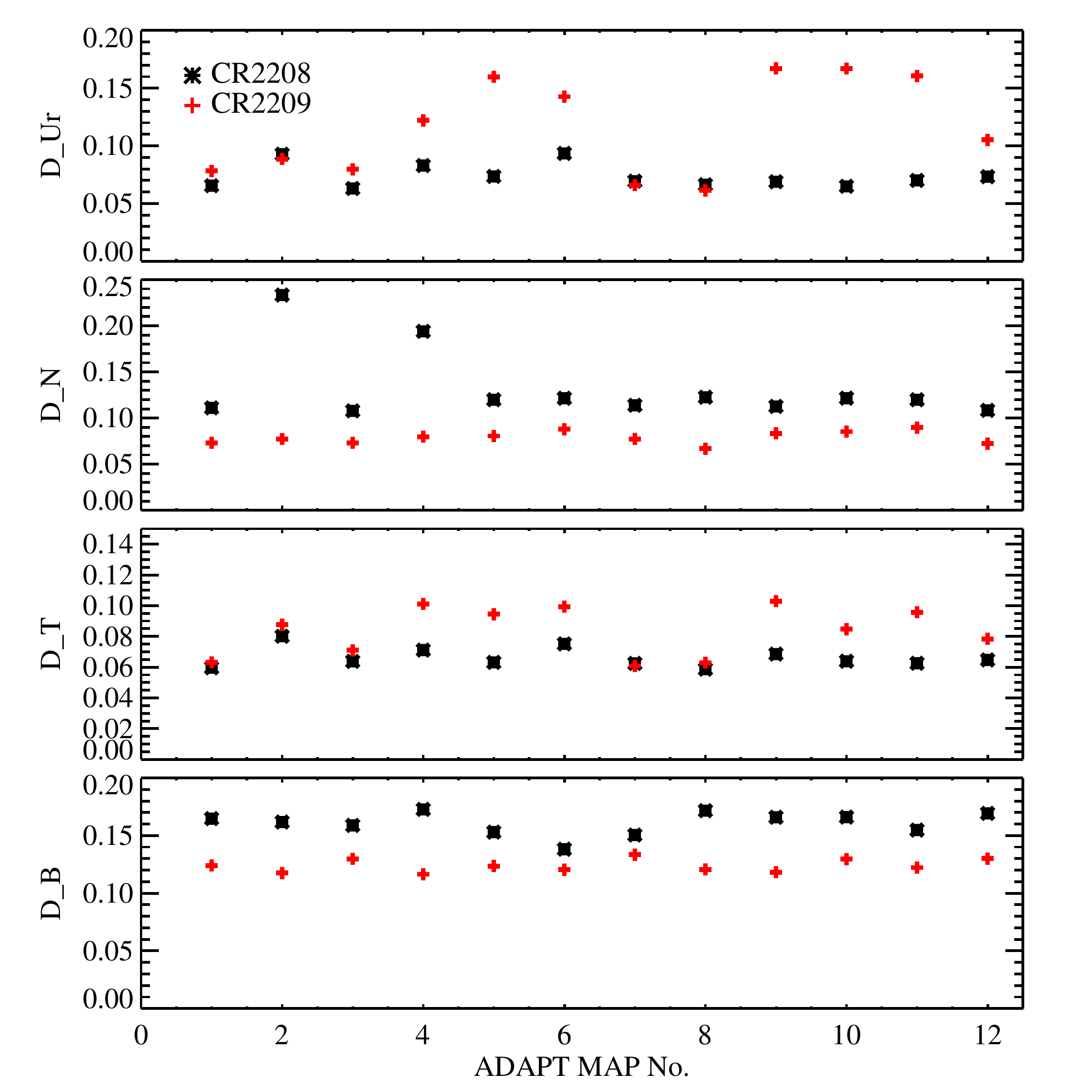}{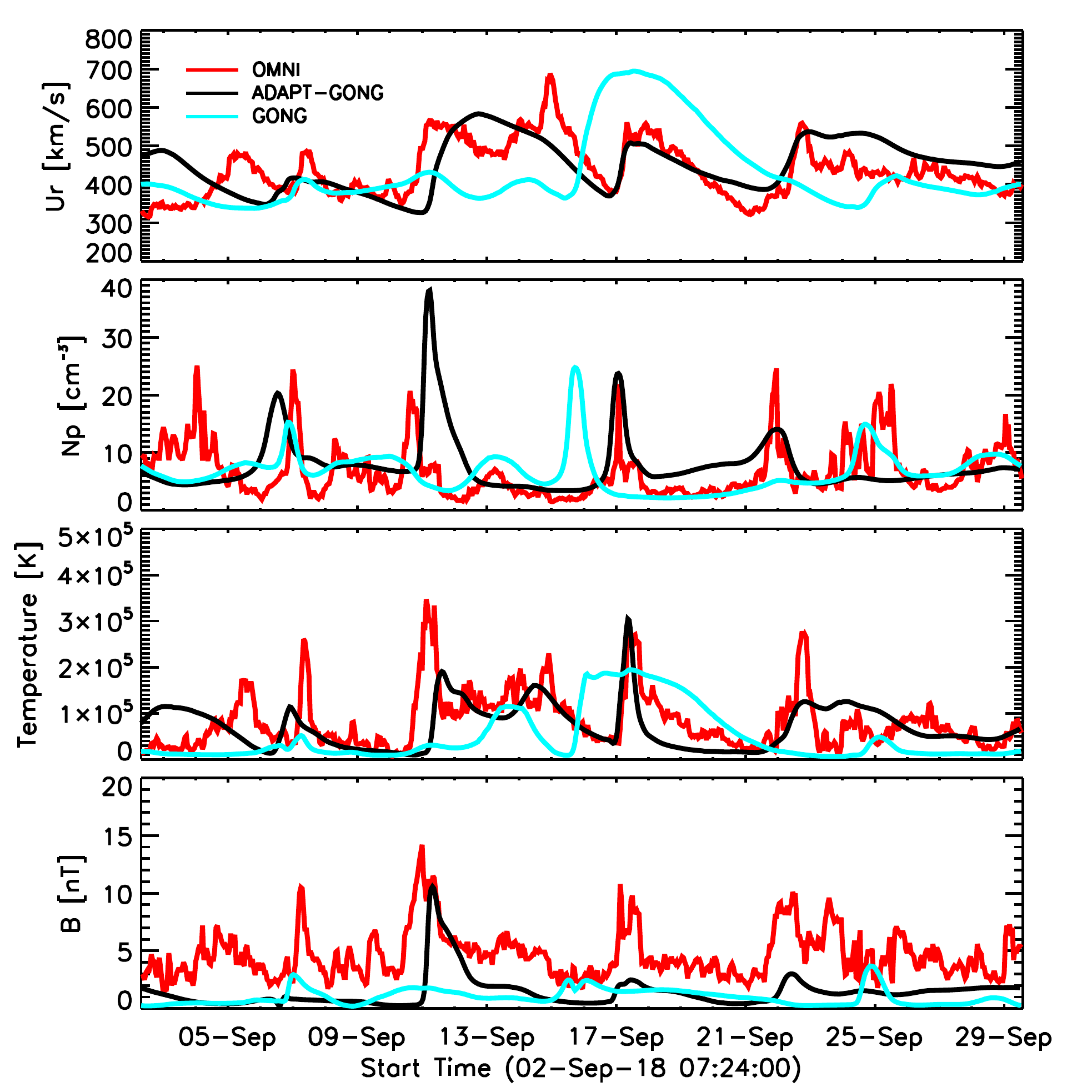}
\caption{Left panel: Distance $D$ between observations and model results for each ADAPT map realization for both CR2208 (black) and CR2209 (red). Right panel: OMNI data (red) and AWSoM simulated solar wind parameters at 1 AU for CR2208 using GONG magnetogram (cyan) and the `best' ADAPT-GONG map (black) as inputs.
}
\label{fig:ADAPT_compare}
\end{figure*}

The left panel of Figure \ref{fig:ADAPT_compare} plots the errors $D$ between the OMNI observations and the AWSoM model results for each plasma parameter for all 12 ADAPT-GONG map realizations for CR2208 (black) and CR2209 (red). 
%For completeness, we interchange the curves and normalize the sum of the Norms with sum of lengths of the two curves divided by 2. This is done for each solar wind parameter at 1 AU, using all ADAPT-GONG realizations for both Carrington rotations.
Tables \ref{tab:table1} and \ref{tab:table2} list the correlation values between the errors of all parameter pairs for CR2208 and CR2209, respectively. We find that the distances $D$ for solar wind velocity (Ur), temperature (T) and density (N) are strongly correlated within each Carrington rotation, in other words the success or failure of the model in reproducing these parameters is highly correlated. Interestingly, the errors of these three plasma parameters do not correlate with the magnetic field error. We do not find a strong correlation between the errors of the corresponding ADAPT map realizations for the two consecutive Carrington rotations either, as shown in Table \ref{tab:table3}. That is, based on the two rotations that we study in this work, it cannot be said with any certainty that a particular ADAPT realization in one rotation that produces the best results will also be the best choice for a subsequent rotation.

\begin{deluxetable}{c|c|c|c}
\tablecaption{Correlation for errors ($D$) between solar wind parameters for CR2208 \label{tab:table1}}
\tablehead{\colhead{CR2208}  & \colhead{Np} & \colhead{T} & \colhead{B}}
\startdata
Ur  & 0.69 & 0.89 & -0.36 \\
Np  & \nodata &0.74 & 0.22 \\
T   & \nodata & \nodata & -0.22 \\
\enddata
\end{deluxetable}

\begin{deluxetable}{c|c|c|c}
\tablecaption{Correlation for errors ($D$) between solar wind parameters for CR2209 \label{tab:table2}}
\tablehead{\colhead{CR2209} & \colhead{Np} & \colhead{T} & \colhead{B}}
\startdata
Ur & 0.84 & 0.83 & -0.25 \\
Np &  \nodata & 0.75 & -0.19 \\
T  &  \nodata & \nodata & -0.60 \\
\enddata
\end{deluxetable}
%%%%%%% TABLE FOR CROSSCORRELATIONS
\begin{deluxetable}{c|c|c|c}
\tablecaption{Correlation for errors between CR2208 and CR2209\label{tab:table3}}
\tablehead{\colhead{Ur} & \colhead{Np} & \colhead{T} & \colhead{B}}
\startdata
0.11 & 0.59 & 0.01 & -0.16 \\ 
\enddata
\end{deluxetable}

Finally, we compare the performance of AWSoM with ADAPT-GONG map and GONG synoptic map. The right panel of \ref{fig:ADAPT_compare} shows the 1 AU OMNI data (red) comparisons of simulation results for CR2208 using GONG synoptic map (cyan) and one realization of the ADAPT-GONG synchronic map (black). It is clearly seen that by using the ADAPT-GONG maps AWSoM is able to capture much more faithfully many features of the observational data time-series at 1 AU, which is its ultimate goal.

\section{Summary and Discussion}\label{sec:Summary}
In this study, we show the AWSoM simulation results for CR2208 and CR2209 (representing solar minimum conditions) and compare them to observations. The simulations cover the domain from the solar chromosphere to the 1 AU heliosphere. We compare our simulation results with a diverse set of observations ranging from low corona, into the inner heliosphere up to 1 AU. These multi-spacecraft observations include data from SDO/AIA and STEREO-A/EUVI near the sun, SOHO/LASCO, IPS and OMNI. As a result, we show comparisons at various heliocentric distances from as low as 1.055-1.205\,\Rs~using EUV tomographic data, at 4 and 5\,\Rs~using LASCO tomographic reconstructions, at 20\,\Rs, 100\,\Rs~and 1 AU using IPS reconstructions along with OMNI data at 1 AU.
%\textcolor{red}{The AWSoM material does not coincide with the introductory sentence of the summary. What follows from composition should be model/observation comparison details. The material works as a stand along paragraph, where the ADAPT paragraph naturally follows.} 
The key features of AWSoM include the following: (1) non-linear interaction of forward propagating and partially reflected Alfv\'{e}n waves leading to coronal heating due to turbulent cascade dissipation. The balanced turbulence at the apex of the closed field lines is also accounted for. (2) The model allows for anisotropic ion temperatures and isotropic electron temperature. (3) It uses the linear wave theory and nonlinear stochastic heating \citep{Cha2011} to distribute the turbulence dissipation to the coronal heating of these three temperatures. (4) For the isotropic electron temperature, the collisionless heat conduction is also included. There are no ad-hoc heating functions.

In the past, HMI and GONG maps were used to provide the magnetic field input to the solar wind models. Here, we use the ADAPT-GONG maps, which are obtained by data assimilation that includes physical transport processes on the Sun. As a result, ADAPT maps provide more realistic estimates of the photospheric magnetic fields especially in the polar regions. We find that near the Sun, the location and extent of coronal holes and active regions are reproduced reasonably well by the model as shown in the synthesized EUV images. The average brightness of the synthetic and observed EUV images are also comparable.

Moving outwards into the corona, we compare our model to 3D DEMT reconstructions of the coronal density and temperature. Here, we find that at the lowest heights (r$\approx$1.025 \Rs), the predicted density is elevated as an artificial extension of the transition region. However, we get excellent agreement (within $\pm 30\%$) for electron density and temperature at heights above 1.055\,\Rs. At heights, r=4\,\Rs~and 5\,\Rs, we compare the model with the 3D electron density provided by SRT using LASCO-C2 observations. Here, we find AWSoM densities accurately match the reconstructions in the northern hemisphere, while the southern hemisphere densities are significantly higher. Further into the heliosphere, the solar wind speed predicted by our model is is found to be within $\pm 20\%$ of the IPS reconstructed speeds at 20 \Rs, 100 \Rs, and 1 AU. 

We show the plasma parameters as predicted by our model for each of the ADAPT-GONG realizations used as input. For both Carrington rotations, the proton density and temperature and solar wind speed are well-predicted by the model. However, we see that the magnetic fields are under-estimated \citep{Lin2017}, and contrary to observations, the solar wind continues to accelerate in the inner heliosphere, even up to 1 AU. This may be due to an overestimation of wave energy in our model. \citet{Suz2006} describe the Alfv\'{e}n wave dissipation by addressing the mode conversion of Alfv\'{e}n waves into slow (magnetoacoustic) waves. In our present model, we do not include this mode conversion and put the energy back into Alfv\'{e}n waves. This can result in excess Alfv\'{e}n wave energy which can lead to too much acceleration of the solar wind in the inner heliosphere.

%We also see the the solar wind parameters (e.g., velocity, density, temperature and magnetic field) based on any given individual ADAPT-GONG map realization over the time scales of a rotation. %We also see that the evolution of the performance (e.g., magnetic field and velocity) of any given individual realization changing on time scales of a rotation, that is, depending on the current sheet configuration. 

%The excellent agreement of the model with observations and reconstructed physical quantities covering orders of magnitude is a significant achievement for the AWSoM model and the physical theories upon which it is based. (said well enough below)

We have shown the success of our model in reproducing the solar minimum conditions throughout the corona and inner heliosphere. These encouraging results with the AWSoM model show it to be a valuable tool to simulate solar minimum conditions. This work represents the achievement of the theoretical turbulence-based model, where self-consistent treatment of the physical processes can reproduce coronal and heliospheric observations over a tremendous range of conditions spanning orders of magnitude in density, temperature and field strength. While this work describes the solar minimum conditions, our next validation work will focus on solar maximum conditions.
%This work is focused on solar minimum conditions. 

%x\section*{Acknowledgements}
This work was primarily supported by the NSF PRE-EVENTS grant no. 1663800. W.M. was supported by  NASA grants 80NSSC18K1208 NNX16AL12G 80NSSC17K0686. We acknowledge high-performance computing support from Cheyenne (doi:10.5065/D6RX99HX) provided by NCAR's Computational and Information Systems Laboratory, sponsored by the National Science Foundation and Frontera from NSF. W.M. acknowledges the NASA supercomputing system Pleiades. This work utilizes data produced collaboratively between Air Force Research Laboratory (AFRL) and the National Solar Observatory. The ADAPT model development is supported by AFRL. B. J. and H.-S Y. thank M. Tokumaru and the ISEE, Japan, IPS group for making their data available for use in comparisons. B. J. is funded by NASA LWS grant 80NSSC17K0684 and AFOSR FA9550-19-1-0356 to UCSD. D.G.Ll. acknowledges CONICET doctoral fellowship (Res. Nro. 4870) to IAFE that supported his participation in this research. A.M.V. acknowledges ANPCyT grant 2012/0973 to IAFE that partially supported his participation in this research.

%\newpage

\end{document}